\RequirePackage{rotating}
\documentclass[usenatbib]{mnras}
\usepackage{newtxtext,newtxmath}
\usepackage[T1]{fontenc}
\DeclareRobustCommand{\VAN}[3]{#2}
\let\VANthebibliography\thebibliography
\def\thebibliography{\DeclareRobustCommand{\VAN}[3]{##3}\VANthebibliography}
\usepackage{graphicx}	
\usepackage{amsmath}	
\usepackage{orcidlink}
\usepackage{gensymb}
\usepackage{longtable}
\usepackage{multirow}
\usepackage{float}
\usepackage{enumitem}
\usepackage{caption}
\usepackage{multirow}
\usepackage{siunitx}
\usepackage{pdflscape}
\usepackage{capt-of}

\newcommand{\teff}{$T_{\rm eff}$}

\newcommand{\rsun}{$R_{\odot}$}
\newcommand{\msun}{$M_{\odot}$}

\newcommand{\rearth}{$R_{\earth}$}

\newcommand{\kepler}{\textsl{Kepler} }
\newcommand{\gaia}{\textsl{Gaia} }

\title[Mutual Inclinations of Planet-Hosting Triples]{Orbital Architectures of Planet-Hosting Binaries III. Testing Mutual Inclinations of Stellar and Planetary Orbits in Triple-Star Systems \thanks{Based partly on observations obtained with the Hobby-Eberly Telescope, which is a joint project of the University of Texas at Austin, the Pennsylvania State University, Ludwig-Maximilians-Universität München, and Georg-August-Universität Göttingen.}}

\author[E. Evans et al.]{Elise L. Evans$^{1}$\thanks{E-mail: elise.evans@ed.ac.uk}\,\orcidlink{0009-0009-6539-2432},
Trent J. Dupuy$^{1}$\,\orcidlink{0000-0001-9823-1445},
Kendall Sullivan$^{2}$\,\orcidlink{0000-0001-6873-8501},
Adam L. Kraus$^{3}$,
Daniel Huber$^{4,5}$\,\orcidlink{0000-0001-8832-4488}, \newauthor
Michael J. Ireland$^{6}$,
Megan Ansdell$^{7}$\,\orcidlink{0000-0003-4142-9842},
Rajika L. Kuruwita$^{8}$\,\orcidlink{0000-0002-9236-2919},
Raquel A. Martinez$^{9}$\,\orcidlink{0000-0001-6301-896X}, \newauthor
Mackenna L. Wood$^{10}$\,\orcidlink{0000-0001-7336-7725}
\\
$^{1}$Institute for Astronomy, University of Edinburgh, Royal Observatory, Blackford Hill, Edinburgh, EH9 3HJ, UK\\
$^{2}$Department of Astronomy and Astrophysics, University of California Santa Cruz, Santa Cruz, CA, 95064, USA \\
$^{3}$Department of Astronomy, The University of Texas at Austin, 2515 Speedway C1400, Austin, TX 78712, USA\\
$^{4}$Institute for Astronomy, University of Hawaii, 2680 Woodlawn Drive, Honolulu, HI 96822, USA\\
$^{5}$Sydney Institute for Astronomy (SIfA), School of Physics, University of Sydney, NSW 2006, Australia\\
$^{6}$Research School of Astronomy and Astrophysics, Australian National University, Canberra 2611, Australia\\
$^{7}$NASA Headquarters, 300 E Street SW, Washington, DC, 20546, USA\\
$^{8}$Heidelberg Institute for Theoretical Studies, Schlo{\ss}-Wolfsbrunnenweg 35, 69118, Heidelberg, Germany\\
$^{9}$Department of Physics and Astronomy, 4129 Frederick Reines Hall, University of California, Irvine, CA 92697, USA\\
$^{10}$MIT Kavli Institute for Astrophysics and Space Research Massachusetts Institute of Technology, Cambridge, MA 02139, USA
}

\date{3 September 2024}
\pubyear{2024}
\begin{document}
\label{firstpage}
\pagerange{\pageref{firstpage}--\pageref{lastpage}}
\maketitle

\begin{abstract}
Transiting planets in multiple-star systems, especially high-order multiples, make up a small fraction of the known planet population but provide unique opportunities to study the environments in which planets would have formed. Planet-hosting binaries have been shown to have an abundance of systems in which the stellar orbit aligns with the orbit of the transiting planet, which could give insights into the planet formation process in such systems. We investigate here if this trend of alignment extends to planet-hosting triple-star systems. We present long-term astrometric monitoring of a novel sample of triple-star systems that host \kepler transiting planets. We measured orbit arcs in 21 systems, including 12 newly identified triples, from a homogeneous analysis of our Keck adaptive optics data and, for some systems, \gaia astrometry. We examine the orbital alignment within the nine most compact systems ($\lesssim500$\,au), testing if either (or both) of the stellar orbits align with the edge-on orbits of their transiting planets. Our statistical sample of triple systems shows a tendency toward alignment, especially when assessing the alignment probability using stellar orbital inclinations computed from full orbital fits, but is formally consistent with isotropic orbits. Two-population tests where half of the stellar orbits are described by a planet-hosting-binary-like moderately aligned distribution give the best match when the other half (non-planet-hosting) has a Kozai-like misaligned distribution. Overall, our results suggest that our sample of triple-star planet-hosting systems are not fully coplanar systems and have at most one plane of alignment. 
\end{abstract}


\section{Introduction}

Over 5000 exoplanets have been discovered so far,\footnote{https://exoplanetarchive.ipac.caltech.edu/docs/counts\_detail.html} and it is becoming clear that they are widespread with a minimum frequency of around one per star for a wide range of stellar masses \citep{Cassan2012}. Multiple star systems are also common, with over half of solar-type stars having at least one stellar companion and younger stars having an even greater multiplicity fraction (e.g., \citealt{Raghavan2010, Duchene2013, Moe2017}). Investigating planet-hosting multiple star systems is therefore important to gain a more complete picture of exoplanets and their characteristics in our galaxy.

Due to the observational difficulties that close binaries present, many planet-searching surveys have focused on stars that are either single or where any stellar companions are very widely separated. Recent transit surveys have provided data that is not as biased against the stellar multiplicity of the targets and thus allows planets in multiple star systems to be studied. However, planet properties estimated from transits in systems with multiple stellar components can be inaccurate due to additional stellar flux diluting the transits, especially if the planet actually orbits the secondary (or tertiary) star. Both of these scenarios result in an underestimation of planet radius (e.g., \citealt{Furlan2017, Sullivan2022b, Sullivan2023}). Characterising the components of multiple star systems that host transiting planets is therefore vital in understanding the properties of planets within these systems \citep{Fontanive2021, Cadman2022}.

Accurate planet properties can give an insight into the formation mechanisms of planets within multiple star systems. Theoretically, stellar companions should have a significant influence on the formation pathways of planets. Additional stars have been shown to produce hostile environments for planet formation by dynamically affecting the protoplanetary disks with processes such as truncation or misalignment (e.g., \citealt{Artymowicz1994, Kraus2012, Martin2014, Jang-Condell2015}). Even if the formation of a planet could be achieved, the interaction of the multiple stars in the system can negatively influence the overall stability of the orbital paths of the planet causing unstable states, collisions or ejections \citep{Holman1999, Haghighipour2006, Kratter2012, Kaib2013}. This is especially true for triple-star systems. With two stellar orbital planes to consider, dynamical interactions can become more prevalent causing an increase in scattering and destructive collisions \citep{Domingos2015}. 

These dynamical barriers to formation are therefore thought to have an impact on the distribution and characteristics of planet-hosting multiple-star systems. Wide binaries with a separation of over 1000\,au do not seem to impact the occurrence rate of planets and therefore the planet formation process \citep{Deacon2016}. Observational data collected has indicated that this, however, does not apply to close binaries (a<100\,au) which shows a lack of close stellar companions to transiting planet hosts (e.g., \citealt{Bergfors2013,Wang2014a,Kraus2016,Fontanive2019,Moe2021,Lester2021, Fontanive2021, Ziegler2021, Clark2022}). This suppression agrees with theoretical models of the formation of close binaries which favours the disk fragmentation model \citep{Lee2019, Tokovinin2020} with the addition of stellar companions causing the disruption and truncation of the protoplanetary disk \citep{Duchene2010}.  Observationally, the protoplanetary disks are not persistent with approximately two-thirds of close binaries dispersing their disks within $\sim 1$\,Myr after formation \citep{Kraus2012, Barenfeld2019}. 

Despite the formation hurdles, exoplanets have been found in close binary systems (e.g., \citealt{Hatzes2003, Dupuy2016, Winters2022}) suggesting that there are some pathways to successful planet formation in these hostile environments. In total, over 200 binary systems with exoplanets have been discovered \citep{Fontanive2021}, while currently there are only 30 planet-hosting triple and quadruple systems combined \citep{Cuntz2022}. As higher-order multiple systems are uncommon it means that while some individual systems have been well studied, population studies have been impossible thus far. 

Recent transit surveys using space-based telescopes, including the Kepler mission \citep{Koch2010} and the Transiting Exoplanet Survey Satellite (TESS; \citealt{Ricker2014}) have been the main contributors to the profusion of planet-hosting visual binaries. This has provided the opportunity to begin studying the orbital architectures of such systems. Following the work of \citet{Dvorak1982}, the orbit of planets around binaries can be split into two categories. First, S-type orbits where the planet is orbiting just one of the stars in the binary and P-type orbits where the planet orbits both stars in a circumbinary orbit. The transiting planets within this work are exclusively planets with S-type orbits, orbiting one host star.

One property that can be tested observationally is the alignment of the stellar orbital plane and the plane of the planet, as the transiting planets have the distinctive characteristic that their orbits are nearly edge-on and therefore have orbital inclinations of close to $90\degree$. \citet{Dupuy2022} observed 45 binary systems that host Kepler planets and from the orbital motions of the stellar companions concluded that there was an overabundance of mutually aligned systems, ruling out randomly orientated orbits at 4.7$\sigma$. \citet{Christian2022} performed a similar study using both planet-hosting wide binaries and a field control sample of wide binaries, both from Gaia. Using a control sample allowed them confidence that any features discovered were astrophysical and not a result of selection effects. By deriving limits on the inclinations of both samples they concluded that there was again a surplus of aligned systems in the planet-hosting subset, with a probability of 0.0037 of both samples being drawn from the same underlying distribution. Studies using other sources of transiting planets around visual binaries such as TESS candidates or K2 candidates have found similar results that point to planet-binary orbital alignment \citep{Behmard2022, Lester2023, Zhang2024}. This orbit-orbit alignment can also be investigated as a joint distribution with spin-orbit alignment. For example, \citet{Rice2024} had a sample of 40 planet-hosting binaries and found eight systems that each exhibit evidence of joint spin-orbit and orbit-orbit alignment. One triple star system within their sample, V1298 Tau, hosts a spin-orbit aligned planet as well as exhibiting orbit-orbit alignment between the primary and secondary but not the tertiary. They also found a trend in the stellar binary inclinations that strongly peaked toward alignment rather than an isotropic distribution.

Triple star systems have an additional orbital plane to consider due to the third star in the system which means not only the alignment of the planet's orbit can be tested but also the alignment of the stellar companions. Observational evidence implies that there is a tendency for triple star systems to have mutually aligned stellar orbits \citep{Worley1967}. \citet{Tokovinin2017} investigated the orbital alignment of 54 hierarchical field triples with visual orbits by calculating the mutual inclination between the orbit of the inner binary and the orbit of the outer companion relative to the barycentre of the binary. They concluded that there was a strong tendency for the triple systems to be coplanar in compact systems ($<50$\,au), especially for low-mass primaries ($M<1$\,\msun) where the average mutual inclination angle was $18\degree$. However, for systems where the outer companion was separated by more than 1000 au, isotropic orientations were found. \citet{Borkovits2016} had also previously studied the stellar alignment of 62 Kepler triple systems containing an eclipsing binary. They found 47\% of these systems to be coplanar, resulting in a distribution of mutual inclination angles that had a large peak at $<10 \degree$, although the distribution was bimodal with a secondary peak around $40 \degree$ which they attributed to Kozai-Lidov cycles.

Individual planet-hosting triples have been observed and studied but the majority of known systems are oriented such that alignment tests are not possible. For example, the nearest star system to the Sun is a planet-hosting hierarchical triple containing the binary $\alpha$~Cen~AB and its outer companion Proxima Centauri \citep{Innes1915}. This system contains one confirmed planet in the habitable zone of its host star, Proxima~b \citep{Anglada2016}, and two candidate planets, Proxima~c  \citep{Damasso2020} and Proxima~d \citep{Faria2022}. As all three planets were discovered using radial velocities, very little is known about their inclinations. Transiting planets in triple-star systems provides the unique opportunity to study the orbital alignment of both the stellar planes and the planetary planes. The M dwarf triple star system LTT 1445 is an example of such a system. The primary, LTT 1445~A, hosts two transiting planets and one non-transiting planet \citep{Winters2019, Winters2022, Lavie2023} as well as a binary pair, LTT 1445~BC, at a separation of $\sim 7\arcsec$ \citep{Dieterich2012, Rodriguez2015}. \citet{Zhang2024} used a combination of RVs, proper motion anomalies and astrometric measurements of the three stellar components to fit the orbit of both the BC binary around the host and the orbit of C around B. They obtained a mutual inclination between these two orbits of $2.88\,\degree \pm 0.63\,\degree$ and therefore concluded that LTT 1445~ABC is a coplanar system. 

In this work, we present 12 years of Keck AO and non-redundant aperture masking (NRM) astrometric monitoring of a sample of triple systems, both compact systems and those with wider companions identified with Gaia astrometry. The main sample consists of nine compact triple systems including Kepler-13 and Kepler-444 which both contain an unresolved companion. This sample also includes previously identified triple systems KOI-0005, KOI-0652, KOI-2032 and KOI-3497 \citep{Kraus2016}, KOI-2626 \citep{Gilliland2014}, as well as two newly identified triple systems KOI-0854 and KOI-3444. We derive individual stellar parameters for the 7 fully resolved triple systems and reassess the false positive probability of both the candidate and confirmed transiting planets hosted by these systems. We measure precise orbit arcs which allowed us to fit full orbits to both the inner binary's orbit and outer stellar companion's orbit relative to the barycentre of the binary for the majority of the triples in the sample. We use two different methods to constrain the alignment of both the stellar orbits to the edge-on planetary orbits, one using the partial orbital arcs and one using full orbital analysis. We find that both methods cannot rule out underlying isotropic orbits at a statistical level. While we find that the alignment in the triple systems are not consistent with the low mutual inclination trends seen in previous binary samples, there is some tentative evidence using both methods of some broad alignment, more than what would be expected for random orbits.

\section{Observations}

\subsection{Sample Selection}

As of 2023-12-06, \kepler has identified 2741 confirmed planets and 1984 candidate planets totalling 4725 planets around 2957 host stars.\footnote{https://exoplanetarchive.ipac.caltech.edu}. As part of our ongoing survey of planet-hosting binaries using the Keck-II telescope and its facility adaptive-optics imager NIRC2, we have observed in total 977 KOI systems. These have been prioritised from the complete list of KOI host stars, focusing on systems that are not false positives with RUWE > 1.2 and distance < 1.2kpc. The survey is described in further detail in \citep{Kraus2016} and Kraus et~al. (in prep.).

Observations were taken using the smallest pixel scale camera using the laser guide star (LGS) AO system \citep{Wizinowich2006, vanDam2006}. We used the broadband $K'$ filter (2.12\,$\mu$m, FWHM=0.35\,$\mu$m) for the majority of the imaging and the narrowband $K_{\rm cont}$ filter (2.17\,$\mu$m, FWHM = 0.03\,$\mu$m) for bright stars that would saturate in $K'$. For the systems with the tightest separations, we acquired both AO imaging and NRM interferograms using the 9-hole aperture mask installed in one of the filter wheels on NIRC2. 

From these observations, we initially removed systems where the closest companion was separated by more than 1000\,au. This narrowed down the sample to 580 systems that all contain a candidate close stellar companion. From there the aim was to identify systems with a second stellar companion, either from the observations or from additional methods.

For the visual triples, we identified systems with a second stellar companion in the observations with a magnitude difference of less than 6. This is to ensure that likely background stars are not included in the sample. 15 visual triple systems were identified using this method. While 9 of these triples have been previously published, 6 of them are newly identified here as candidate triples. Of these systems, 4 of them have the outer companion previously identified and for two systems we present both the inner and outer companion as newly identified components in the candidate triple systems.

We also compared the list of candidate binary systems to the wide-binary catalogue compiled by \citet{El-Badry2021}, again making the 6 magnitude difference cut for the inner companion. The wide-binary catalogue uses parallax and proper motion measurements from \gaia~eDR3 to identify candidate companions that are likely to be physically associated. This method revealed 10 candidate triple systems with a wide outer companion. We also performed our own independent search for wide companions to the binary candidates to ensure we include all possible candidates. We queried \gaia~DR3 within 2 arcminutes of each KOI binary candidate to identify wide companions that have a similar parallax and proper motion to the primary. Such similarity is a likely indicator that the wide companion is physically associated. We identified 4 further candidate systems this way. One of these (KOI-1615) appears to have two wide, bound companions in \gaia~DR3 which, along with our observations of a close stellar companion to the primary, would make it a quadruple system. We retain the system in our sample for completeness but the analysis of quadruple and higher-order multiples is beyond the scope of this paper.

Finally, we also searched the literature for any known unresolved companions to KOIs in our AO imaging sample, which resulted in adding KOI-0013 and KOI-3158 to our sample.

KOI-0013 has historically been known as the proper-motion binary BD+46~2629~AB \citep{Aitken1904}, consisting of two A-type stars Kepler-13~A and Kepler-13~B, with a separation of approximately 1$\arcsec$ \citep{Szabo2011}. A third low-mass stellar component was discovered orbiting Kepler-13~B using radial velocities in an eccentric orbit \citep{Santerne2012}. The planet identified with Kepler transits \citep{Barnes2011} has later been shown to be a highly irradiated gas giant orbiting the primary star \citep{Howell2019}.

KOI-3158 consists of a K0 dwarf (Kepler-444~A) and a tight M-type spectroscopic binary separated by $\sim 0.3$\,au from each other and about 66\,au from the primary \citep{Lillo2014, Dupuy2016}. Five transiting planetary candidates are orbiting Kepler-444~A in a compact system with separations up to 0.08\,au and sub-Earth radii of 0.4--0.7\,\rearth\ discovered using Kepler light-curve data \citep{Campante2015, Buldgen2019}. \citet{Dupuy2016} constrained the orbit of Kepler-444~BC relative to A using a combination of adaptive optics (AO) imaging and radial velocities and found an eccentric, edge-on orbit that from dynamical considerations was found to have a high probability of being aligned with the planetary orbit.  Using additional AO imaging, as well as RV measurements and Gaia astrometry of the primary,  \citet{Zhang2023} further constrained the outer orbit and derived a consistent result of the minimum misalignment being $1.6 \degree$--$4.6 \degree$.

In total, we have identified 31 candidate triple and quadruple systems. 

\subsection{Confirming physical association}

In order to quantitatively determine the likelihood of spatially resolved companions being bound together, we calculate the probability that each is co-moving with the primary, using methods established by \cite{Deacon2016} and rooted in the similar concept of open-cluster membership probabilities (e.g., \citealt{Sanders1971}; \citealt{Francic1989}). Our implementation will be further described by Kraus et al. (in 

\begin{landscape}
\begin{table}
\caption{Summary of the triple star systems in the sample.}
\label{tab:sample_table}
\begin{tabular}{lccccccccccc}
\hline
KOI & Kepler &  Planet  & Configuration & Outer Gaia DR3 ID & Distance & Inner sep. & Outer sep.  & Inner Field & Outer Field & Inner Ref. & Outer Ref. \\
&  & Disposition   &  &  & (pc) & (au) & (au)  & Probability & Probability & &  \\
\hline
\multicolumn{12}{c}{Compact triples (within 600 au)} \\
\hline
0005 & -- & 1 PC, 1 FP & \framebox{AB -- C} & -- & $559_{-23}^{+21}$  & $16.0_{-0.7}^{+0.6}$ & 78 $\pm$ 3 & \num{6.3e-7} & \num{9.0e-6} & 9 & 15,9 \\
0013 & 13 & 1 CP & \framebox{A* -- BC} & 2130632159130638464 & $488_{-8}^{+10}$ & -- & $576_{-10}^{+11}$ & -- & $< 10^{-9}$ & 12 & 13 \\
0652 & 636 & 1 CP & \framebox{A -- BC} & 2077382707923763328 & $469_{-4}^{+3}$ & 38.4 $\pm$ 0.3 & 565 $\pm$ 4 & \num{1.9e-6} & \num{7.4e-4} & 9 & 9 \\
0854 & 705 & 1 CP & \framebox{AB -- C} & -- & 274 $\pm$ 2 & $4.4 \pm 0.3 $ & $42.3_{-0.3}^{+0.4}$ & \num{5.6e-7} & \num{3.5e-2} & 9 & 9 \\
2032 & 1063 & 1 CP & \framebox{A -- BC} & 2052800269327720832 & $517_{-18}^{+25}$ & $33.5_{-1.2}^{+1.6}$ & $561_{-20}^{+27}$ & \num{4.4e-9} & \num{4.3e-4} & 9 & 9 \\
2626 & 1652 & 1 CP & \framebox{A -- BC} & -- & 233.3\dag & 37.6 & 46.9 & \num{4.4e-6} & \num{2.1e-5} & 4 & 4 \\
3158 & 444 & 5 CP & \framebox{A* -- BC} & 2101486923382009472 & 36.52 $\pm$ 0.02 & $\lesssim 0.37$ & $67.4 \pm 0.7 $ & -- & $< 10^{-9}$ & 3,4 & 11\\
3444 & -- & 4 PC & \framebox{A -- BC} & 2073740060272520320 & 91.3 $\pm$ 0.2 & 4.93 $\pm$ 0.01 & $98.6 \pm 0.2$ & -- & $< 10^{-9}$ & 1 & 10\\
3497 & 1512 & 1 CP & \framebox{A -- BC} & -- & $278_{-10}^{+11}$& $23.4_{-0.8}^{+0.9}$ & $234_{-8}^{+9}$ & \num{8.1e-6} & \num{1.5e-4} & 9 & 9 \\
\hline
\multicolumn{12}{c}{Wide triples (Outer separation > 600 au) } \\
\hline
0288 & 1714 & 1 CP & \framebox{AB} -- C & 2128069433752982144 & $432_{-5}^{+4}$& $150.0_{-1.6}^{+1.5}$ & $14710_{-150}^{+140}$ & \num{5.0e-5} & -- & 9 & 6 \\
0307 & 520 & 2 CP & \framebox{AB} -- C & 2077597765529420416 & $810_{-80}^{+140}$& $61_{-6}^{+10}$ & $183000_{-18000}^{+31000}$ & \num{4.4e-5} & -- & 7 & 6 \\
1613 & 907 & 1 CP, 2 PC & \framebox{AB} -- C & 2103879696905184896 & $610_{-150}^{+190}$ & $134_{-33}^{+42}$ & $44600_{-4400}^{+7600}$ & \num{1.5e-6} & -- & 10 & 14 \\
1961 & 1027 & 1 CP, 1 FP & \framebox{AB} -- C & 2102588737113424512 & $402_{-4}^{+5}$& $13.9_{-0.1}^{+0.2}$ & $7644_{-67}^{+88}$ & \num{9.4e-4} & -- & 9 & 5, 14\\
2117 & 1795 & 1 CP & \framebox{BC} -- A & 2100440257031482112 & $475_{-59}^{+78}$& $157_{-20}^{+26}$ & $120000_{-15000}^{+20000}$ & \num{1.3e-5} & -- & 16 & 6 \\
2517 & 1264 & 1 CP & \framebox{AC} -- B & 2127065583934822528 & $759_{-18}^{+20}$& $146_{-3}^{+4}$ & $50800_{-1200}^{+1300}$ & \num{8.2e-4} & -- & 1 & 6\\
2971 & --  & 2 PC & \framebox{AC} -- B & 2073646700553637248 & $783_{-6}^{+7}$& $236_{-2}^{+7}$ & $6105_{-45}^{+55}$ & \num{8.6e-4} & -- & 2 & 6\\
3196 & --  & 2 PC & \framebox{AC} -- B & 2106994445843732096 & $320.2_{-1.1}^{+1.4}$& $40.5_{-0.1}^{+0.2}$ &$3198_{-11}^{+13}$ & \num{6.3e-4} & -- & 1 & 6\\
4329 & --  & 1 PC & \framebox{AB -- C} & 2133440621069065344 & $631_{-8}^{+7}$ & $21.8_{-0.3}^{+0.2}$ & $1218_{-15}^{+14}$ & \num{1.1e-7} & -- & 1 & 2\\
4661 & 1966 & 1 CP & \framebox{AC -- B} & 2053614045374562304 & $487_{-12}^{+13}$& $74.7_{-1.8}^{+1.9}$ &$1914_{-47}^{+50}$ & \num{6.7e-6} & -- & 1 & 16\\
5581 & 1634 & 1 CP & \framebox{AC} -- B & 2127951889088883584 & 602 $\pm$ 6& $100.4_{-1.1}^{+0.9}$ & $90420_{-960}^{+840}$ & 0.024 & -- & 1 & 6\\
7842 & --  & 1 PC & \framebox{AB} -- C & 2102811731815500288 & $926_{-17}^{+14}$& $72.0_{-1.3}^{+1.1}$ & $32600_{-590}^{+490}$ & \num{1.3e-3} & -- & 1 & 6\\
\hline
\multicolumn{12}{c}{Candidate Triple Systems} \\
\hline
4528 & --  & 1 PC & \framebox{AB -- C} & -- & $534_{-35}^{+33}$& 36 $\pm$ 2 & 97 $\pm$ 6  & --  & -- & 1 & 1\\
4759 & --  & 1 PC & \framebox{A -- BC} & -- & 912 $\pm$ 24 & $70.5_{-1.9}^{+1.8}$ & 649 $\pm$ 17  & --  & -- & 1 & 1\\
5930 & --  & 1 PC & \framebox{AB} -- C & 2134911973789149952	& $227_{-4}^{+3}$& 16.8 $\pm$ 0.3 & 320 $\pm$ 5  & --  & -- & 1 & 6\\
\hline
\multicolumn{12}{c}{Candidate Quadruple Systems} \\
\hline
1615 & 908 & 1 CP & \framebox{AB} -- CD & 2076194101491522304 & $294.3_{-1.2}^{+1.1}$ & $9.4 \pm 0.5 $ & $2422_{-10}^{+9}$ & \num{4.7e-5} & -- & 9 & 14\\
 &  &  & & 2076194101502796032 &  & 256.2 $\pm$ 1.0 & & --  & --& 14 & \\
\hline
\end{tabular}
\begin{list}{}{}
\item[Note.] The designation of the planets is either CP (Confirmed Planet), PC (Candidate Planet) or FP (False Positive) taken from the Exoplanet Follow-up Observing Program (ExoFOP), https://exofop.ipac.caltech.edu/tess/. The configuration column shows the hierarchical pairing of stars by brightness, and the box around it indicates the stellar components that are within 1 \kepler pixel and therefore shows which components could host the transiting planets. Components indicated with * are definitively known as the host star of the planet. The distances are all taken from \citet{BailerJones2021} apart from when indicated with \dag which is taken from \citet{Kraus2016} as there is no parallax available from \gaia~DR3. 

\item[References.] (1) This work; (2) \cite{Baranec2016}; (3) \cite{Campante2015}; (4) \cite{Dupuy2016}; (5) \cite{Dupuy2022};  (6) \cite{El-Badry2021}; (7) \cite{Furlan2017}; (8) \cite{Gilliland2014}; (9) \cite{Kraus2016}; (10) \cite{Law2014}; (11) \cite{Lillo2014}; (12) \cite{Santerne2012}; (13) \cite{Szabo2011}; (14) \gaia~DR3 \cite{Gaia2023}; (15) \cite{Wang2014a}; (16) \cite{Ziegler2017};
\end{list}
\end{table}
\end{landscape}

prep). This probability is based on the relative linear motion, relative separation, and difference in magnitude (potentially in multiple filters) of each star. 

We create a model for the field star population by querying Gaia for all sources within $\rho < 1\degree$, including their relative proper motions and parallaxes, and then computing their stellar parameters ($M$ and $T_{eff}$). Depending on what Gaia measurements are available, in order of preference, we compute the parameters of these stars from the absolute $M_G$ magnitude (as computed from the parallax and apparent $G$ magnitude), the $B_p-R_p$ colour, the $G-R_p$ colour, or as a last resort by assuming a temperature of $T_{\rm eff} = 4500$\,K as is typical for faint Gaia sources that do not have colours. In all cases, we interpolate the mass-$T_{\rm eff}$-color-magnitude relations of \citet{Pecaut2016} \footnote{Retrieved on 2024 May 17 from \url{http://www.pas.rochester.edu/~emamajek/EEM_dwarf_UBVIJHK_colors_Teff.txt}} to determine the stellar properties of each field star. We then use this set of unrelated field stars to forward-model a population of field interlopers with values of projected separation, relative brightness in each filter where a contrast is available, relative proper motion, and relative parallax. We then also create a corresponding model for the population of all binary companions that is based on the demographics of \citet{Raghavan2010}, again computing their projected separations, relative proper motions, relative parallaxes, and contrasts in the filters where observations are available. 

Finally, we use KDEs to smooth the empirical field-star population and the synthetic binary population and produce continuous probability density functions, and use the relative densities of the field and binary populations at the phase-space location of each candidate companion to estimate its probability of being drawn from either the binary posterior or the field interloper posterior. The kernel widths were chosen to be much larger than the typical distance between adjacent simulated binaries, but not larger than the typical extent of the population: 0.2\,dex in $\log{\rho}$, the candidate's observational uncertainty plus 0.2 magnitudes for each contrast, the quadratic sum of the observational uncertainty and predicted orbital motion for the relative proper motion, and the observational uncertainty for the relative parallax. Of the 31 candidate triples in our sample, 23 of them had probabilities of both components being bound of $> 99$\%. KOI-4759 had a probability of both companions being bound of 78\%. This system is retained in the sample as a candidate requiring follow-up observation. Two wide systems identified by \citet{El-Badry2021} (KOI-4407 and KOI-5943) had probabilities $< 0.001$\% of the inner component from AO imaging being bound, and one candidate system identified from \citet{Kraus2016} (KOI-2813) had both inner and outer components likely to not be bound with probabilities of  $< 0.001$\%. Finally, for three systems (KOI-0387, KOI-2059, KOI-2733) the inner pair was likely to be bound ($> 99$\%), but the outer pair was not ($< 0.001$\%), meaning that they can be retained for future work on binary systems, but we exclude them here.

Three of the triple candidates (KOI-4528, KOI-4759, and KOI-5930) only have one epoch of observations, so they cannot yet be proper-motion confirmed, and we note them here as candidate triple systems. This leaves a total of 21 confirmed triple systems in our sample, of which 9 have been previously identified in the literature and 12 are presented here as newly recognised planet-hosting triple star systems. Table~\ref{tab:sample_table} provides a summary of the sample of triples described here including the probability of each companion being a background field object.

Close stellar companions ($\rho<1000$\,au) have been shown to impact the formation and evolution of planets (e.g., \citealt{Fontanive2021}) while wider companions have little effect on planets orbiting the host star (e.g., \citealt{Deacon2016, Kraus2016, Christian2022}). For this reason we choose to focus on the closely separated systems with outer separations of $\rho<1000$\,au. Our sample of triples contains only systems with an outer companion separation of either $\rho>1200$\,au or $\rho<600$\,au with no systems residing between these two limits. This means that the effective cutoff separation for compact systems in our sample is 600\,au. There are 9 compact triple systems that meet the criteria of having both their stellar companions within 600\,au of the primary. In 7 of the 9 systems, the primary is an individual star orbited by a close binary (A-BC) and the remaining 2 systems are close binaries with a wide tertiary companion (AB-C). 

\subsection{Astrometry from AO imaging}

\setcounter{table}{1}
\begin{table*}
\caption{Relative astrometry measurements of the sample of KOIs from our Keck/NIRC2 adaptive optics imaging and aperture-masking interferometry.}
\label{tab:astrometry}
\begin{tabular}{lcccccl}

\hline
Name & \multicolumn{2}{c}{Epoch} & Separation & Position Angle & $\Delta m$ & Filter \\
& (UT) & (MJD) & (mas) & ($\degree$) & (mag) & \\

\hline

KOI-0005 AB & 2012-08-14 & 56153.45 & 28.1 $\pm$ 1.5 & 142.8 $\pm$ 0.9 & 0.20 $\pm$ 0.09 & $K'$ \\
KOI-0005 AB & 2013-08-20 & 56524.42 & 29.6 $\pm$ 1.5 & 146 $\pm$ 4 & 0.34 $\pm$ 0.09 & $K_\text{cont}$ \\
KOI-0005 AB & 2014-07-28 & 56866.45 & 31.1 $\pm$ 1.3 & 151 $\pm$ 3 & 0.34 $\pm$ 0.10 & $K'$ \\
KOI-0005 AB & 2015-07-22 & 57225.43 & 31.6 $\pm$ 2.2 & 149.6 $\pm$ 2.0 & 0.42 $\pm$ 0.08 & $K'$ \\
KOI-0005 AB & 2017-06-28 & 57932.40 & 32.02 $\pm$ 0.16 & 155.9 $\pm$ 1.9 & 0.26 $\pm$ 0.08 & $K'$ \\
KOI-0005 AB & 2019-06-12 & 58646.35 & 30.0 $\pm$ 0.4 & 156.9 $\pm$ 2.0 & 0.22 $\pm$ 0.08 & $K'$ \\
KOI-0005 AB & 2020-06-18 & 59018.58 & 30.1 $\pm$ 1.0 & 162.3 $\pm$ 1.9 & 0.30 $\pm$ 0.06 & $K'$ \\
KOI-0005 AC & 2012-08-14 & 56153.45 & 120.2 $\pm$ 1.3 & 305.8 $\pm$ 1.5 & 1.800 $\pm$ 0.021 & $K'$ \\
KOI-0005 AC & 2013-08-20 & 56524.42 & 125.6 $\pm$ 1.4 & 305.7 $\pm$ 0.4 & 1.97 $\pm$ 0.04 & $K_\text{cont}$ \\
KOI-0005 AC & 2014-07-28 & 56866.45 & 127.0 $\pm$ 0.9 & 305.1 $\pm$ 0.8 & 1.98 $\pm$ 0.08 & $K'$ \\
KOI-0005 AC & 2015-07-22 & 57225.43 & 128.4 $\pm$ 1.4 & 306.04 $\pm$ 0.26 & 2.00 $\pm$ 0.07 & $K'$ \\
KOI-0005 AC & 2017-06-28 & 57932.40 & 130.4 $\pm$ 0.7 & 305.8 $\pm$ 0.3 & 1.93 $\pm$ 0.04 & $K'$ \\
KOI-0005 AC & 2019-06-12 & 58646.35 & 134.5 $\pm$ 1.2 & 306.66 $\pm$ 0.22 & 1.90 $\pm$ 0.04 & $K'$ \\
KOI-0005 AC & 2020-06-18 & 59018.58 & 137.7 $\pm$ 0.5 & 306.75 $\pm$ 0.23 & 1.97 $\pm$ 0.03 & $K'$ \\
\hline & \\[-1.5ex]
\end{tabular}
\begin{list}{}{}
\item[Note.] A full version of this table is available at the end of this preprint. 
\end{list}
\end{table*}

The AO imaging observations we use here were taken over 55 nights spanning from 2012~Jul~6 UT to 2023~Jun~9 UT. We used the same reduction pipeline as described in \citet{Kraus2016} to produce calibrated images, using techniques such as flat-fielding and dark subtraction, but we performed our own astrometric analysis. This analysis follows the methods described in \citet{Dupuy2019}, adapted for use on triple star systems. Briefly, for the majority of our triple systems, we fitted an empirical template PSF that was computed from the image itself using StarFinder \citep{Diolaiti2000}. This PSF was fitted to each component to derive ($x$,$y$) NIRC2 positions for each star, iterating and updating both the PSF and the stellar parameters until a stable solution was reached. For the tightest systems, where the PSFs are not sufficiently separated for Starfinder to compute a PSF, we instead fitted an analytic PSF to each component, similar to our previous work \citep{Liu2006,Dupuy2009}. This PSF was the sum of three concentric 2D Gaussians, each with different free parameters for FWHM, ellipticity, orientation, and amplitude, which we determined simultaneously with the binary parameters.

From the pixel coordinates, we computed angular separations and position angles (PAs) to measure the relative astrometry. To do this, the pixel scale as well as the orientation of NIRC2 and the nonlinear distortion must be accounted for. For data taken prior to 2015 Apr 13, when the AO system was realigned, we used the astrometric calibration of 9.952 ± 0.002 mas/pix \citep{Yelda2010}, and we used \citet{Service2016} for data collected afterwards. These calibrations provide uncertainty for the linear terms on the pixel scale (fractional error of $4\times10^{-4}$) and orientation (0.02$\degree$).  We measured the astrometry for individual images on a given night and then computed the mean to provide the relative astrometry for each epoch. The uncertainty in the results is a quadrature combination of the rms of the astrometry for the individual images and the calibration uncertainty for the separation and position angle. The uncertainty in the magnitude difference ($\Delta m$) is the rms of the individual measurements. Table~\ref{tab:astrometry} reports the complete set of binary parameters measured from both our AO images and NRM data. 

There is also an uncertainty on the nonlinear distortion term of the calibration, but we neglect this error in this analysis. Distortion is expected to be correlated at small pixel scales ($\sim$10 pixels) for our tight inner binaries. For the outer companions, the distortion uncertainty would be more significant (up to 1.5\,mas), however, these companions have larger errors due to the linear calibration term uncertainties, which dominate for wider binaries. 

The NRM data was reduced following the technique of \citet{Kraus2016}. Briefly, the frames are Fourier-transformed and the squared visibility and closure phase is extracted for each baseline before being calibrated against the instrumental squared visibilities and closure phases, estimated from calibrator stars observed in the same night. These closure phases can then be used to fit for a binary solution, by searching a grid to find the minimum $\chi^2$ for the separation and position angle assuming the star is a binary. The uncertainties in the fit are then increased so that the resulting reduced $\chi^2$ is equal to 1.

\subsection{Observations with the Hobby-Eberly Telescope}
For 6 of the 9 close triples, we obtained moderate-resolution, red-optical spectra using the red arm of the second-generation Low-Resolution Spectrograph (LRS2-R; \citealt{Chonis2014, Chonis2016}) at the Hobby-Eberly Telescope (HET) at McDonald Observatory as part of a program to spectroscopically survey planet-hosting multiple stars. Because the HET is queue scheduled, not all of our targets were observed, but a majority were. LRS2 is a moderate-resolution (R$\sim$1700) continuously-tiled integral field spectrograph (IFS) with two settings, each with two channels. LRS2-R is the redder arm of the instrument and covers the red and far-red channels, which cover $6500 < \lambda < 8470$ \AA\ and $8230 < \lambda < 10500$ \AA, respectively. Although our observations included both channels simultaneously, the far-red channel had severe telluric contamination and low S/N, so we restricted our analysis to only the red channel. 

Our observing strategy and the instrument details are described in \citet{Sullivan2022c}. To briefly summarize, the observations were typically taken during grey or bright times, with an upper limit on the seeing of $\sim 2\farcs5$. These conditions were acceptable because our systems were unresolved and we only required a single composite spectrum. Our exposure times were either 300\,s or a time sufficient to achieve an S/N $>$ 100 for the primary star. After data reduction, the source was extracted using an aperture clipped at 2.5 times the seeing, calculated in the wavelength frame with the highest S/N.

\section{Revised Stellar Parameters}

For the full orbital analysis of the triples in our sample as well as to locate the barycentre of the binary component in each system, accurate stellar masses are required. To obtain constraints on the component masses for our 7 fully-resolved triples, we retrieved the individual stellar parameters for each component using the method presented in \citet{Sullivan2022b} and modified for HET data in \citet{Sullivan2022c} and \citet{Sullivan2023} but adjusted slightly to account for the third star in the system. We summarize the method here for completeness, with an emphasis on changes in methods between \citet{Sullivan2023} and this work. 

Briefly, we assembled data including spectra from HET/LRS2-R (when available), unresolved $r'i'z'JHK_{s}$ broadband photometry from the Kepler Input Catalog (KIC; \citealt{Brown2011}) and the 2-Micron All-Sky Survey (2MASS; \citealt{Skrutskie2006}), and high-resolution adaptive optics imaging from NIRC2 on Keck. The majority of our triples had a contrast in a single photometric band from our AO imaging, but one system (KOI-2626) had optical speckle imaging reported in \citet{Furlan2017} with both components of the triple resolved. When analyzing KOI-2626, we included the speckle measurements in our fit along with the NIR AO contrasts. We fit the data set with a three-component spectral model using the BT-Settl stellar atmosphere models \citep{Allard2013, Rajpurohit2013, Allard2014, Baraffe2015} with the \citet{Caffau2011} linelist. 

To perform the fitting we used a custom-modified Gibbs algorithm, then used \texttt{emcee} \citep{Foreman-Mackey2013} to assess the statistical spread of the retrieved values. We retrieved a set of 8 stellar parameters: the individual stellar component \teff and radius values, the best-fit distance to the system, and the extinction. We placed a prior on the stellar radii using stellar radii derived using the MESA Isochrones and Stellar Tracks (MIST) evolutionary models \citep{Paxton2011, Paxton2013, Paxton2015, Choi2016, Dotter2016} at an age of 1 Gyr.

\begin{table*}
\caption{Stellar parameter fit results for all KOIs in our sample}
\label{tab:star results}
\begin{tabular}{cccccccccccc}
\hline
KOI & $T_{{\rm eff},1}$ & $T_{{\rm eff},2}$ & $T_{{\rm eff},3}$ & $T_{\rm Kepler}$ & $R_{1}$ & $R_{2}/R_{1}$ & $R_{3}/R_{1}$ & $R_{\rm Kepler}$ & f$_{\rm corr, p}$ & f$_{\rm corr, s}$ & f$_{\rm corr, t}$\\

& (K) &  (K) & (K) & (K) & (\rsun) & & & (\rsun) &  & & \\

\hline
0005 & $6659^{+58}_{-54}$ & $6327^{+58}_{-60}$ & $4585^{+91}_{-80}$ & $5945\pm119$ & $1.396^{+0.032}_{-0.028}$ & $0.886^{+0.006}_{-0.005}$ & $0.507^{+0.005}_{-0.005}$ & $2.24^{+0.14}_{-0.13}$ & $1.01^{+0.08}_{-0.07}$ & $1.13^{+0.09}_{-0.09}$ & $2.44^{+0.23}_{-0.20}$\\
0652 & $5517^{+94}_{-73}$ & $4583^{+115}_{-90}$ & $3880^{+45}_{-32}$ & $5945\pm119$ & $0.858^{+0.027}_{-0.020}$ & $0.799^{+0.008}_{-0.011}$ & $0.600^{+0.004}_{-0.005}$ & $2.24^{+0.14}_{-0.13}$ & $1.17^{+0.12}_{-0.10}$ & $1.91^{+0.22}_{-0.18}$ & $2.91^{+0.36}_{-0.29}$\\
0854 & $3667^{+12}_{-11}$ & $3615^{+15}_{-14}$ & $2904^{+6}_{-3}$ & $5305\pm106$ & $0.417^{+0.006}_{-0.006}$ & $0.956^{+0.024}_{-0.024}$ & $0.298^{+0.006}_{-0.006}$ & $1.15^{+0.05}_{-0.05}$ & $1.15^{+0.08}_{-0.07}$ & $1.22^{+0.08}_{-0.07}$ & $3.01^{+0.21}_{-0.19}$\\
2032 & $6067^{+53}_{-56}$ & $5868^{+57}_{-65}$ & $5661^{+64}_{-78}$ & $3593\pm72$ & $1.040^{+0.027}_{-0.029}$ & $0.923^{+0.004}_{-0.004}$ & $0.853^{+0.004}_{-0.004}$ & $0.56^{+0.02}_{-0.02}$ & $1.01^{+0.22}_{-0.15}$ & $1.08^{+0.23}_{-0.16}$ & $1.16^{+0.26}_{-0.16}$\\
2626 & $3648^{+12}_{-12}$ & $3515^{+12}_{-9}$ & $3376^{+15}_{-16}$ & $3554\pm 80$ & $0.401^{+0.010}_{-0.011}$ & $0.857^{+0.004}_{-0.003}$ & $0.699^{+0.004}_{-0.003}$ & $0.398^{+0.049}_{-0.054}$ & $1.36^{+0.19}_{-0.16}$ & $1.54^{+0.23}_{-0.17}$ & $1.82^{+0.26}_{-0.20}$\\
3444 & $3790^{+7}_{-9}$ & $3500^{+1}_{-0}$ & $3500^{+1}_{-0}$ & $3705\pm 74$ & $0.494^{+0.004}_{-0.004}$ & $0.268^{+0.002}_{-0.002}$ & $0.236^{+0.001}_{-0.001}$ & $0.529^{+0.030}_{-0.038}$ & $0.98^{+0.08}_{-0.06}$ & $1.24^{+0.18}_{-0.35}$ & $1.21^{+0.17}_{-0.62}$\\
3497 & $4756^{+52}_{-44}$ & $3881^{+31}_{-21}$ & $3701^{+19}_{-18}$ & $3665\pm73$ & $0.737^{+0.014}_{-0.014}$ & $0.745^{+0.006}_{-0.006}$ & $0.618^{+0.003}_{-0.004}$ & $0.52^{+0.01}_{-0.01}$ & $2.38^{+0.41}_{-0.32}$ & $4.45^{+0.89}_{-0.58}$ & $5.30^{+0.99}_{-0.72}$\\
\hline
\end{tabular}
\begin{list}{}{}
\item[Note.] The \kepler values for the composite stellar system properties have been taken from \citet{Berger2018} apart from for KOI-2626 and KOI-3444 which used values from \citet{Mathur2017}. The revised stellar temperatures, radii and the planetary radius correction factor if the primary, secondary, or tertiary star is the host is also shown. 
\end{list}
\end{table*}

When available, we placed a Gaussian prior on the parallax using the Gaia DR3 \citep{Gaia2023} parallax. We set a Gaussian prior on the extinction using the mean and standard deviation of the predicted extinction at the appropriate distance and location using the 3D Bayesian dust map \texttt{bayestar} \citep{Green2019} implemented in the \texttt{dustmaps} package\footnote{\url{https://dustmaps.readthedocs.io/en/latest/index.html}}. With the best-fit \teff, we used a MIST isochrone at an age of 1 Gyr to infer a mass for each star in every system. Table~\ref{tab:star results} summarises the retrieved stellar parameters for the components of each system. Table~\ref{tab:star results} also summarizes the planet radius correction factor for each star in the system, which is the multiplicative factor by which to correct the reported Kepler planetary radii as defined in \citet{Ciardi2015} and \citet{Furlan2017}. Table~\ref{tab:masses} summarises the masses derived for each component in the 7 visual triples. The masses for the components in the unresolved triples, KOI-0013 and KOI-3158, have been taken from the literature.  

\begin{table}
\renewcommand{\arraystretch}{1.4}
\caption{Masses for the stellar components in the compact triple systems.}
\label{tab:masses}
\begin{tabular}{lcccc}
\hline
KOI & $M_A$ & $M_B$ & $M_C$ & Ref.\\
& (\msun)  & (\msun) & (\msun) & \\
\hline
0005 & $ 1.302_{-0.020}^{+0.020}$ & $1.186_{-0.022}^{+0.020}$ & $ 0.735_{-0.018}^{+0.020}$ & 5 \\
0013 & $2.05$ & $1.95$ & $>0.4,<1$ & 1,2\\
0652 & $ 0.939_{-0.019}^{+0.024}$ & $ 0.734_{-0.020}^{+0.025}$ & $ 0.550_{-0.014}^{+0.014}$ & 5\\
0854 & $ 0.450_{-0.007}^{+0.006}$ & $ 0.419_{-0.009}^{+0.009}$ & $ 0.104_{-0.001}^{+0.001}$ & 5\\
2032 & $ 1.096_{-0.018}^{+0.019}$ & $ 1.035_{-0.019}^{+0.019}$ & $ 0.974_{-0.019}^{+0.020}$ & 5 \\
2626 & $ 0.440_{-0.007}^{+0.007}$ & $ 0.355_{-0.006}^{+0.008}$ & $ 0.280_{-0.008}^{+0.008}$ & 5\\
3158 & $0.75 \pm 0.03$ & $ 0.307_{-0.008}^{+0.009}$ & $ 0.296_{-0.008}^{+0.008}$ & 3,4 \\
3444 & $ 0.516_{-0.168}^{+0.078}$ & $ 0.438_{-0.092}^{+0.231}$ & $ 0.348_{-0.002}^{+0.174}$ & 5\\
3497 & $ 0.772_{-0.010}^{+0.011}$ & $ 0.550_{-0.009}^{+0.011}$ & $ 0.469_{-0.009}^{+0.010}$ & 5\\
\hline
\end{tabular}
\begin{list}{}{}
\item[References.] (1) \cite{Szabo2011}; (2) \cite{Santerne2012}; (3) \cite{Buldgen2019}; (4) \cite{Zhang2023}; (5) This work.
\end{list}
\end{table}

\section{Planetary parameters}
\subsection{False positive analysis}
\label{sec:FPA}

Multiple star systems cause complications when determining planet parameters from transits due to the extra flux provided by the stellar companions. This means that multiple-star systems are likely to harbour false positives (FPs). Erroneous classifications of FPs are also expected due to centroid offsets as transiting planets around stellar companions may cause such offsets. In our sample, there are 16 candidate or confirmed planets in 9 systems. KOI-0005 is the only system to have an FP (KOI-0005.02), but we retained it in the sample due to the candidate planet KOI-0005.01. None of the planets in the sample have a centroid-offset flag, which if present could indicate that the planet does not orbit the primary. KOI-0013 and KOI-3158, the only two triple systems that have an unresolved stellar companion, have both been definitively shown to have their planets orbiting the primary star \citep{Howell2019, Buldgen2019}. They are therefore not included in our false positive re-analysis. For the remaining seven systems, we reassess their false positive probabilities for each stellar component to confirm their status as validated planets and potentially identify which star they orbit.

\begin{table}
\centering
\caption{Probabilities of the primary, secondary or tertiary being the host star for each planetary companion based on stellar densities.}
\begin{tabular}{lccc}
\hline 
KOI & Primary  & Secondary  & Tertiary \\
 & Probability & Probability & Probability \\
\hline
0005.01 & 0.6323 & 0.3677 & 0.0000 \\
0652.01 & 0.6837 & 0.3163 & 0.0000 \\
0854.01 & 0.1328 & 0.8670 & 0.0002 \\
2032.01 & 0.4253 & 0.3482 & 0.2265 \\
2626.01 & 0.0780 & 0.2922 & 0.6298 \\
3444.01 & 0.9977 & 0.0016 & 0.0007 \\
3444.02 & 0.9984 & 0.0010 & 0.0006 \\
3444.03 & 0.9283 & 0.0579 & 0.0138 \\
3444.04 & 0.8682 & 0.0817 & 0.0502 \\
3497.01 & 1.0000 & 0.0000 & 0.0000 \\
\hline
    \end{tabular}
    \label{tab:prob_triples_host_star}
\end{table}

For the remaining 10 planets around 7 host systems, we use the mass and radius of each stellar component as described above to calculate stellar mean density distributions. Using the measured transit durations, we then also calculated the expected stellar density for each component assuming the planet was around each one. We excluded grazing transits by assuming a uniform impact parameter from 0 to $R_p/R_\star$, assumed a Rayleigh distribution for the eccentricities with a mean of 0.05 \citep{VanEylen2015}, and accounted for dilution from the additional stellar flux by correcting the factor of $(R_p/R_s)^{3/2}$ from the assumed $R_p$ and $R_s$ to the calculated one above \citep{Seager2003}.  Comparing these two stellar density distributions for each component gives an indication of which star the planet could orbit, following the method in \citet{Gaidos2016} which calculates a probability from Monte-Carlo simulations. The results of these probabilities are shown in Table~\ref{tab:prob_triples_host_star}. We found that out of the 10 planets, 4 were consistent with orbiting any star within their triple system with a probability of >0.1\% (KOI-2032.01, KOI-2626.01, KOI-3444.03, and KOI-3444.04). KOI-2626.01 has the highest probability of being around the tertiary, and the remaining three all favoured the primary, with the two planets around KOI-3444 having a high probability of being hosted by the primary (>86\%). KOI-0005.01, KOI-0652.01, KOI-0854.01, KOI-3444.01 and KOI-3444.02 were consistent with orbiting the primary or secondary but not consistent with orbiting the tertiary. KOI-0854.01 is the only one of these 5 systems to favor the secondary. Finally, KOI-3497.01 was the only planet with a high probability of $> 99.99$\% of orbiting the primary and was inconsistent with orbiting the secondary or tertiary. None of the 10 planets show evidence of being FPs as each one had at least one acceptable match between the stellar density distributions for the primary, secondary or tertiary. The four planets around KOI-3444 as well as KOI-0005.01 are the only candidate planets in our sample. As their host star probabilities were all consistent with their candidate planet status we retain them in our sample as planetary candidates. KOI-0005 has an additional observation from TESS showing a consistent planetary detection for KOI-0005.01.

\begin{landscape}
\begin{table}
\centering
\caption{Planet parameter fit results for all planets in the close visual triple KOI sample.}
\label{tab:planet results}
\begin{tabular}{ccccccccccccc}
\hline
KOI & R$_{\rm p, pri}$ & R$_{\rm p, sec}$ & R$_{\rm p, tri}$ & R$_{\rm Kep}$ & T$_{\rm eq,pri}$ & T$_{\rm eq,sec}$ & T$_{\rm eq,tri}$ & T$_{\rm eq, Kep}$ & S$_{\rm pri}$ & S$_{\rm sec}$ & S$_{\rm tri}$ & S$_{\rm Kep}$\\
 & ($R_{\earth}$) & ($R_{\earth}$) & ($R_{\earth}$) & ($R_{\earth}$) & (K) & (K) & (K) & (K) & (S$_{\earth}$) & (S$_{\earth}$) & (S$_{\earth}$) & (S$_{\earth}$) \\
 \hline
0005.01* & $7.26^{+0.76}_{-0.73}$ & $8.05^{+0.83}_{-0.83}$ & $17.56^{+1.96}_{-1.93}$ & $7.14\pm0.52$ & $1434^{+58}_{-58}$ & $1282^{+53}_{-53}$ & $703^{+32}_{-32}$ & $1441$ & $862.40^{+57.25}_{-56.45}$ & $569.17^{+43.47}_{-45.87}$ & $73.36^{+7.20}_{-7.21}$ & $1020.05\pm223.43$\\
0652.01 & $5.26^{+0.68}_{-0.68}$ & $8.58^{+1.20}_{-1.19}$ & $12.90^{+1.88}_{-1.84}$ & $4.43\pm0.38$ & $666^{+37}_{-38}$ & $491^{+28}_{-28}$ & $363^{+22}_{-24}$ & $614$ & $42.70^{+2.57}_{-5.44}$ & $14.66^{+1.66}_{-1.66}$ & $6.00^{+-0.45}_{-1.23}$ & $33.64\pm9.19$\\
0854.01 & $2.24^{+0.20}_{-0.20}$ & $2.37^{+0.20}_{-0.21}$ & $5.89^{+0.54}_{-0.53}$ & $1.94\pm0.12$ & $220^{+8}_{-8}$ & $212^{+8}_{-8}$ & $95^{+4}_{-4}$ & $233$ & $0.58^{+0.02}_{-0.02}$ & $0.50^{+0.02}_{-0.02}$ & $0.06^{+0.00}_{-0.00}$ & $0.69\pm0.15$\\
2032.01 & $2.16^{+0.51}_{-0.52}$ & $2.32^{+0.57}_{-0.57}$ & $2.52^{+0.61}_{-0.63}$ & $2.08\pm0.36$ & $805^{+72}_{-71}$ & $749^{+67}_{-67}$ & $693^{+64}_{-62}$ & $931$ & $95.09^{+7.45}_{-6.84}$ & $73.87^{+5.98}_{-5.75}$ & $56.74^{+5.11}_{-4.71}$ & $177.81\pm80.90$\\
2626.01 & $2.17^{+0.39}_{-0.39}$ & $2.48^{+0.44}_{-0.44}$ & $2.90^{+0.53}_{-0.53}$ & $1.58\pm0.20$ & $250^{+17}_{-17}$ & $223^{+15}_{-15}$ & $194^{+14}_{-13}$ & $242$ & $0.92^{+0.03}_{-0.03}$ & $0.64^{+0.02}_{-0.02}$ & $0.43^{+0.02}_{-0.02}$ & $0.81\pm0.30$\\
3444.01* & $0.75^{+0.06}_{-0.07}$ & $0.94^{+0.17}_{-0.26}$ & $0.81^{+0.25}_{-0.37}$ & $0.76\pm0.04$ & $399^{+14}_{-14}$ & $191^{+7}_{-7}$ & $180^{+6}_{-8}$ & $404$ & $6.24^{+2.79}_{-3.61}$ & $4.34^{+1.70}_{-1.74}$ & $7.34^{+6.57}_{-4.74}$ & $6.31\pm1.31$\\
3444.02* & $4.93^{+0.44}_{-0.46}$ & $6.11^{+1.11}_{-1.70}$ & $5.29^{+1.63}_{-2.45}$ & $4.98\pm0.28$ & $238^{+8}_{-8}$ & $114^{+4}_{-4}$ & $107^{+3}_{-4}$ & $240$ & $0.78^{+0.35}_{-0.45}$ & $0.54^{+0.21}_{-0.22}$ & $0.92^{+0.82}_{-0.59}$ & $0.79\pm0.16$\\
3444.03* & $0.49^{+0.04}_{-0.05}$ & $0.61^{+0.11}_{-0.16}$ & $0.53^{+0.16}_{-0.24}$ & $0.50\pm0.03$ & $675^{+24}_{-25}$ & $323^{+12}_{-13}$ & $305^{+9}_{-13}$ & $682$ & $50.64^{+22.61}_{-29.31}$ & $35.19^{+13.75}_{-14.09}$ & $59.58^{+53.27}_{-38.46}$ & $51.07\pm10.65$\\
3444.04* & $0.74^{+0.07}_{-0.07}$ & $0.91^{+0.18}_{-0.25}$ & $0.79^{+0.25}_{-0.37}$ & $0.74\pm0.05$ & $386^{+14}_{-14}$ & $185^{+7}_{-7}$ & $174^{+6}_{-8}$ & $390$ & $5.39^{+2.41}_{-3.12}$ & $3.74^{+1.46}_{-1.50}$ & $6.34^{+5.67}_{-4.09}$ & $5.44\pm1.14$\\
3497.01 & $1.95^{+0.41}_{-0.42}$ & $3.72^{+0.83}_{-0.82}$ & $4.31^{+0.86}_{-0.91}$ & $0.80\pm0.12$ & $573^{+45}_{-47}$ & $402^{+31}_{-31}$ & $348^{+26}_{-27}$ & $276$ & $13.40^{+0.16}_{-1.23}$ & $3.75^{+0.23}_{-0.18}$ & $2.47^{+0.10}_{-0.13}$ & $1.38\pm0.58$\\
\hline
\end{tabular}
\begin{list}{}{}
\item[Note. ]The revised and \citet{Thompson2018} planetary radii, instellations, and equilibrium temperatures. Candidate planets are indicated with *.
\end{list}
\end{table}
\end{landscape}

\subsection{Revised Planet Parameters}

Table~\ref{tab:planet results} presents revised planetary radii, equilibrium temperatures, and instellation fluxes for the planets in the sample of triples. Values for whether the primary, secondary, or tertiary star is the host are all presented for completeness. Almost all of the systems undergo significant revisions to their radii, equilibrium temperatures, and instellation fluxes based on their original \kepler parameters regardless of which star in the triple is assumed to be the planet host. Based on the \kepler measured radii, our planetary sample contains four rocky planets (0.5--1\,\rearth), one super-Earth (1.0--1.75\,\rearth), two sub-Neptunes (1.75--3.5\,\rearth), two sub-Jovians (3.5--6\,\rearth) and one Jovian (6\,\rearth--14.3\,\rearth). With our revised planetary radii one of the rocky planets (KOI-3497.01) is a sub-Neptune regardless of which star is the host and the super-Earth (KOI-2626.01) is either a sub-Neptune if hosted by the primary or a sub-Jovian if hosted by the secondary or tertiary star. KOI-0854.01 only had a change of classification from sub-Neptune to sub-Jovian if the tertiary was assumed to be the host star, and KOI-3444.02 was revised from a sub-Jovian to a Jovian if hosted by the secondary star. While the remaining six planets had revisions to their radii, these corrections did not result in a classification change regardless of which star was the host.

\begin{table*}
\centering
\setcounter{table}{6}
\renewcommand{\arraystretch}{1.4}
\caption{Linear fits for the inner and outer stellar companions of the compact triple-star systems}
\label{tab:linear_fits}
\begin{tabular}{lcccccccccc}
\hline
System & Inner/Outer & $t_0$  & $\Delta t$  & $\rho_0$  & $\theta_0$  & $\dot{\rho}$  & $\dot{\theta} \rho_0$ &\multicolumn{2}{c}{Gamma Value}\\ 
& & $(\mathrm{MJD})$ & $(\mathrm{yr})$ &$(\mathrm{mas})$ & $\left({ }^{\circ}\right)$ & $\left(\text{mas } \mathrm{yr}^{-1}\right)$ & $\left(\text{mas } \mathrm{yr}^{-1}\right)$ & Median $\pm 1\sigma$ & 95.4\% c.i. \\ \hline

KOI-0005 & I & 56940 & 4.87 & 30.40 $\pm$ 0.53 &  148.52 $\pm$ 0.80 &  0.60 $\pm$ 0.20 & 1.40 $\pm$ 0.21 &$66.8_{-7.7}^{+7.3}$ & 51.3, 81.6\\
KOI-0005 & O & 57480 & 7.84 & 142.32 $\pm$ 0.49 &  308.54 $\pm$ 0.19 &  1.63 $\pm$ 0.16 & 0.81 $\pm$ 0.16 &$26.5_{-5.1}^{+4.9}$ & 16.6, 36.8\\
\hline
KOI-0013 & I & -- & -- & -- & -- & -- & -- & -- & --  \\ 
KOI-0013 & O & 57850 & 8.10 & 1156.39 $\pm$ 0.16 &  99.91 $\pm$ 0.01 &  -0.40 $\pm$ 0.04 & -0.28 $\pm$ 0.06 &$35.2_{-6.1}^{+5.7}$ & 22.8, 46.7\\

\hline
KOI-0652 & I & 57392 & 6.00 & 65.47 $\pm$ 0.25 &  290.81 $\pm$ 0.17 &  0.14 $\pm$ 0.09 & 0.30 $\pm$ 0.06 &$65.2_{-14.0}^{+14.3}$ & 42.4, 90.0\\
KOI-0652 & O & 58239 & 8.99 & 1237.89 $\pm$ 0.54 &  93.29 $\pm$ 0.01 &  0.33 $\pm$ 0.14 & 0.22 $\pm$ 0.07 &$34.5_{-12.5}^{+16.8}$ & 8.5, 70.5\\

\hline
KOI-0854 & I & 57070 & 3.18 & 17.70 $\pm$ 0.58 &  222.20 $\pm$ 2.84 &  1.01 $\pm$ 0.37 & 2.56 $\pm$ 0.55 &$68.3_{-8.7}^{+8.1}$ & 49.9, 84.4\\
KOI-0854 & O & 57070 & 9.90 & 157.56 $\pm$ 2.05 &  180.88 $\pm$ 0.59 &  0.80 $\pm$ 0.56 & -0.45 $\pm$ 0.43 &$31.2_{-21.2}^{+30.2}$ & 0.0, 81.1\\

\hline
KOI-2032 & I & 58137 & 10.82 & 57.82 $\pm$ 0.74 &  124.62 $\pm$ 0.36 &  -1.26 $\pm$ 0.16 & -0.81 $\pm$ 0.07 &$32.6_{-3.8}^{+4.4}$ & 25.0, 41.4\\
KOI-2032 & O & 58137 & 10.82 & 1116.24 $\pm$ 0.62 &  318.08 $\pm$ 0.03 &  -0.00 $\pm$ 0.15 & -0.41 $\pm$ 0.15 &$75.6_{-15.9}^{+10.0}$ & 44.3, 90.0\\

\hline
KOI-2626 & I & 57211 & 5.93 & 96.97 $\pm$ 0.21 &  82.65 $\pm$ 0.06 &  -1.07 $\pm$ 0.11 & -0.51 $\pm$ 0.05 &$25.6_{-2.7}^{+3.2}$ & 20.3, 32.0\\
KOI-2626 & O & 57211 & 5.93 & 179.11 $\pm$ 0.16 &  21.69 $\pm$ 0.11 &  -0.97 $\pm$ 0.08 & -0.77 $\pm$ 0.17 &$38.6_{-7.2}^{+5.9}$ & 24.3, 50.4\\

\hline
KOI-3158 & I & -- & -- & -- & -- & -- & -- & -- & --  \\ 
KOI-3158 & O & 58030 & 4.89 & 1840.46 $\pm$ 0.17 &  252.90 $\pm$ 0.01 &  -0.82 $\pm$ 0.10 & 1.34 $\pm$ 0.17 &$58.5_{-4.7}^{+4.3}$ & 49.0, 67.0\\

\hline
KOI-3444 & I & -- & -- & -- & -- & -- & -- & -- & --  \\ 
KOI-3444 & O & 57945 & 1.84 & 1084.64 $\pm$ 0.32 &  190.23 $\pm$ 0.01 &  2.80 $\pm$ 0.50 & -0.17 $\pm$ 0.31 &$5.0_{-3.5}^{+5.4}$ & 0.0, 15.1\\

\hline
KOI-3497 & I & 57945 & 7.95 & 74.10 $\pm$ 0.20 &  25.63 $\pm$ 0.11 &  -2.49 $\pm$ 0.08 & 2.20 $\pm$ 0.05 &$41.5_{-1.1}^{+1.1}$ & 39.4, 43.7\\
KOI-3497 & O & 57945 & 7.95 & 808.22 $\pm$ 0.32 &  355.23 $\pm$ 0.02 &  0.00 $\pm$ 0.11 & 0.63 $\pm$ 0.08 &$83.4_{-7.4}^{+4.6}$ & 70.0, 90.0\\

\hline
\end{tabular}
\end{table*}

\section{Measuring Orbital arcs}
\label{sec:orbital_arcs}

For small regions of an orbit, the motion of a stellar companion relative to either the primary star or the barycenter of the inner binary is expected to appear to be linear. The observed astrometric motion of our companions can therefore be approximated as such, given that the separations in Table~\ref{tab:sample_table} imply an average orbital period of 1700 years. 

For inner orbits, we measured the astrometry for the fainter star relative to the brighter one. Linear models were fitted to the separations and PAs as a function of time in order to measure the instantaneous orbital motion at the mean epoch. We subtracted the mean epoch from each observational epoch so that the zeroth order coefficient in each fit provides a measurement for the separation and PA at this mean epoch ($\rho_0,\theta_0$). The first-order coefficients would therefore be equivalent to the linear motion per year ($\dot{\rho}, \dot{\theta}$). We convert the angular linear motion from degrees per year to mas per year by using the separation at the mean epoch ($\rho_0$) so that both first-order coefficients are in units of mas/yr. The values and errors for these coefficients have been calculated using the python package \textsc{numpy.polyfit}. For systems that had more than two epochs, we calculated the $\chi^2$ value for both the separation and PA linear fits and the probability of achieving this $\chi^2$ assuming the orbital motion was linear. Values of $p(\chi^2) < 0.05$ could indicate that a linear model is not a good fit to the data. We iterated this process starting with two epochs and adding an additional one until the probability was less than 0.05 or all epochs had been added. Only two systems showed evidence of non-linear motion: KOI-0005 and KOI-0652. KOI-0005~AB has a time baseline of approximately 18\% of the estimated orbital period, so our linear fit analysis uses the first 5 epochs out of a possible 7 ($\approx$16\% of its orbit). KOI-0652~BC has a total time baseline of approximately 10\% of the estimated orbital period, so our linear fit analysis uses the first 4 epochs out of a possible 6 ($\approx$7\% of its orbit).

For outer orbits, the astrometry in Table~\ref{tab:astrometry} is measured for the outer single star relative to the brightest star in the inner binary. In order to fit for astrophysically meaningful linear motion, this astrometry needs to be determined relative to the barycenter on the inner binary. The relative position of the inner binary is typically measured $\sim$10$\times$ more precisely than the outer companion and therefore the errors on the inner astrometry can be approximated as negligible. Assuming this, the position of the outer companion relative to the barycenter of the inner binary can be written as:

\begin{align}
    \label{eq:outer}
\Delta {\alpha^*}_{  \text{3-1}} + [M_\text{2} / (M_\text{1} + M_\text{2}) \times \Delta {\alpha ^*}_{\text{1-2}}] &= \Delta {\alpha^*}_{  \text{3-12}} + \mu_{{\alpha^*}, \text{3-12}} \times t \\
\Delta {\delta}_{  \text{3-1}} + [M_\text{2} / (M_\text{1} + M_\text{2}) \times \Delta {\delta}_{\text{1-2}}] &= \Delta {\delta}_{  \text{3-12}} + \mu_{{\delta}, \text{3-12}} \times t ,
\end{align}

where subscripts 1 and 2 represent the brighter and fainter star of the inner binary, subscript 12 indicates their barycentre, and subscript 3 is for the outer companion \citep{Dupuy2022a}. $\Delta {\alpha^*}$ is the relative right ascension equal to $\Delta\alpha\cos{\delta}$, $\Delta {\delta}$ is the relative declination, and $\mu$ is the linear motion at each epoch $t$, and $M$ are the stellar masses from Table~\ref{tab:masses}. The astrometry on the left-hand side can be found in Table~\ref{tab:astrometry}. We performed $10^4$ Monte-Carlo trials by randomly selecting the astrometry measurements from a Gaussian distribution with the same mean and standard deviation as the observations and randomly selecting stellar masses from the posterior distribution. The mean and standard deviation of these trials for each observational date was then used as the relative astrometry measurements within the linear motion calculations described above for the inner binaries. Table~\ref{tab:linear_fits} shows the results of the linear motion for both the inner binary and the outer companion relative to the barycentre of the inner binary. As the separations for the outer companion are in general much farther than the inner binary separation, the motion recorded in the same time baseline is often much smaller, and as such, none of these outer companions show evidence of non-linear motion. 

\begin{figure}
    \centering
    \includegraphics[width=\columnwidth]{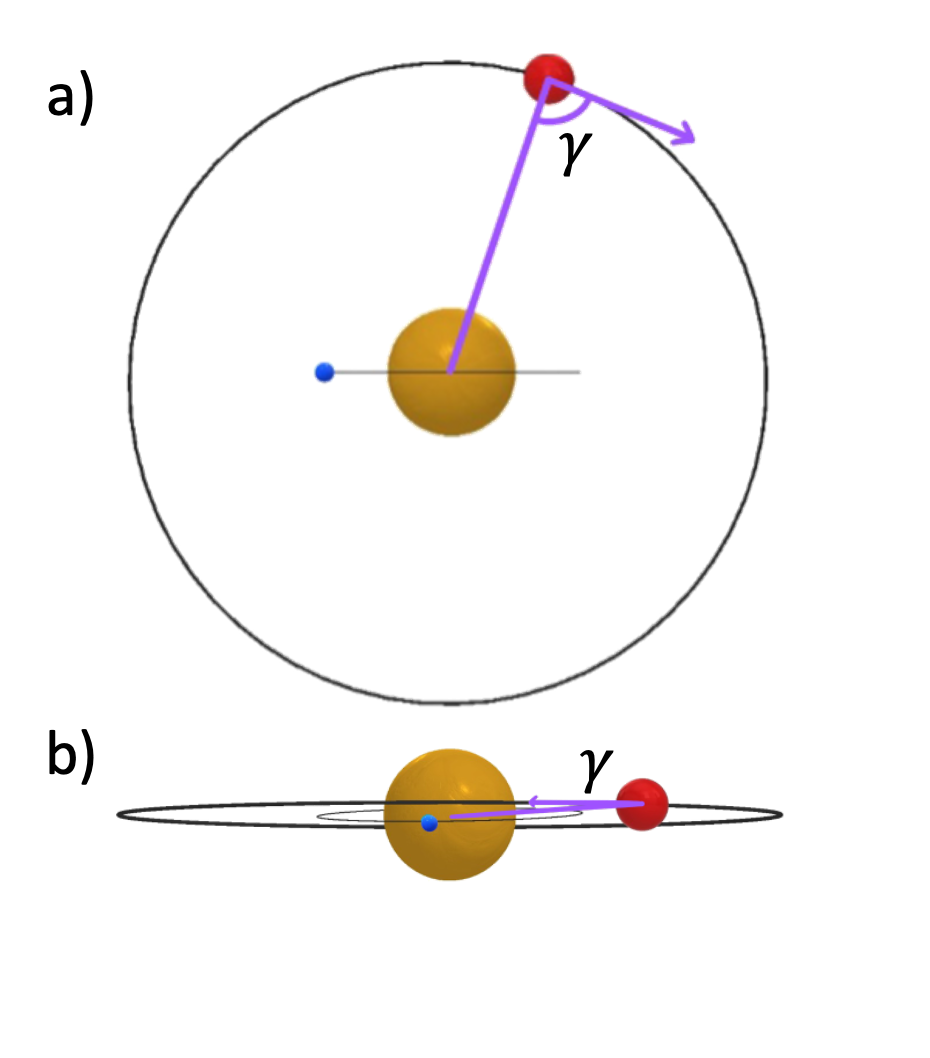}
    \caption{Pictogram depicting a planet (\textit{blue}) orbiting a primary star (\textit{yellow}), with a stellar companion (\textit{red}). Face-on orbits have $\gamma \sim 90 \degree$ (a) while edge-on orbits have $\gamma \sim 0 \degree$ (b). }
    \label{fig:gamma_pictogram}
\end{figure}

\section{Test of Orbital Alignment}

\begin{figure}
    \centering
    \includegraphics[width=\columnwidth]{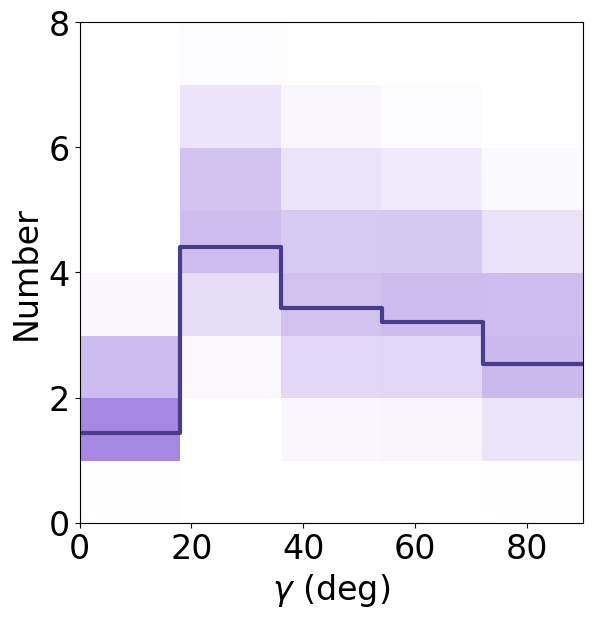}
    \caption{Histogram of $\gamma$ for both stellar orbits in our sample of 9 compact triples (15 total measurements). We account for the uncertainty in each measurement by drawing $10^5$ Monte Carlo samples, with the shaded regions depicting the fraction of trials that resulted in each bin as an indication of the spread of the histogram. The purple line corresponds to the average over all trials.}
    \label{fig:gamma_histograms}
\end{figure}

Building on work done previously investigating orbital arcs of wide binaries by \citet{Tokovinin2015}, we use the angle $\gamma$ as a test of orbital alignment. This is the angle between the line joining two stars and the star’s velocity vector. All systems in our sample have transiting planets that therefore have a nearly edge-on orbit ($83\degree < i < 90\degree$). If the planet orbits on the same plane as the stellar companion, and hence there is mutual alignment in the system, low values of $\gamma$ are expected. Figure~\ref{fig:gamma_pictogram} is a pictogram depicting $\gamma$ values close to $90 \degree$ for a face-on orbit where all the motion is in the $\theta$ direction and close to $0 \degree$ for an edge-on aligned orbit where all the motion is in the $\rho$ direction. However, other factors such as eccentricity and viewing angle can give a small value of $\gamma$ so the alignment can only be tested by a statistical sample. The angle $\gamma$ is computed using the equation:

\begin{align}
    \gamma  \equiv \arctan{(|\dot{\theta}|, |\dot{\rho}|)},
\end{align}

where the absolute value of the orbital motion has been used in order to limit $\gamma$ to 0--90$\degree$. We calculated  $\gamma$ for both the fainter companion in the inner binary (depicted in Figure~\ref{fig:gamma_pictogram}) and the outer companion relative to the barycenter of the inner binary. We propagated the errors of the linear motion using $10^5$ Monte-Carlo trials, randomly selecting values for the linear motion for each trial from Gaussian distributions with the same mean and standard deviations as the linear motion parameters. Table~\ref{tab:linear_fits} reports $\gamma$ for the 9 triples, including 1$\sigma$ and 2$\sigma$ confidence intervals from the Monte Carlo trials.   

\begin{figure*}
    \centering
    \includegraphics[width=\textwidth]{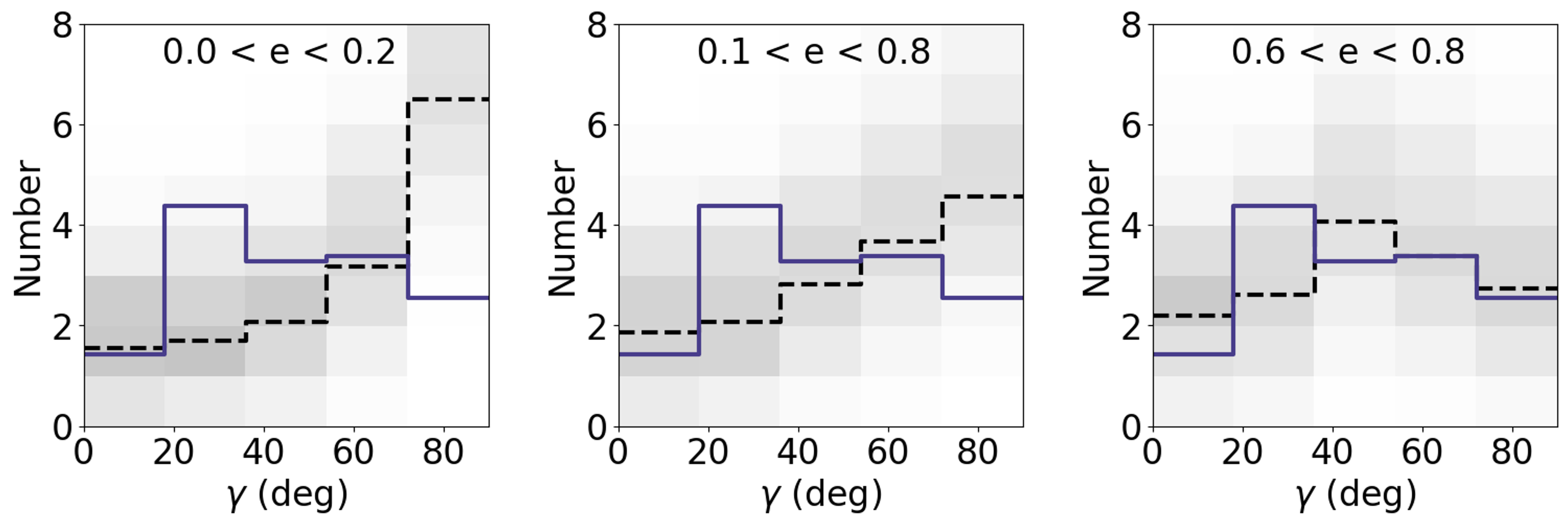}
    \caption{Histogram of the simulated angle $\gamma$ for orbits with isotropic inclinations, equivalent to the stellar orbit being uncorrelated to the planet's orbit. The eccentricity distributions have been varied from low (\textit{left}), field binary-like (\textit{middle}), and high (\textit{right}). The dashed line is the average over $10^5$ Monte Carlo trials, with the shading indicating what fraction of the trials have resulted in a value for that bin. The purple lines are the average measured gamma distributions shown in Figure~\ref{fig:gamma_histograms}.}
    \label{fig:isotropic_histograms}
\end{figure*}

Figure~\ref{fig:gamma_histograms} shows a histogram of all $\gamma$ values. Six of the 9 triples have both inner and outer $\gamma$ measurements, and 3 have outer $\gamma$ measurements only, totalling 15 angles altogether. There is a lack of systems with the lowest values of $\gamma$ (0$\degree$--18$\degree$) and an apparent peak at $18\degree$--$36\degree$. There is no evidence here for alignment within the systems, although, at angles $>20\degree$ there is a downward trend. This is not the expected result for the inner binaries, as previous work (e.g., \citealt{Dupuy2022, Christian2022, Behmard2022}) would suggest there should be some evidence of alignment in these systems. Small number statistics could, however, be distorting the distribution of $\gamma$ values measured. Another point to consider is that it is unknown for most systems which star the planet is around and therefore approximately half of the $\gamma$ angles do not correspond to the orbits of stars that host transiting planets. 

\subsection{Comparing to simulated orbit arcs}
\label{sec:one-pop}

\begin{figure*}
    \centering
    \includegraphics[width=\textwidth]{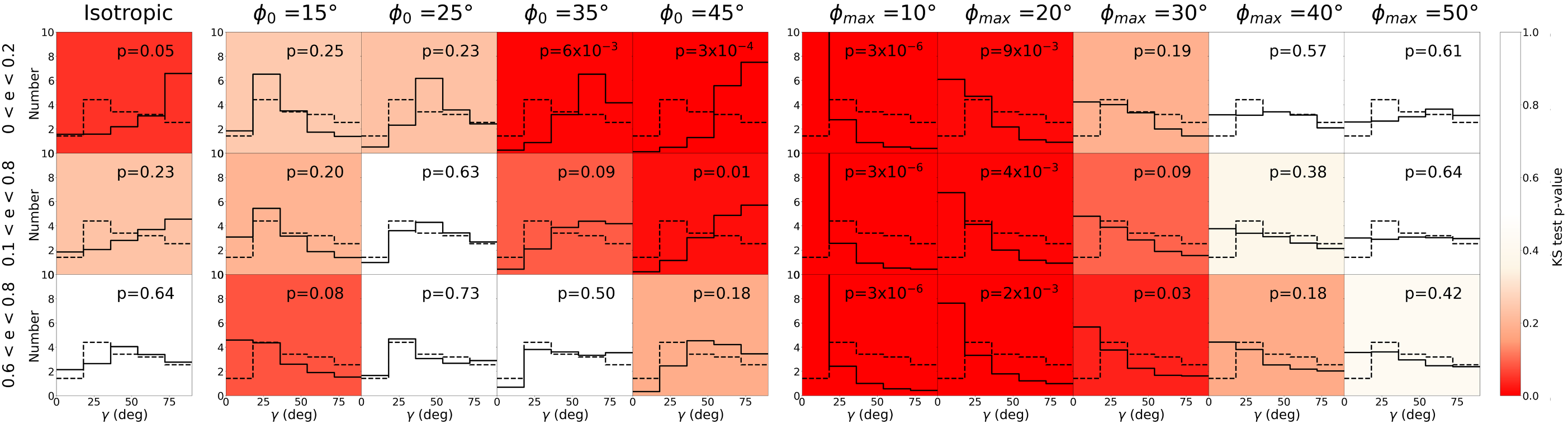}
    \caption{Histograms of simulated $\gamma$ values (\textit{solid}). The background colour of each plot indicates the p-value for a K-S test between the simulated cdf of $\gamma$ values against the measured $\gamma$ distribution (\textit{dashed}), averaged over $10^5$ Monte Carlo trials. Each plot represents a different simulation, where each row in each block has a different eccentricity distribution and each column has a different alignment scenario (isotropic, $\phi_0$, and $\phi_{\text{max}}$).}
    \label{fig:KS}
\end{figure*}

To provide a comparison to our measured $\gamma$ distribution, we simulated orbital arcs calculated following the technique of \citet{Dupuy2022}. Briefly, the orbital parameters to describe a complete orbit are chosen randomly from prior distributions. The argument of periastron ($\omega$) and PA of the ascending node ($\Omega$) were drawn from uniform distributions from 0--360$\degree$ while the period and semimajor axis were fixed at 1 and the time of periastron set to 0. The inclination is chosen from a distribution equal to $\arccos(\mathcal{U})$, where $\mathcal{U}$ is uniform from 0 to 1, simulating isotropic viewing angles. Regarding eccentricity, three distinct cases are considered: low eccentricity (uniform from 0 to 0.2), field binary (uniform from 0.1 to 0.8; \citealt{Raghavan2010}), and high eccentricity (uniform from 0.6 to 0.8). In order to calculate $\gamma$ for each synthetic orbit, random observation times were chosen from a uniform distribution from 0 to 1. At this time, the separation and position angle are calculated along with two other times of $\pm0.01$\% of the period before and after the random observation time. The linear motion is then computed as the average in the difference of the motion. 

The results from these simulations, where the orbital plane of the transiting planet is independent of the orbital plane of the stars, are shown in Figure~\ref{fig:isotropic_histograms} along with the measured average $\gamma$ distribution from Figure~\ref{fig:gamma_histograms}. Both the low-eccentricity and field-binary eccentricity distributions have maxima at high values of $\gamma$, rising up to $90\degree$. This is unlike the observed $\gamma$ distribution which peaks at $\approx$20--40$\degree$. However, the high eccentricity, isotropic scenario appears to better recreate the observed distribution. Based on the field-binary eccentricity distribution \citep{Raghavan2010}, high eccentricity systems are expected to be rare. While this eccentricity distribution gave the best fit, it is unlikely that our sample is comprised of only highly eccentric systems. 

To quantitatively compare our measured $\gamma$ distribution to the simulated results, we computed the Kolmogorov-Smirnov (K-S) statistic of the corresponding cumulative distribution function (cdf). This test calculates a probability ($p$-value) of the null hypothesis that two distributions were drawn from the same parent population. We calculated the K-S statistic for each of the $10^5$ Monte Carlo trials for the measured $\gamma$ distributions compared to the average cdf from our simulated orbits. We find no simulations with isotropic inclinations that reject the null hypothesis at $p<0.05$. The high-eccentricity distribution gave the best K-S statistic ($p=0.64$) while the low eccentricity scenario gave a $p$-value of 0.05, and the field binary eccentricity gave a $p$-value of 0.23. 

Instead of assuming isotropic viewing angles, we can simulate a range of mutual inclinations with respect to the inclination of the planet. Assuming the planet has an inclination $\approx90 \degree$, the mutual inclination between a stellar orbit and the planet's orbit ($\phi$) can be written:
\begin{equation}
\begin{aligned}
\label{eq:alignment_planet_1}
\cos{\phi_{\star{\rm -p}}} = \cos{(90\degree - i_\star)}\cos{(\Omega_\star- \Omega_{\text{p}})},
\end{aligned}
\end{equation}
where $i_\star$ is the inclination of the stellar orbit and $\Omega_\star$, $\Omega_{\text{p}}$ are the longitude of the ascending node for the stellar orbit and the planet's orbit respectively. A more thorough discussion of this equation is given in Section~\ref{planet_inc}. Rearranging this equation gives:
\begin{equation}
\begin{aligned}
\label{eq:inclination}
i_\star = \arcsin{\left (\frac{\cos{\phi_{\star{\rm -p}}}}{\Omega_\star- \Omega_{\text{p}}}\right )}
\end{aligned}
\end{equation}
$\Omega_{\text{p}}$ is an unknown quantity and therefore we simulate values for $\Omega_\star- \Omega_{\text{p}}$ directly by assuming a uniform distribution between 0$\degree$ to 360$\degree$, and ensuring that the absolute value of the cosine of this angle is larger than $|\cos{\phi_{\star{\rm -p}}}|$ so that the arcsin can be computed.

For the distribution of $\phi_{\star{\rm -p}}$, we look at two distinct cases. The first is that $\phi_{\star{\rm -p}}$ values are chosen randomly from a uniform distribution from $0\degree < \phi_{\star{\rm -p}} < \phi_{\text{max}}$. This is equivalent to star-planet alignment within $\phi_{\text{max}}$, where here we have chosen $\phi_{\text{max}}$ values from $10\degree$ to $50 \degree$ in $10\degree$ intervals. The second case is where the star and planet are misaligned by a narrowly specific amount, where $|\phi_{\star{\rm -p}} - \phi_0| < 5\degree$. We have tested $\phi_0$ values from $15\degree$ to $45 \degree$ in $10\degree$ intervals. A $\phi_0$ value of $5\degree$ would be equivalent to alignment with $\phi_{\text{max}}$ of $10 \degree$, so this duplicate is not included. Testing each of these mutual inclination scenarios with each of the three eccentricity distributions described above (low, field binary-like and high) results in 27 additional simulations, combined with the isotropic viewing angle simulations giving 30 unique simulations to compare against the observed $\gamma$ distribution. 

Figure~\ref{fig:KS} provides an overview of the results of these comparisons. The high eccentricity distribution with the narrowly distributed mutual inclinations from $20\degree < \phi_{\star{\rm -p}} < 30\degree$ ($\phi_0 = 25\degree$) gave the best K-S statistic ($p=0.73$). Similarly, going to a field binary-like distribution with the same mutual inclinations also provided an acceptable fit ($p=0.63$). While the high eccentricity with $\phi_0=35\degree$ also gave a relatively good K-S statistic ($p=0.51$), moving the other way to $\phi_0 = 15\degree$ gave a $p$-value of only 0.08. Even with high eccentricities, small and narrowly distributed mutual inclinations are not a good fit for the observed distribution. However, we rule out at the $p<0.05$ level scenarios with low eccentricity distributions and mutual inclinations between $\phi_0 = 35\degree$--$45\degree$ ($p$-values of 0.006 and 0.0003 respectively). We also rule out field binary-like eccentricity distributions and $\phi_0 =45\degree$ with a $p$-value of 0.01.

For the cases instead where the mutual inclinations were sampled from $0\degree$ to a maximum angle $\phi_{\text{max}}$, the scenarios with a higher value of $\phi_{\text{max}}$ gave a better fit to the observed distribution. A field binary-like eccentricity distribution with $\phi_{\text{max}} = 50\degree$ gave the second-best fit overall with a $p$-value of 0.64. Acceptable simulations also extended to both the low ($p=0.61$) and high ($p=0.42$) eccentricity distributions with $\phi_{\text{max}} = 50\degree$. Seven simulations produced poor fits with the observed distributions and can be ruled out at $p<0.05$. For every eccentricity distribution, the lowest mutual inclinations ($\phi_{\text{max}} = 10\degree$--$20\degree$) all produced low K-S statistics ($3\times10^{-6} < p < 9\times10^{-3}$). We also find that $\phi_{\text{max}}=30\degree$ is ruled out for high eccentricities. Thus, the $\gamma$ distribution for our sample is not a good match to scenarios where the mutual inclination distribution spans coplanar to low-$\phi_{\star{\rm -p}}$ orbits for all eccentricity distributions. 

Overall our simulations resulted in $p$-values ranging from $3\times10^{-6}$ to 0.73, with the best matches being either high eccentricity distributions with narrowly distributed mutual inclinations (any $\phi_0$), isotropic mutual inclinations, or low/field binary-like eccentricities with moderately misaligned mutual inclinations ($\phi_0=15\degree$--35$\degree$ or $\phi_{\rm max}=30\degree$--50$\degree$). We rule out 10 of our 30 simulations at the  $p<0.05$ level, including 3 narrowly distributed mutual inclination scenarios ($\phi_0 = 35\degree$ or $\phi_0 = 45\degree$) and 7 low mutual inclination simulations across all eccentricity distributions. Ruling out the low mutual inclination scenarios suggests that the triples within the sample are not fully coplanar systems. 

\begin{figure}
    \centering
    \includegraphics[width=\columnwidth]{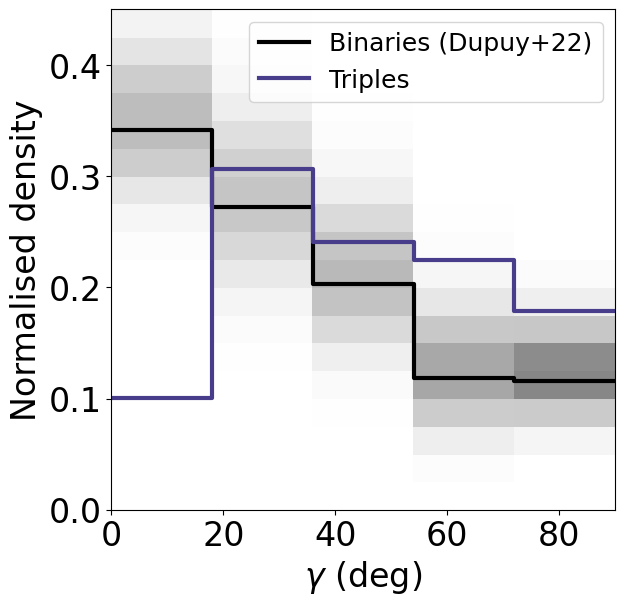}
    \caption{Normalised histogram of our $\gamma$ measurements compared to the sample of 42 binary KOIs from \citet{Dupuy2022}. The solid line corresponds to the average over $10^5$ Monte Carlo trials, and shading corresponds to the fraction of those trials resulting in that number for each bin. The \citet{Dupuy2022} distribution peaks at $\gamma < 15 \degree$, implying an abundance of edge-on orbits. This peak is not seen in the $\gamma$ distribution of our sample of triple systems (purple line).}
    \label{fig:binaries_histogram}
\end{figure}

\begin{figure*}
    \centering
    \includegraphics[width=\textwidth]{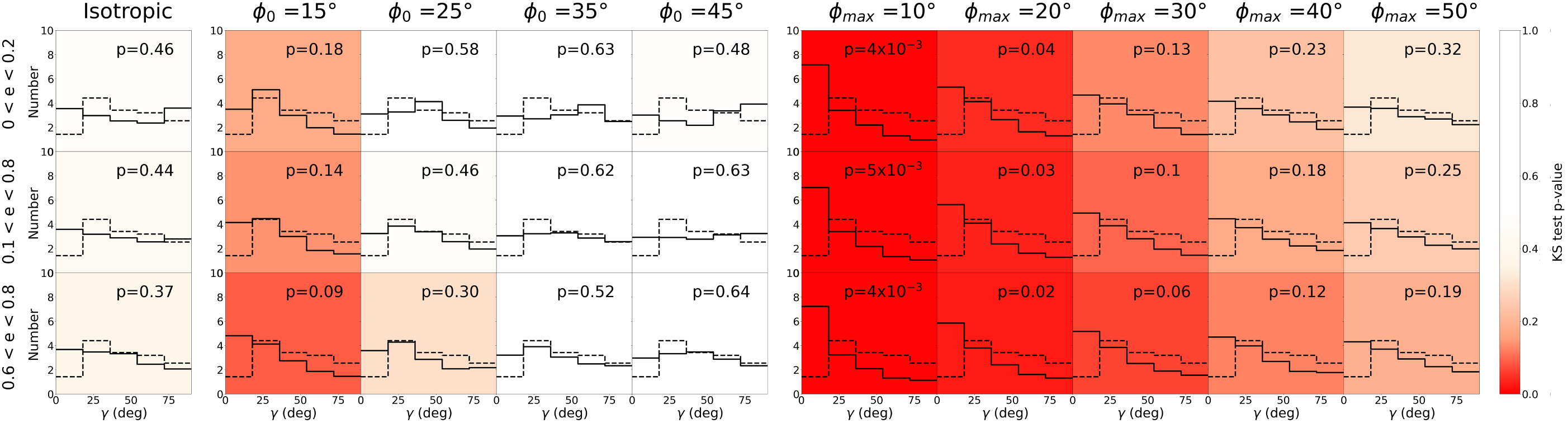}
    \caption{Histograms of simulated $\gamma$ values (\textit{solid}) for the two-population test. 60\% of each simulation is from the assumed planet-hosting distribution where $\phi_{\text{max}}=30\degree$ and the eccentricity distribution is field binary-like. For each plot, the other 40\% of the simulation has one of the 30 unique test distributions. Each row in each block has a different eccentricity distribution and each column has a different alignment scenario (isotropic, $\phi_0$ and $\phi_{\text{max}}$).
    The background colour of each plot indicates the p-value for a K-S test between the simulated cdf of $\gamma$ values against the measured $\gamma$ distribution (\textit{dashed}). }
    \label{fig:KS_binary}
\end{figure*}

\subsection{Comparing to planet-hosting binaries}

Previous work by \citet{Dupuy2022} performed an analysis on 45 KOI binaries with similar separations to our sample. KOI-0854, KOI-3158, and KOI-3444 were included in the \citet{Dupuy2022} binary sample and therefore have been removed from the binaries we compare to given that they are in our sample of triples. Figure~\ref{fig:binaries_histogram} shows the comparison of our distribution of $\gamma$ to their 42 binaries, similar to Figure~\ref{fig:gamma_histograms} but with a normalised density to show the comparison between the distribution for binaries and triples. 

\citet{Dupuy2022} found that there was an overabundance of low $\gamma$ values that, from similar simulations as described above, could only be explained by low mutual inclinations between the planet and stellar orbit. The simulated orbits that they found best matched their data used a field binary-like eccentricity distribution and uniform inclinations between $0\degree$ and $30\degree$ ($p=0.81$). In comparison, our sample of triples gives  $p=0.09$ for the same orbit simulation. It appears therefore that the $\gamma$ distribution for binaries does not match our distribution for the triple systems well. However, a 2-sample K-S test over $10^5$ Monte Carlo trials results in a $p$-value of 0.18 so we cannot rule out that these two data sets come from the same underlying distribution.

\subsection{Two-Population Tests}
\label{sec:twopop}

In the previous work of \citet{Dupuy2022}, all the $\gamma$ angles within the binary sample represent stellar orbits containing at least one star known to host a planet. This is not the case for our sample of triple systems. For the 6 visual triples where we have measured orbital motion for both the inner and outer companion, we have calculated 12 values of $\gamma$ in total. As each of these systems only has one known planet, these values represent an equal mix of orbits that include a planet-hosting star and orbits that only include non-planet-hosting stars. KOI-3444 contributes one $\gamma$ value for the outer orbit of the primary star relative to the inner binary. All 4 planets around KOI-3444 have been shown in Section~\ref{sec:FPA} to have a high probability of orbiting the primary so we class this as a $\gamma$ value associated with a planet-hosting stellar companion. For KOI-0013 and KOI-3158, their planets are known to be hosted by the primary and so the one visual orbit is also associated with the planet-hosting star. In total, 40\% (=6/15) of the $\gamma$ values are associated with orbits from non-planet-hosting stellar pairs, and 60\% (=9/15) are from stellar pairs that are planet-hosting. 

We consider therefore that a two-population model might be required to explain our observed distribution. As 60\% of the distribution is a result of planet-hosting stellar companions, we set 60\% of our combined model to have a field binary-like eccentricity distribution and a mutual inclination up to $\phi_{\text{max}}=30\degree$ to match the best case produced by previous work done by \citet{Dupuy2022}. For the remaining 40\% of the model distribution, we test each pairing of eccentricity and mutual inclination distribution as described above, resulting in 30 additional unique tests. 

Figure~\ref{fig:KS_binary} is a summary of these two-population models, plotted with the original $\gamma$ distribution for the triple systems. None of the models provided a better match for the observed distribution than the single population model (0.6 < e < 0.8 and $\phi_0 =25\degree$). However, overall they do give more acceptable simulations than the single-population models. For example, for isotropic orbits, a high eccentricity distribution was needed for a good match ($p>0.3$) for the single-population models, whereas all three tested eccentricity distributions for the two-population model resulted in acceptable matches ($0.37<p<0.46$) for isotropic orbits. In total, for the single-population models 9/30 (=30\%) of the tests resulted in a $p$-value of greater than 0.3, whereas the two-population models resulted in 12 (=40\%). The best fitting two-population model (60\% field binary-like eccentricities and $\phi_{\rm max} = 35\degree$) used non-planet-hosting orbits with narrow mutual inclinations of $\phi_0=35\degree$--$45\degree$ ($p=0.62$--0.64). These results again suggest that the triples within the sample are not completely coplanar, but instead are consistent with having at most one plane of alignment.

Triple star systems are, in general, not expected to all be mutually aligned. One of the major formation pathways to hierarchical triples is thought to be the separate formation of an inner binary and outer companion, with the gravitational interactions between the two systems forming a bound system. The introduction of a third stellar companion to the stable binary can cause Kozai-Lidov cycles where the inclination between the orbit of the inner binary and the orbit of the outer companion relative to the barycentre can vary periodically \citep{Toonen2016}. Short-period triples have been shown to have mutual inclinations that peak at $\sim 40\degree$ due to Kozai-Lidov cycles, which would correspond to the simulations where $\phi_0 = 35\degree$--$45\degree$ \citep{Fabrycky2007}. Unlike previous work on planet-hosting binaries (e.g., \citealt{Dupuy2022}), we exclusively consider Kozai-Lidov cycles as a result of stellar-stellar interactions and not planetary-stellar interactions. The longest Kozai-Lidov timescale in our sample is from the widest system, KOI-0652, which is $\sim 0.4$\,Myr. As this is less than the approximate age of the stars in our sample ($\sim 5$\,Gyr), it is feasible that Kozai-Lidov cycles could be operating in any of the triple systems. It is therefore interesting that the best match for the two-population model where the planet-hosting orbits are aligned within $30\degree$ is the non-planet hosting orbit being misaligned with exactly this range of mutual inclinations. While this is the two-population simulation that gives the best $p$-value, we note it is not a well-matched distribution to the observed histogram. While we are limited by the sample size, it is clear that a more complex model is needed to fully explain the shape of the observed distribution.

\section{Full orbital analysis}

\begin{table*}
    \centering
    \caption[]{Priors for both the \textsc{orvara} orbital fits and the \textsc{lofti} orbital fits.}
    \begin{tabular}{l|ll}
Parameter & \textsc{orvara} Prior & \textsc{lofti} Prior \\
\hline & \\[-1.5ex]
Eccentricity ($e$) & Uniform [0,1] & Uniform [0,1]\\
Inclination ($i$) & $\sin{(0\degree < i < 180\degree)} $ & $\sin{(0\degree < i < 180\degree)} $ \\
Argument of Perihelion ($\omega$) & N/A & Uniform [0,$2\pi$] \\
$\sqrt{e}\sin{\omega}$ & Uniform & N/A \\
$\sqrt{e}\cos{\omega}$ & Uniform & N/A \\
Semi-major axis (a) & 1/a (Log-flat) & N/A \\
Longitude of ascending node ($\Omega$) & Uniform [$-\pi$,$3\pi$]& N/A \\
Mean longitude at the ref. epoch of 2010.0 ($\lambda_{\text{ref}}$) & Uniform & N/A \\
Orbit Phase ($t_{\text{periastron}} - t_{\text{ref}})/\text{Period}$ & N/A & Uniform [0,1] \\
Total Mass ($M_T$) & N/A & Normal [$M_T$, $\sigma_{M_T}$]\\
Individual component masses ($M_{1/2}$) & 1/M (Log-flat) & N/A \\
Distance & N/A & Normal [$D_0,D_{std}$]\\
Parallax ($ \varpi $) & Normal [$\varpi_0,\varpi_{std}$] & N/A \\
\hline
    \end{tabular}
    \label{tab:priors}
\end{table*}

\begin{figure*}
    \centering
    \includegraphics[width=\textwidth]{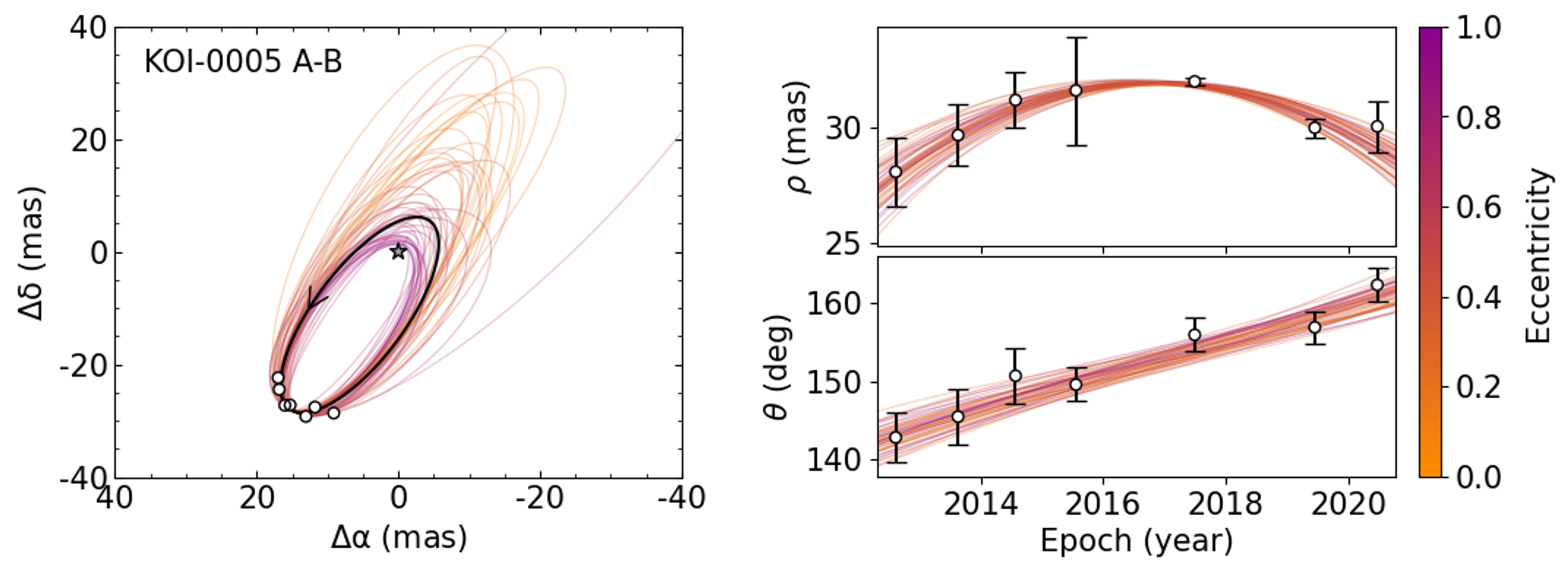}
    \caption{\textit{Left:} 50 random orbits from the posterior sample for the \textsc{orvara} fit for the inner binary of KOI-0005. The colour of the orbit indicates the eccentricity and the positions of the companion KOI-0005~B, relative to the primary KOI-0005~A (shown with a black star) are marked with white circles. \textit{Right:} The measured astrometry for the position angle and relative separation over time, overlaid by 50 possible orbital solutions.}
    \label{fig:5_orvara_orbit}
\end{figure*}

An additional method can be used to assess the alignment between the planetary and stellar orbital planes, separate from the $\gamma $ analysis. With accurate distances available for our sample, and our newly derived stellar parameters, we can perform a full Keplerian orbital analysis for these triple systems. From these complete orbits, we can use the inclination constraints to investigate the planetary-stellar alignment. The full orbital analysis also allows the alignment of the two stellar planes to be assessed which was not possible within the $\gamma$ analysis. As all systems are hierarchical triples, our orbital analysis is separated into the inner binary and the outer companion relative to the barycenter of the inner binary.

For the inner binary, we fit the relative orbit using \textsc{orvara} (v1.1.4; \citealt{Brandt2021}), a Markov chain Monte Carlo (MCMC) orbit fitter with an efficient eccentric anomaly solver. Although \textsc{orvara} has the capability to fit both RV and Hipparcos–Gaia Catalog of Accelerations (HGCA) astrometry, here we only have the astrometry in Table~\ref{tab:astrometry} available, apart from KOI-3158 which is discussed separately below. The posteriors of the orbital parameters are calculated using the affine-invariant \citep{Goodman2010} MCMC sampler \textsc{emcee} \citep{Foreman-Mackey2013} with parallel-tempering \citep{Vousden2016}. We fitted eight orbital parameters: eccentricity $e$, inclination $i$, argument of periastron $\omega$, position angle of the ascending node $\Omega$, semi-major axis $a$, mean longitude at the reference epoch of 2010.0 $\lambda_{\text{ref}}$, and masses of the components of the binary pair $M_1$ and  $M_2$. The default priors have been used for all of these parameters, shown in Table~\ref{tab:priors}, apart from the component masses for which we have adopted a Gaussian prior based on the masses in Table~\ref{tab:star results}. A Gaussian prior for the parallax has also been imposed, in which all the parallax measurements are based on \gaia~DR3 results apart from KOI-2626 which in the absence of \gaia\ data we have adopted a distance from \cite{Kraus2016}. The \textsc{orvara} orbital fits results are based on fits with 100 walkers and $5\times10^6$ steps for the MCMC, 5 temperatures for parallel tempering, 75\% burn-in and thinning to retain every 50th step. To ensure convergence we used the minimum steps and burn-in needed for the median and standard deviation were stable to within at least 20\% for all the systems.

Figure~\ref{fig:5_orvara_orbit} is an example of a sky-projected orbit fit for the inner binary of KOI-0005, showing the measured astrometry and 50 random accepted orbits. For KOI-3444, the tertiary component is resolved in only one epoch out of six. We therefore treat the centre of light for the unresolved component as the centre of mass in the remaining five epochs and fit the orbit using \textsc{orvara}. We also take this approach for the unresolved binary in KOI-0013. The orbit fits for these two systems, and the remaining five inner binaries, can be found in appendix~\ref{sec:appendix}. 

The orbits of the outer companions relative to the barycenter of the inner binary have all been fitted using the python package \textsc{lofti\_gaia} \citep{Pearce2020}, based on Orbits-For-The-Impatient (\textsc{ofti}; \citet{Blunt2017}). \textsc{lofti\_gaia} was designed to fit the orbital parameters of binaries that are resolved in Gaia using proper motions. This has been adapted instead to use the linear motions calculated in Section~\ref{sec:orbital_arcs} at the mean epoch for each system. We again used the stellar masses in Table~\ref{tab:star results} and the parallax to constrain the total mass and distance to the system. \textsc{ofti} using a rejection sampling method by computing orbits from random values of four orbital parameters ($e$, $\Omega$, $i$, and the orbital phase relative to time of periastron $\tau$) drawn from distributions of the priors shown in Table~\ref{tab:priors}. By scaling the semi-major axis and rotating the longitude of the ascending node to match the input parameters, the trial orbit is either rejected or accepted based on how well its linear motion matches the input linear motion. 

\begin{figure}
    \centering
    \includegraphics[width=\columnwidth]{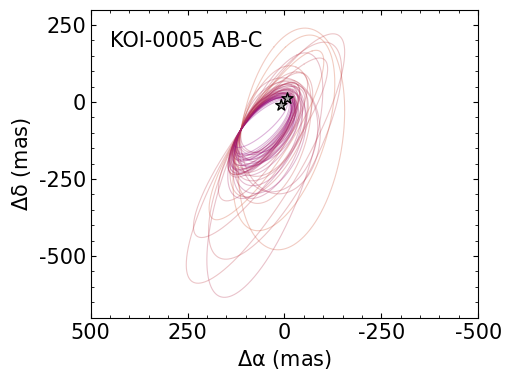}
    \caption{50 orbits from the posterior sample for the \textsc{lofti} fit for the outer companion KOI-0005~C relative to the host binary (black stars; separation of AB relative to C is not to scale). The colour of the orbit indicates the eccentricity. }
    \label{fig:5_lofti_orbit}
\end{figure}

Figure~\ref{fig:5_lofti_orbit} is an example of one \textsc{lofti} fit for the outer companion of KOI-0005 showing 50 random accepted orbits after running the fit until $10^6$ orbits were accepted. The orbits for the remaining 5 triples can be found in appendix~\ref{sec:appendix}. Table~\ref{tab:orbits} shows the orbital parameters calculated as a result of both the orbital analyses of the inner binary and the outer companion.

\begin{table*}
\centering
\renewcommand{\arraystretch}{1.4}
\caption{Orbital parameters from \textsc{orvara} and \textsc{lofti} orbit fits for both the inner binary and the outer stellar companion.}
\label{tab:orbits}
\begin{tabular}{lcccccccc}
\hline
System & Inner/Outer &  $a$ & $e$ & $i$ & $\omega$ & $\Omega$ & $P$ & $t_p$ \\
 &  & (au) &  & $(\degree)$ & $(\degree)$ & $(\degree)$ & (Yr) & (Yr) \\ [-0.5ex]
\hline & \\[-3.5ex]
KOI-0005 & I &  $13_{-3}^{+6}$ & $0.47_{-0.34}^{+0.39}$ & $68_{-28}^{+6}$ & $6_{-50}^{+36}$ & $152_{-7}^{+7}$ & $31_{-10}^{+21}$ & $2029_{-3}^{+10}$ \\
KOI-0005 & O &  $83_{-20}^{+68}$ & $0.74_{-0.20}^{+0.12}$ & $52_{-30}^{+17}$ & $4_{-15}^{+26}$ & $158_{-9}^{+9}$ & $430_{-140}^{+620}$ & $1943_{-59}^{+22}$ \\
\hline
KOI-0013 & I &  -- & -- & -- & -- & -- & -- & -- \\
KOI-0013 & O &  $560_{-180}^{+430}$ & $0.55_{-0.34}^{+0.33}$ & $107_{-6}^{+19}$ & $145_{-23}^{+33}$ &$113_{-13}^{+23}$ & $6200_{-2700}^{+8300}$ & $4000_{-1000}^{+5300}$ \\
\hline
KOI-0652 & I &  $22_{-5}^{+11}$ & $0.58_{-0.4}^{+0.33}$ & $73_{-26}^{+7}$ & $172_{-30}^{+25}$ & $111_{-5}^{+5}$ & $92_{-27}^{+74}$ & $2048_{-5}^{+18}$ \\
KOI-0652 & O &  $630_{-220}^{+580}$ & $0.55_{-0.38}^{+0.30}$ & $69_{-20}^{+9}$ & $42_{-58}^{+54}$ & $108_{-17}^{+26}$ & $11000_{-5000}^{+18000}$ & $-720_{-9000}^{+1300}$ \\
\hline
KOI-0854 & I &  $4.9_{-1.5}^{+5.2}$ & $0.77_{-0.38}^{+0.12}$ & $72_{-17}^{+58}$ & $93_{-57}^{+42}$ &$44_{-19}^{+11}$ & $12_{-5}^{+23}$ & $2015_{-3}^{+11}$ \\
KOI-0854 & O &  $45_{-16}^{+44}$ & $0.60_{-0.39}^{+0.29}$ & $106_{-17}^{+23}$ & $121_{-62}^{+55}$ & $172_{-42}^{+14}$ & $310_{-150}^{+550}$ & $1930_{-270}^{+40}$ \\
\hline
KOI-2032 & I &  $22_{-3}^{+10}$ & $0.7_{-0.33}^{+0.21}$ & $116_{-10}^{+29}$ & $153_{-51}^{+25}$ & $126_{-9}^{+6}$ & $68_{-15}^{+53}$ & $2037_{-2}^{+6}$ \\
KOI-2032 & O &  $580_{-180}^{+560}$ & $0.50_{-0.31}^{+0.26}$ & $122_{-14}^{+19}$ & $83_{-48}^{+41}$ & $135_{-30}^{+18}$ & $7900_{-3400}^{+14000}$ & $-1220_{-7040}^{+1450}$ \\
\hline
KOI-2626 & I &  $18_{-4}^{+9}$ & $0.78_{-0.42}^{+0.16}$ & $100_{-4}^{+12}$ & $116_{-36}^{+22}$ & $85_{-11}^{+8}$ & $93_{-29}^{+81}$ & $2046_{-8}^{+45}$ \\
KOI-2626 & O &  $40_{-14}^{+55}$ & $0.69_{-0.34}^{+0.23}$ & $107_{-8}^{+20}$ & $67_{-34}^{+53}$ & $39_{-17}^{+24}$ & $250_{-120}^{+650}$ & $1900_{-310}^{+50}$ \\
\hline
KOI-3158 & I &  -- & -- & -- & -- & -- & -- & -- \\
KOI-3158 & 0 &  $52_{-3}^{+3}$ & $0.550_{-0.048}^{+0.049}$ & $85.48_{-0.38}^{+0.35}$ & $226.9_{-5.2}^{+6.4}$ & $250.79_{-0.20}^{+0.18}$ & $324_{-26}^{+31}$ & $2233_{-23}^{+29}$ \\
\hline
KOI-3444 & I &  -- & -- & -- & -- & -- & -- & -- \\
KOI-3444 & 0 &  $92_{-30}^{+49}$ & $0.65_{-0.38}^{+0.29}$ & $93_{-2}^{+5}$ & $14_{-48}^{+47}$ & $8.9_{-3.8}^{+1.8}$ & $810_{-360}^{+820}$ & $2400_{-140}^{+440}$ \\
\hline
KOI-3497 & I &  $48_{-13}^{+15}$ & $0.18_{-0.12}^{+0.28}$ & $74_{-4}^{+3}$ & $12_{-46}^{+30}$ & $168_{-7}^{+5}$ & $290_{-110}^{+150}$ & $2100_{-40}^{+160}$ \\
KOI-3497 & O &  $210_{-70}^{+270}$ & $0.70_{-0.36}^{+0.14}$ & $65_{-19}^{+9}$ & $87_{-40}^{+33}$ & $98_{-75}^{+73}$ & $2300_{-1000}^{+5600}$ & $1100_{-2900}^{+360}$ \\
\hline
\end{tabular}
\end{table*}

\subsection{Kepler-444 (KOI-3158)}

\begin{figure*}[h]
    \centering
    \includegraphics[width=\textwidth]{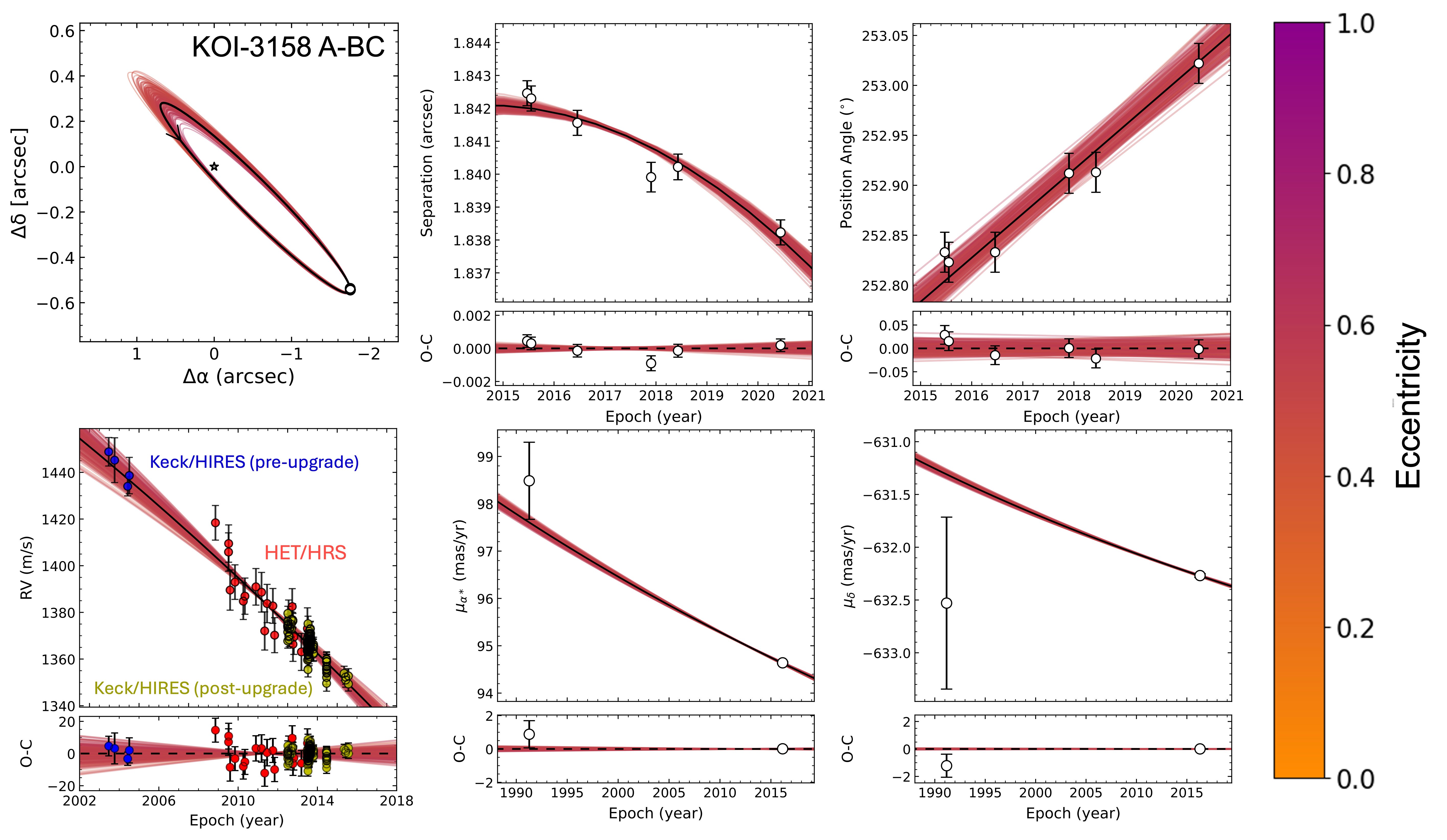}
    \caption{\textit{Top left:} 50 random orbits from the posterior sample for the \textsc{orvara} fit for the visual components of KOI-3158. The colour of the orbit indicates the eccentricity and the positions of the unresolved binary, KOI-3158 BC, relative to the primary (shown with a black star) are marked with white circles. \textit{Top middle and right:} The measured astrometry for the position angle and relative separation over time, overlaid by 50 possible orbital solutions. \textit{Bottom left:} Multi-epoch RVs from HET/HRS and Keck/HIRES overlaid by 50 possible orbital solutions. \textit{Bottom middle and right}: The absolute astrometry from Hipparcos (J1991.25) and Gaia EDR3 (J2016) overlaid by 50 possible orbital solutions.} 
    \label{fig:3158_orvara_orbit}
\end{figure*}

The orbit of the unresolved binary Kepler-444~BC around the planet-hosting primary Kepler-444~A has previously been studied by \citet{Dupuy2016}, \citet{Stalport2022} and \citet{Zhang2023}. Here, we provide a new, independent analysis of the orbit using specialized AO imaging astrometry measurements. Our data for this target span the time before and after the most recent Keck~II AO system realignment in 2015. When we observed this target, we ensured that the orientation of NIRC2 was fixed (north up) and that the primary was at the same $(x,y)$ pixel location on NIRC2. Given the nonlinear distortion of NIRC2 has never been shown to be variable over time, this observing strategy should enable higher accuracy astrometry than is normally possible. Thus, we used only observations of Kepler-444 after the 2015 NIRC2 realignment and also where the primary is within the box of $x=510$--520 and $y=520$--530 on NIRC2 in full frame mode coordinates. Our astrometry errors are thereby only limited by the error on the PA and pixel scale, which is 0.004 mas/pix \citep{Service2016}. 

Our orbital analysis uses other published data for the system. 
\citet{Dupuy2016} obtained spectra of KOI-3158A from 2012 July to 2015 July including three epochs of spectra of the companion KOI-3158BC using the HIRES spectrometer. \citet{Zhang2023} also obtained spectra from 2008 November to 2013 July of KOI-3158 A with the High Resolution Spectrograph (HRS) on the Hobby-Eberly Telescope (HET), including one epoch of the companion binary. 167 previously published RVs of the primary were also collated. \citet{Zhang2023} re-analysed the spectra of KOI-3158BC previously published by \citet{Dupuy2016}, including their additional spectra. They combined a system velocity ($\text{RV}_{\text{BC}}$ = -124.35 $\pm$ 0.11 km s$^{-1}$) with the known RV of the primary to derive a $\Delta$RV of -3.1 $\pm$ 0.2 km s$^{-1}$ at the epoch 2456783.1\,JD. 

In the \textsc{orvara} fit we combine the relative astrometry measurements, KOI-3158 A's multi-epoch RVs, the single-epoch relative RV and Hipparcos-Gaia absolute astrometry. We also impose a Gaussian prior on the mass of the primary of $M_A = 0.75 \pm 0.03 $\,\msun, following the method of \citet{Zhang2023}. Figure~\ref{fig:3158_orvara_orbit} presents a sky-projected orbit fit along with the separation, PA, absolute astrometry from Hipparcos and \gaia and the multi-epoch RVs as a function of time with these orbit solutions overlaid. The fit is run using 100 walkers, 5 temperatures for parallel tempering, $10^5$ steps with thinning to retain every 50th step and 10\% burn-in. Significantly less steps were needed for the solution to converge in comparison to the orbit fits described above that are based only on our astrometry. The fitted orbital characteristic solutions of this highly eccentric orbit are shown in Table~\ref{tab:orbits}. Our parameters and uncertainties are comparable to the results of orbital fit by \citet{Zhang2023}. Despite the uncertainties on the astrometry being on average $\sim4\times$ smaller than the astrometry used previously, our astrometry covers a smaller time baseline of five years in comparison to the previously used nine years. However, \citet{Zhang2023} showed that the orbital fit was dominated by the $\Delta$ RV and so the new astrometry has made very little impact on the measured orbital parameters.

\subsection{Mutual inclination between the stellar orbits}

The true mutual inclination of the orbital plane of the inner binary and the outer star can be measured using the equation:
\begin{equation}
\begin{aligned}
\label{eq:mut_align}
\cos{i_{\text{I-O}}} = \cos{i_{\text{I}}}\cos{i_{\text{O}}} + \sin{i_{\text{I}}}\sin{i_{\text{O}}}\cos{(\Omega_{\text{I}}- \Omega_{\text{O}})}
\end{aligned}
\end{equation}
where $i_{\text{I}}$ and $i_{\text{O}}$ are the inclinations of the inner binary and the outer component, $\Omega_{\text{I}}$ and $\Omega_{\text{O}}$ are the longitude of the ascending nodes for the inner binary and the outer component, and $i_{\text{I-O}}$ is the misalignment between these orbital planes \citep{Tokovinin2017}, equivalent to $\phi_{\star-\star}$.

To measure the inclinations and longitudes of the ascending nodes, complete orbits need to be fitted to the relative astrometry measurements. Visual orbits result in a $180 \degree$ ambiguity in $\Omega$ as they do not distinguish between the ascending and descending nodes without auxiliary information. The values in Table~\ref{tab:orbits} from the orbital analysis are quoted in the range of 0-180\degree apart from for KOI-3158 which due to the radial velocities does not have this uncertainty. The ambiguity in $\Omega$ is equivalent to a $\pm180\degree$ in the last term of Equation~\ref{eq:mut_align} and results in two values for the mutual inclination for each system, $\phi^+_{\star-\star}$ and $\phi^-_{\star-\star}$. 

Mutual inclination values for the triple systems have been calculated using values of $i$ and $\Omega$ from the orbit fits in Table~\ref{tab:mut_inc}. The posteriors for $i$ and $\Omega$ have been taken to produce $10^5$ values for the mutual inclination for each system. The median $\pm 1 \sigma$ values are shown in Table~\ref{tab:mut_inc}. There is no clear pattern of low values of mutual inclination between the stellar orbits, suggesting no preference for alignment. Due to the lack of precision in the orbital fits for the outer companion due to the small time baseline in the astrometry compared to the orbital periods, the values of the mutual inclination between the stellar orbital planes typically have large distributions and therefore large uncertainties. This, combined with the ambiguity in the mutual inclination angle and the small number of triples with visual orbits for both the inner and outer companions, means the alignment of the stellar orbits has a large uncertainty and we do not attempt a quantitative assessment. 

\citet{Tokovinin2017} found a strong preference for alignment in systems that had an outer component with a separation of $<50$\,au. Formation theories suggest that this is approximately the scale of the circumstellar disks which would have driven the evolution of the systems. For systems with a wider separation of larger than 1000 au, they found no tendency for alignment. Only 2 of our triple systems, KOI-0854 and KOI-2626, fall into the regime where the outer component is separated by less than 50\,au, so with such a small sample of extremely compact triples it is therefore unsurprising that we do not find any evidence of mutual stellar alignment.

\begin{table}
    \centering
    \renewcommand{\arraystretch}{1.5}
    \caption[]{Values for the mutual inclination between the stellar orbit of the inner binary and the outer companion relative to the barycenter.}
    \begin{tabular}{ccc}
System &\multicolumn{1}{c}{${\phi}^+_{\star-\star} (\degree)$ } &\multicolumn{1}{c}{${\phi}^-_{\star-\star} (\degree)$ } \\
& Median $\pm 1\sigma$ & Median $\pm 1\sigma$ \\
\hline
KOI-0005 & $24_{-14}^{+29}$ &  $111_{-39}^{+25}$ \\
KOI-0652 & $23_{-13}^{+28}$ &  $131_{-37}^{+19}$ \\
KOI-0854 & $83_{-34}^{+37}$ &  $103_{-39}^{+37}$ \\
KOI-2032 & $40_{-22}^{+33}$ &  $98_{-31}^{+28}$ \\
KOI-2626 & $48_{-22}^{+19}$ &  $120_{-22}^{+22}$ \\
KOI-3497 & $66_{-53}^{+50}$ &  $98_{-57}^{+41}$ \\
\hline
    \end{tabular}
    \label{tab:mut_inc}
\end{table}

\subsection{Mutual inclination between the stellar and planetary orbits}
\label{planet_inc}

\begin{figure*}
    \centering
    \includegraphics[width=\textwidth]{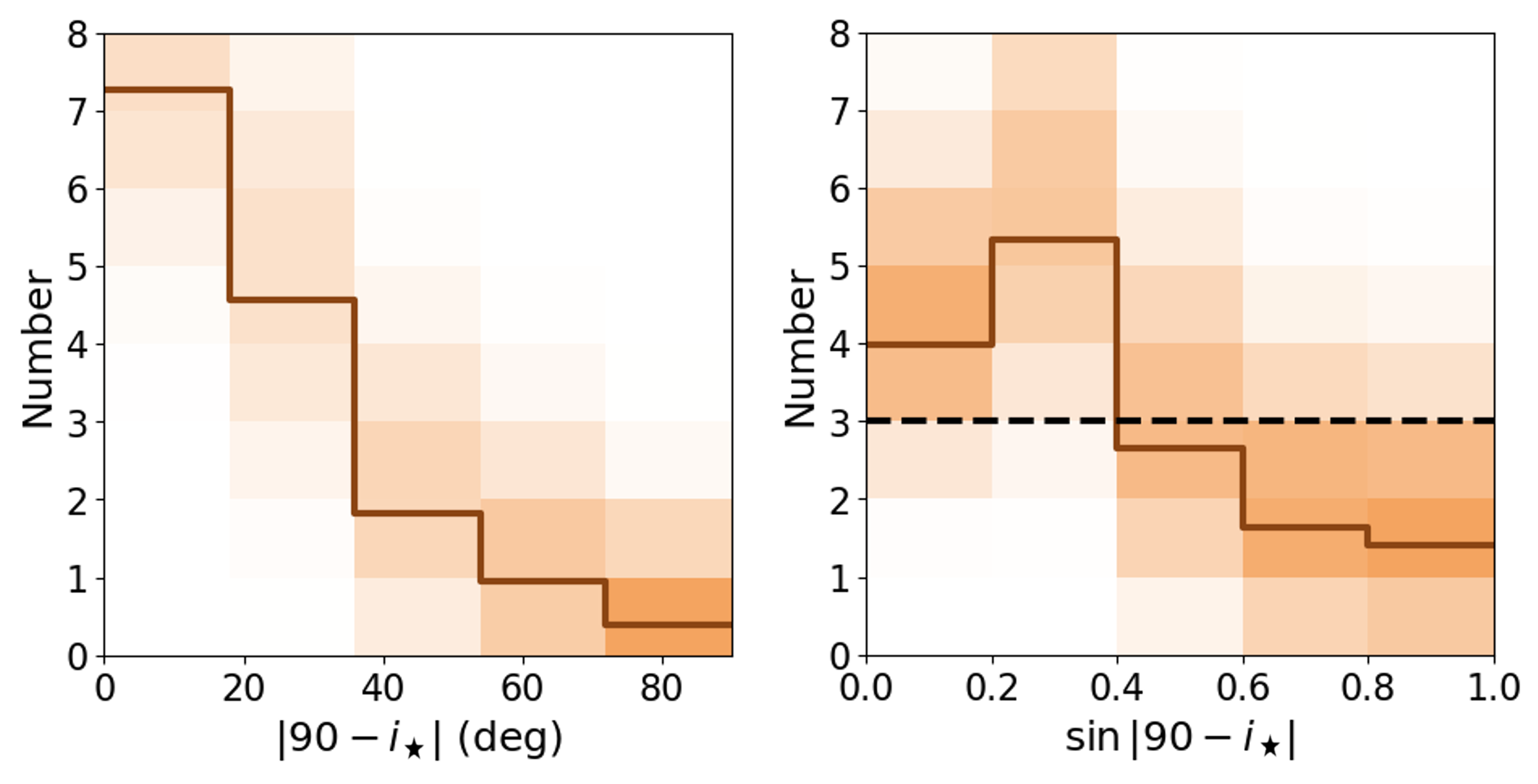}
    \caption{Histogram of $|90 - i_\star|$ (\textit{left}) and $\sin{(|90 - i_\star|})$ (\textit{right}) for all stellar orbital inclinations (both inner and outer pairs). We account for the uncertainty in each measurement by drawing $10^5$ Monte Carlo samples, with the shaded regions depicting the fraction of trials that resulted in each bin as an indication of the spread of the histogram. The solid line corresponds to the average over all trials. The dashed line indicates the distribution expected if the stellar orbital inclinations are drawn from an isotropic distribution. There is an overabundance of systems where stellar orbit inclination is close to the inclination of the planet (i $\approx 90 \degree$) implying that overall there is a tendency towards alignment between the stellar and planetary orbital planes in these systems.}
    \label{fig:90minusi}
\end{figure*}

In addition to studying the alignment of the stellar planes, the alignment of the stellar orbits and the planetary orbit can also be investigated. As it is not known which component hosts the planets, the alignment of both the inner binary and the outer companion's orbit against the planet's edge-on orbit can be measured.  

Equation~\ref{eq:mut_align} can be rewritten to investigate the planet alignment as follows:
\begin{equation}
\begin{aligned}
\label{eq:planet_middle}
\cos{i_{\star-\text{p}}} = \cos{i_\star}\cos{i_{\text{p}}} + \sin{i_\star}\sin{i_{\text{p}}}\cos{(\Omega_\star- \Omega_{\text{p}})},
\end{aligned}
\end{equation}
where $i_{\star-\text{p}}$ is the misalignment between the orbital plane of the planet and the host star ($\phi_{\star{\rm -p}}$), $i_\text{P}$ is the inclination of the planet, $i_\star$ is the inclination of the stellar orbit, $\Omega_{\text{p}}$ is the longitude of the ascending node for the planet and $\Omega_\star$ is the longitude of the ascending node for the stellar orbit. For the triple systems, the stellar orbits can either be the inner binary or the outer companion relative to the barycenter of the binary. 
 
The transiting planet must have an inclination of close to $90 \degree$ and so this equation simplifies to Equation~\ref{eq:alignment_planet_1}, where $\cos{\phi_{\star{\rm -p}}} \propto \cos{(90\degree - i_\star)}$.
In this case, the longitude of the ascending node for the planet is unknown so the true mutual misalignment $i_{\star-\text{p}}$ cannot be measured directly. Instead, $|90 - i_\star|$ is used as an equivalent to the minimum misalignment. If the longitude of the ascending node for the planet and stellar orbits were equal then $|90 - i_\star|$ would be equal to the true misalignment. Due to this, the mutual alignment between the planet and stellar orbits can only be investigated statistically. 

High values of $|90 - i_\star|$ from large relative inclinations are a result of misaligned systems. However, due to the unknown longitude of the ascending node, low values of $|90 - i_\star|$ do not necessarily correspond to aligned systems. As the longitude of the ascending node is expected to be distributed randomly, an overabundance of low $|90 - i_\star|$ values would suggest that there is more alignment in the systems than would be expected for random orbits.

Figure~\ref{fig:90minusi} shows a histogram for the minimum misalignment between the planet's orbit and both corresponding stellar orbits in each system, plotted both as $|90 - i_\star|$ and $\sin{(|90 - i_\star|})$. If the stellar orbital inclinations were drawn from an isotropic distribution, they would produce a flat distribution in $\sin{(|90 - i_\star|})$ space. There is an apparent overdensity of inclinations close to $90 \degree$ resulting in values of $\sin{(|90 - i_\star|})$ < 0.4 to be more common than expected from a flat distribution. While individual inclination measurements close to $90 \degree$ do not directly imply those systems are aligned due to the unknown longitude of the ascending node for the planet, a significant overdensity like this would imply that there is more alignment between the orbit of the planet and the stellar orbits than there would be if the orbits are random. However, performing a K-S test between the observed distribution of $\sin{(|90 - i_\star|})$ and the expected flat distribution for isotropic orbits resulted in a $p$-value of 0.085 and therefore we cannot rule out an underlying isotropic distribution of inclinations. 

These results are in broad agreement with the results from the $\gamma$ distribution. Both methods cannot rule out underlying isotropic orbits at a significant statistical level. However, in the full orbital analysis there is tentative evidence for an overabundance of aligned orbits seen in a peak of small values of $\sin{(|90 - i_\star|})$. From the $\gamma$ distribution there is also tentative evidence for alignment. For the one-population tests described in Section \ref{sec:one-pop}, all of the cases where the mutual inclination was less than $50\degree$, $40\degree$ or $30\degree$ (apart from the high eccentricity case) could not be ruled out. These scenarios are more aligned than what would be expected for isotropic orbits, so while we rule out highly aligned scenarios we again see tentative evidence for minor alignment in the planet-hosting triples. 

Many previous works have shown alignment between the stellar orbit and the planetary orbit in binaries (e.g., \citealt{Dupuy2022, Christian2022, Behmard2022, Lester2023}), and our results using two different methods provide tentative evidence for similar alignment in triples with an abundance of systems with inclinations close to $90 \degree$. The limiting factor in both methods is the sample size which may not be large enough to detect this alignment at the 2$\sigma$ level and therefore we cannot rule out isotropic orbits. Another point to consider is that it is unknown which star the planet is orbiting and therefore our distributions of $\sin{(|90 - i_\star|})$ include stellar pairs that host a planet as well as stellar pairs with no transiting planets. This combination could mean orbits of non-planet-hosting companions are attenuating the peak toward mutual alignment. As discussed in Section~\ref{sec:twopop}, Kozai-Lidov cycles may cause the mutual inclination of triple systems to vary periodically, and hence cause misalignment between the non-planet hosting stellar orbit and the edge-on orbit of the planet. With this possible pathway to misalignment of stellar orbits, it would then be unsurprising that the sample of stellar orbits as a whole does not result in significant evidence of alignment.

\section{Summary}

We present results from 12 years of astrometric orbit monitoring of 24 candidate triple star systems that host \kepler planets, including 9 compact systems where all three stellar components are within 600 au. Seven of the compact triple systems are fully spatially resolved, and two more, KOI-0013 and KOI-3158, have an unresolved inner companion. The goal of our observations is to determine the stellar orbital parameters and thereby statistically assess the alignment between the edge-on orbits of the transiting planets, the orbital planes of the inner stellar binaries, and the orbital planes of the outer stellar companions in these hierarchical triple systems. 

Our full sample includes compact visual triples identified with AO imaging as well as stellar pairs resolved in AO imaging that have an outer component identified with \gaia astrometry. We use Keck LGS AO imaging and non-redundant aperture masking of our sample of triple systems over multiple epochs to measure the separation, position angle, and magnitude of each component relative to the primary star. From this, we derive stellar parameters, including masses, and update the planetary radii from the initial one derived from the \kepler measurements assuming the star was single. We also rule out three candidate triples as the chance association of a physically bound binary and a background star. 

For the 7 fully resolved compact triple systems within the sample, we compare the stellar density distribution calculated from the stellar parameters to the distribution derived from the transit parameters to constrain which of the three stars in each system could be the host star. We find that only one planet is most likely to be hosted by the tertiary and one planet is most likely to be hosted by the secondary. The remaining planets were all most likely to be hosted by the primary, with one planet being consistent with only the primary. All of the planets in the sample were consistent with being hosted by at least one of the stellar components.

Using high-precision relative astrometry, we measured the linear motion in each of our systems. From this, we computed the angle $\gamma$ between the vector of orbital motion and the vector of the corresponding stellar pair as a test for alignment. As the transiting planets are in edge-on orbits, if the stellar orbits were aligned they would have motion in predominantly the separation direction and hence would have a small angle of $\gamma$. Our results are based on 15 $\gamma$ angles from 9 triple systems. 

We found that low mutual inclinations ($\phi=0$--$20\degree$) cannot explain the observed results for any of the three tested eccentricity distributions, suggesting that there is not a clear trend of both stellar planes being aligned with the plane of the planet. A single underlying distribution of high eccentricities ($0.6 < e < 0.8$) with a mutual inclination between the planetary and stellar orbit of $20\degree < \phi_{\star-\text{p}} < 30\degree$ was the best match to our observations, but our sample is unlikely to contain exclusively high eccentricity systems. However, a wide range of simulated distributions was consistent with the data, including any eccentricity distribution with $\phi_\text{max}=40\degree$, $\phi_\text{max}=50\degree$ or $\phi_0=25\degree$. Isotropic orbits with either a high eccentricity distribution or a field binary-like distribution were also consistent with the observed data. 

We tested two-population models assuming that each system only has one planet-hosting stellar pair and that their orbits follow an underlying distribution of mutual alignment up to $\phi = 30\degree$ with a field binary-like eccentricity distribution. This is modelled after the best-matching mutual inclination distribution found by \citet{Dupuy2022}. The non-planet-hosting orbits in our two-population tests could have any mutual inclination distribution, and the best fit was $40\degree < \phi < 50\degree $ with a field binary-like eccentricity distribution. These results were consistent with having a combination of stellar orbits aligned with the plane of the planet, and orbits from the non-planet hosting companion (either the outer companion relative to the planet-hosting binary, or the plane of the non-planet hosting binary) being consistent with either isotropic orbits or orbits driven by Kozai-Lidov cycles. These cycles can only influence orbits that are already misaligned, which is consistent with our finding that there is not a tendency for both stellar orbits in triple systems to be aligned with the planetary orbit.   

We used an additional independent method to test the alignment of the triple systems. The relative astrometry was used to fit complete sets of orbital parameters for the visual components of the compact triples. We used the resulting orbital angles ($i$ and $\Omega$) to directly calculate the mutual inclination between the two stellar orbital planes and constrain the mutual inclination between the planetary and stellar orbital planes. The results from this method are in broad agreement with the results from the $\gamma$ analysis. Again, isotropic orbits could not be ruled out at the 2$\sigma$ level, possibly because the sample size is not sufficiently large. The mutual inclination analysis also provided tentative evidence for an abundance of aligned systems, agreeing with the previous results that there is likely a combination of aligned planet-hosting stellar orbits and misaligned non-planet hosting stellar orbits with respect to the edge-on orbit of the transiting planet.

Our observations of multiple-star systems that host \kepler planets are ongoing. We aim to continue to monitor the 9 compact triples presented here to increase our orbital coverage and hence improve the precision of our orbital characteristics. There are also 3 candidate triple systems presented here that currently only have one epoch of observations each. We plan on monitoring these systems further, not only to verify their existence but also to include them in our sample once orbital motion is obtained.

Alignment tests using visual orbits can only be conducted using statistical samples and hence our small sample size limits its statistical power, despite being the largest sample of planet-hosting triple systems analysed to date. Our orbital studies are also severely hindered by the large distances to most of the \kepler planet hosts. At these distances, the spatial resolution of AO imaging results in wide stellar separations and periods that will be impossible to observe more than a fraction of in our lifetime. \textsl{TESS} planet hosts present a solution to both these problems by finding planets around nearby stars. Multiplicity surveys of these planet hosts will reach closer separations and thus should provide a larger sample of compact planet-hosting triple-star systems that will undergo faster orbital motion. Such a sample would allow the alignment of the planetary and stellar orbits in triple systems to be more rigorously investigated.

\section*{Acknowledgements}

We thank the anonymous referee for comments that improved our manuscript. T.~Dupuy acknowledges support from UKRI STFC AGP grant ST/W001209/1. DH acknowledges support from the Alfred P. Sloan Foundation, the National Aeronautics and Space Administration (80NSSC22K0781), and the Australian Research Council (FT200100871).
Some of the data presented herein were obtained at the W.M.\ Keck Observatory, which is operated as a partnership between the California Institute of Technology, the University of California, and NASA. The Observatory was made possible by the generous financial support of the W.M.\ Keck Foundation. 
This work has made use of data from the European Space Agency (ESA) mission Gaia, processed by the Gaia Data Processing and Analysis Consortium (DPAC). Funding for the DPAC has been provided by national institutions, in particular, the institutions participating in the Gaia Multilateral Agreement.
The authors thank Michael C. Liu and Mark W. Phillips for obtaining some of the Keck data presented here. We would also like to thank Lewis Warrey for their graphic design contributions to Figure~\ref{fig:gamma_pictogram}.
The authors wish to recognize and acknowledge the very significant cultural role and reverence that the summit of Maunakea has always had within the indigenous Hawaiian community. We are most fortunate to have the opportunity to conduct observations from this mountain. 

For the purpose of open access, the author has applied a Creative Commons Attribution (CC BY) licence to any Author Accepted Manuscript version arising from this submission.

\section*{Data Availability}
All of our NIRC2 data are available on the Keck Observatory Archive (KOA), which is operated by the W. M. Keck Observatory and the NASA Exoplanet Science Institute (NExScI), under contract with the National Aeronautics and Space Administration.

\onecolumn
\clearpage
\captionsetup{width=\textwidth,labelfont=bf}
\setcounter{table}{1}
\begin{longtable}{lcccccc}

\caption{Relative astrometry measurements of our KOIs with two stellar companions from our Keck/NIRC2 adaptive optics imaging and aperture-masking interferometry.}
\label{tab:astrometryv2}

\\ \hline \\[-1.5ex] \multicolumn{1}{c}{Name} & \multicolumn{2}{c}{Epoch} & \multicolumn{1}{c}{Separation} & \multicolumn{1}{c}{Position Angle}& \multicolumn{1}{c}{$\Delta m$}& \multicolumn{1}{c}{Filter}\\ [1.5ex]
& (UT) & (MJD) & (mas) & ($\degree$) & (mag) & \\ [1.5ex]
\endfirsthead

\multicolumn{3}{c}%
{{ \tablename\ \thetable{} \textit{: continued }}} \\
\\ \hline \\[-1.5ex] \multicolumn{1}{c}{Name} & \multicolumn{2}{c}{Epoch} & \multicolumn{1}{c}{Separation} & \multicolumn{1}{c}{Position Angle}& \multicolumn{1}{c}{$\Delta m$}& \multicolumn{1}{c}{Filter}\\ [1.5ex]
& (UT) & (MJD) & (mas) & ($\degree$) & (mag) & \\ [1.5ex] \hline
\endhead

\hline & \\[-1.5ex]
KOI-0005 AB & 2012-08-14 & 56153.45 & 28.1 $\pm$ 1.5 & 142.8 $\pm$ 0.9 & 0.20 $\pm$ 0.09 & $K'$ \\
KOI-0005 AB & 2013-08-20 & 56524.42 & 29.6 $\pm$ 1.5 & 146 $\pm$ 4 & 0.34 $\pm$ 0.09 & $K_\text{cont}$ \\
KOI-0005 AB & 2014-07-28 & 56866.45 & 31.1 $\pm$ 1.3 & 151 $\pm$ 3 & 0.34 $\pm$ 0.10 & $K'$ \\
KOI-0005 AB & 2015-07-22 & 57225.43 & 31.6 $\pm$ 2.2 & 149.6 $\pm$ 2.0 & 0.42 $\pm$ 0.08 & $K'$ \\
KOI-0005 AB & 2017-06-28 & 57932.40 & 32.02 $\pm$ 0.16 & 155.9 $\pm$ 1.9 & 0.26 $\pm$ 0.08 & $K'$ \\
KOI-0005 AB & 2019-06-12 & 58646.35 & 30.0 $\pm$ 0.4 & 156.9 $\pm$ 2.0 & 0.22 $\pm$ 0.08 & $K'$ \\
KOI-0005 AB & 2020-06-18 & 59018.58 & 30.1 $\pm$ 1.0 & 162.3 $\pm$ 1.9 & 0.30 $\pm$ 0.06 & $K'$ \\
KOI-0005 AC & 2012-08-14 & 56153.45 & 120.2 $\pm$ 1.3 & 305.8 $\pm$ 1.5 & 1.800 $\pm$ 0.021 & $K'$ \\
KOI-0005 AC & 2013-08-20 & 56524.42 & 125.6 $\pm$ 1.4 & 305.7 $\pm$ 0.4 & 1.97 $\pm$ 0.04 & $K_\text{cont}$ \\
KOI-0005 AC & 2014-07-28 & 56866.45 & 127.0 $\pm$ 0.9 & 305.1 $\pm$ 0.8 & 1.98 $\pm$ 0.08 & $K'$ \\
KOI-0005 AC & 2015-07-22 & 57225.43 & 128.4 $\pm$ 1.4 & 306.04 $\pm$ 0.26 & 2.00 $\pm$ 0.07 & $K'$ \\
KOI-0005 AC & 2017-06-28 & 57932.40 & 130.4 $\pm$ 0.7 & 305.8 $\pm$ 0.3 & 1.93 $\pm$ 0.04 & $K'$ \\
KOI-0005 AC & 2019-06-12 & 58646.35 & 134.5 $\pm$ 1.2 & 306.66 $\pm$ 0.22 & 1.90 $\pm$ 0.04 & $K'$ \\
KOI-0005 AC & 2020-06-18 & 59018.58 & 137.7 $\pm$ 0.5 & 306.75 $\pm$ 0.23 & 1.97 $\pm$ 0.03 & $K'$ \\
\hline
KOI-0013 A-BC & 2013-06-13 & 56456.48 & 1157.2 $\pm$ 1.1 & 280.00 $\pm$ 0.08 & 0.16 $\pm$ 0.04 & $K'$ \\
KOI-0013 A-BC & 2013-08-07 & 56511.35 & 1157.87 $\pm$ 0.23 & 279.948 $\pm$ 0.021 & 0.128 $\pm$ 0.007 & $K'$ \\
KOI-0013 A-BC & 2020-06-18 & 59018.57 & 1156.2 $\pm$ 0.5 & 279.870 $\pm$ 0.010 & 0.204 $\pm$ 0.007 & $K'$ \\
KOI-0013 A-BC & 2021-07-19 & 59414.47 & 1154.46 $\pm$ 0.24 & 279.840 $\pm$ 0.008 & 0.242 $\pm$ 0.008 & $K'$ \\
\hline
KOI-0288 AB & 2012-08-14 & 56153.43 & 347.30 $\pm$ 0.09 & 319.38 $\pm$ 0.05 & 3.101 $\pm$ 0.009 & $K'$ \\
KOI-0288 AB & 2014-07-28 & 56866.54 & 347.46 $\pm$ 0.26 & 319.60 $\pm$ 0.06 & 3.083 $\pm$ 0.009 & $K'$ \\
KOI-0288 AB & 2020-06-18 & 59018.64 & 350.8 $\pm$ 1.8 & 319.98 $\pm$ 0.11 & 3.07 $\pm$ 0.04 & $K'$ \\
\hline
KOI-0307 AB & 2019-06-12 & 58646.53 & 68.51 $\pm$ 0.27 & 241.4 $\pm$ 0.6 & 0.17 $\pm$ 0.06 & $K'$ \\
KOI-0307 AB & 2021-07-19 & 59414.38 & 64.1 $\pm$ 0.5 & 239.2 $\pm$ 0.3 & 0.15 $\pm$ 0.04 & $K'$ \\
KOI-0307 AB & 2022-07-05 & 59765.39 & 61.9 $\pm$ 0.5 & 239.70 $\pm$ 0.05 & 0.0600 $\pm$ 0.0006 & $K'$ \\
\hline
KOI-0652 AB & 2014-06-12 & 56820.44 & 1210.2 $\pm$ 0.8 & 272.826 $\pm$ 0.016 & 0.70 $\pm$ 0.04 & $K_\text{cont}$ \\
KOI-0652 AB & 2014-07-18 & 56856.49 & 1209.7 $\pm$ 1.3 & 272.83 $\pm$ 0.05 & 0.77 $\pm$ 0.08 & $K'$ \\
KOI-0652 AB & 2014-08-13 & 56882.35 & 1209.24 $\pm$ 0.17 & 272.876 $\pm$ 0.016 & 0.717 $\pm$ 0.018 & $K'$ \\
KOI-0652 AB & 2020-06-10 & 59010.59 & 1211.3 $\pm$ 0.4 & 272.840 $\pm$ 0.015 & 0.80 $\pm$ 0.04 & $K'$ \\
KOI-0652 AB & 2022-07-06 & 59766.37 & 1213.9 $\pm$ 1.0 & 272.845 $\pm$ 0.023 & 0.811 $\pm$ 0.019 & $K'$ \\
KOI-0652 AB & 2023-06-08 & 60103.61 & 1214.02 $\pm$ 0.16 & 272.903 $\pm$ 0.012 & 0.738 $\pm$ 0.025 & $K'$ \\
KOI-0652 BC & 2014-06-12 & 56820.44 & 65.0 $\pm$ 1.2 & 290.9 $\pm$ 1.2 & 1.09 $\pm$ 0.07 & $K_\text{cont}$ \\
KOI-0652 BC & 2014-07-18 & 56856.49 & 64.9 $\pm$ 0.6 & 289.7 $\pm$ 2.4 & 1.03 $\pm$ 0.08 & $K'$ \\
KOI-0652 BC & 2014-08-13 & 56882.35 & 65.4 $\pm$ 0.4 & 290.43 $\pm$ 0.22 & 1.025 $\pm$ 0.021 & $K'$ \\
KOI-0652 BC & 2020-06-10 & 59010.59 & 66.1 $\pm$ 0.4 & 291.96 $\pm$ 0.24 & 0.983 $\pm$ 0.022 & $K'$ \\
KOI-0652 BC & 2022-07-06 & 59766.37 & 64.40 $\pm$ 0.27 & 292.4 $\pm$ 0.5 & 0.996 $\pm$ 0.011 & $K'$ \\
KOI-0652 BC & 2023-06-08 & 60103.61 & 62.41 $\pm$ 0.27 & 295.4 $\pm$ 0.5 & 1.046 $\pm$ 0.025 & $K'$ \\
\hline
KOI-0854 AB & 2013-07-17 & 56490.53 & 16.1 $\pm$ 1.0 & 209 $\pm$ 5 & 0.30 $\pm$ 0.23 & $K' + 9H$  \\
KOI-0854 AB & 2016-09-20 & 57651.29 & 19.3 $\pm$ 0.6 & 235.4 $\pm$ 2.7 & -0.050 $\pm$ 0.009 & $K' + 9H$ \\
KOI-0854 AC & 2013-07-17 & 56490.53 & 154.6 $\pm$ 0.6 & 181.5 $\pm$ 0.5 & 3.65 $\pm$ 0.11 & $K'$ \\
KOI-0854 AC & 2014-07-29 & 56867.40 & 153.2 $\pm$ 2.6 & 181.4 $\pm$ 0.9 & 3.81 $\pm$ 0.09 & $K'$ \\
KOI-0854 AC & 2017-07-01 & 56867.50 & 162 $\pm$ 4 & 179.5 $\pm$ 2.7 & 3.82 $\pm$ 0.30 & $K'$ \\
KOI-0854 AC & 2023-06-09 & 60104.48 & 159 $\pm$ 7 & 180.0 $\pm$ 1.5 & 3.60 $\pm$ 0.13 & $K'$ \\
\hline
KOI-1613 AB & 2012-08-14 & 56153.39 & 211.69 $\pm$ 0.21 & 184.489 $\pm$ 0.029 & 1.044 $\pm$ 0.010 & $K'$ \\
KOI-1613 AB & 2013-08-25 & 56529.27 & 210.7 $\pm$ 0.4 & 184.45 $\pm$ 0.10 & 1.057 $\pm$ 0.010 & $K'$ \\
KOI-1613 AB & 2014-08-13 & 56882.45 & 209.11 $\pm$ 0.04 & 184.59 $\pm$ 0.05 & 1.0544 $\pm$ 0.0022 & $K'$ \\
KOI-1613 AB & 2015-07-27 & 57230.48 & 206.8 $\pm$ 0.5 & 184.64 $\pm$ 0.06 & 1.14 $\pm$ 0.04 & $K'$ \\
KOI-1613 AB & 2016-06-16 & 57555.57 & 206.31 $\pm$ 0.07 & 184.528 $\pm$ 0.024 & 1.0433 $\pm$ 0.0023 & $K'$ \\
KOI-1613 AB & 2016-07-15 & 57584.48 & 206.15 $\pm$ 0.18 & 184.60 $\pm$ 0.04 & 1.070 $\pm$ 0.005 & $K'$ \\
KOI-1613 AB & 2017-07-07 & 57941.34 & 204.58 $\pm$ 0.15 & 184.43 $\pm$ 0.04 & 1.053 $\pm$ 0.006 & $K'$ \\
KOI-1613 AB & 2018-06-07 & 58276.57 & 203.27 $\pm$ 0.09 & 184.532 $\pm$ 0.021 & 1.053 $\pm$ 0.005 & $K'$ \\
KOI-1613 AB & 2023-06-09 & 60104.35 & 194.87 $\pm$ 0.07 & 184.61 $\pm$ 0.04 & 1.0461 $\pm$ 0.0010 & $K'$ \\
\hline
KOI-1615 AB & 2012-07-06 & 56114.62 & 31.8 $\pm$ 1.6 & 122.0 $\pm$ 1.6 & 1.81 $\pm$ 0.10 & $K' + 9H$ \\
KOI-1615 AB & 2014-07-30 & 56868.55 & 30.2 $\pm$ 2.8 & 138.5 $\pm$ 2.8 & 2.23 $\pm$ 0.20 & $K' + 9H$ \\
KOI-1615 AB & 2014-11-30 & 56991.20 & 23.4 $\pm$ 2.2 & 139 $\pm$ 4 & 1.76 $\pm$ 0.29 & $K' + 9H$ \\
KOI-1615 AB & 2016-09-20 & 57651.27 & 17.5 $\pm$ 0.9 & 146 $\pm$ 3 & 0.81 $\pm$ 0.27 & $K' + 9H$ \\
\hline
KOI-1961 AB & 2014-07-31 & 56869.49 & 34.60 $\pm$ 0.20 & 258.10 $\pm$ 0.20 & 0.155 $\pm$ 0.008 & $K' + 9H$ \\
KOI-1961 AB & 2015-07-21 & 57224.39 & 36.99 $\pm$ 0.15 & 261.55 $\pm$ 0.29 & 0.190 $\pm$ 0.007 & $K' + 9H$ \\
KOI-1961 AB & 2016-11-07 & 57699.21 & 40.1 $\pm$ 0.4 & 263.5 $\pm$ 0.8 & 0.234 $\pm$ 0.028 & $K' + 9H$ \\
KOI-1961 AB & 2017-07-01 & 57935.50 & 42.08 $\pm$ 0.11 & 268.37 $\pm$ 0.29 & 0.160 $\pm$ 0.008 & $K' + 9H$ \\
KOI-1961 AB & 2019-06-12 & 58646.39 & 45.17 $\pm$ 0.15 & 273.51 $\pm$ 0.17 & 0.209 $\pm$ 0.010 & $K' + 9H$ \\
KOI-1961 AB & 2023-03-29 & 60032.60 & 46.52 $\pm$ 0.18 & 282.06 $\pm$ 0.18 & 0.159 $\pm$ 0.013 & $K' + 9H$ \\
\hline
KOI-2032 AB & 2012-08-13 & 56152.42 & 1085.7 $\pm$ 0.5 & 138.39 $\pm$ 0.05 & 0.19 $\pm$ 0.07 & $K'$ \\
KOI-2032 AB & 2014-08-13 & 56882.49 & 1086.3 $\pm$ 0.5 & 138.482 $\pm$ 0.026 & 0.218 $\pm$ 0.007 & $K'$ \\
KOI-2032 AB & 2021-07-19 & 59414.33 & 1091.2 $\pm$ 1.4 & 138.35 $\pm$ 0.04 & 0.216 $\pm$ 0.011 & $K'$ \\
KOI-2032 AB & 2023-06-08 & 60103.63 & 1091 $\pm$ 3 & 138.49 $\pm$ 0.14 & 0.14 $\pm$ 0.10 & $K'$ \\
KOI-2032 AC & 2012-08-13 & 56152.42 & 1149.8 $\pm$ 0.4 & 138.09 $\pm$ 0.03 & 0.34 $\pm$ 0.05 & $K'$ \\
KOI-2032 AC & 2014-08-13 & 56882.49 & 1148.0 $\pm$ 0.5 & 137.900 $\pm$ 0.022 & 0.443 $\pm$ 0.006 & $K'$ \\
KOI-2032 AC & 2021-07-19 & 59414.33 & 1145.7 $\pm$ 1.7 & 137.67 $\pm$ 0.04 & 0.41 $\pm$ 0.04 & $K'$ \\
KOI-2032 AC & 2023-06-08 & 60103.63 & 1140 $\pm$ 6 & 137.5 $\pm$ 0.1 & 0.56 $\pm$ 0.14 & $K'$ \\
KOI-2032 BC & 2012-08-13 & 56152.42 & 63.91 $\pm$ 0.11 & 128.87 $\pm$ 0.26 & 0.167 $\pm$ 0.011 & $K'$ \\
KOI-2032 BC & 2014-08-13 & 56882.49 & 62.84 $\pm$ 0.23 & 128.0 $\pm$ 0.5 & -0.254 $\pm$ 0.027 & $K'$ \\
KOI-2032 BC & 2021-07-19 & 59414.33 & 54.73 $\pm$ 0.21 & 120.2 $\pm$ 1.0 & 0.154 $\pm$ 0.024 & $K'$ \\
KOI-2032 BC & 2023-06-08 & 60103.63 & 49.8 $\pm$ 0.6 & 121.0 $\pm$ 0.9 & 0.33 $\pm$ 0.05 & $K'$ \\
\hline
KOI-2117 AB & 2015-07-25 & 57228.54 & 329.13 $\pm$ 0.18 & 111.11 $\pm$ 0.05 & 0.567 $\pm$ 0.012 & $K'$ \\
KOI-2117 AB & 2019-07-05 & 58669.55 & 328.91 $\pm$ 0.09 & 111.337 $\pm$ 0.022 & 0.575 $\pm$ 0.009 & $K'$ \\
KOI-2117 AB & 2023-06-09 & 60104.54 & 328.60 $\pm$ 0.12 & 111.56 $\pm$ 0.03 & 0.577 $\pm$ 0.007 & $K'$ \\
\hline
KOI-2517 AB & 2019-06-26 & 58660.52 & 192.53 $\pm$ 0.20 & 154.58 $\pm$ 0.06 & 2.477 $\pm$ 0.010 & $K'$ \\
\hline
KOI-2626 AB & 2013-07-06 & 56479.55 & 206.01 $\pm$ 0.16 & 212.88 $\pm$ 0.06 & 0.480 $\pm$ 0.011 & $K'$ \\
KOI-2626 AB & 2013-07-18 & 56491.52 & 205.6 $\pm$ 0.1 & 212.854 $\pm$ 0.010 & 0.4589 $\pm$ 0.0008 & $K'$ \\
KOI-2626 AB & 2014-07-28 & 56866.52 & 203.9 $\pm$ 0.4 & 212.59 $\pm$ 0.06 & 0.464 $\pm$ 0.005 & $K'$ \\
KOI-2626 AB & 2014-07-29 & 56867.37 & 204.41 $\pm$ 0.26 & 212.60 $\pm$ 0.07 & 0.489 $\pm$ 0.020 & $K'$ \\
KOI-2626 AB & 2015-06-21 & 57194.51 & 203.36 $\pm$ 0.21 & 212.26 $\pm$ 0.07 & 0.474 $\pm$ 0.009 & $K'$ \\
KOI-2626 AB & 2017-06-29 & 57933.41 & 200.3 $\pm$ 0.8 & 211.4 $\pm$ 0.4 & 0.50 $\pm$ 0.03 & $K'$ \\
KOI-2626 AB & 2019-06-12 & 58646.61 & 198.56 $\pm$ 0.10 & 211.07 $\pm$ 0.04 & 0.447 $\pm$ 0.004 & $K'$ \\
KOI-2626 AC & 2013-07-06 & 56479.55 & 161.7 $\pm$ 0.3 & 184.79 $\pm$ 0.05 & 1.044 $\pm$ 0.008 & $K'$ \\
KOI-2626 AC & 2013-07-18 & 56491.52 & 161.60 $\pm$ 0.17 & 184.71 $\pm$ 0.04 & 1.023 $\pm$ 0.012 & $K'$ \\
KOI-2626 AC & 2014-07-28 & 56866.52 & 160.2 $\pm$ 0.6 & 184.37 $\pm$ 0.18 & 1.022 $\pm$ 0.015 & $K'$ \\
KOI-2626 AC & 2014-07-29 & 56867.37 & 161.3 $\pm$ 0.3 & 184.77 $\pm$ 0.12 & 1.09 $\pm$ 0.03 & $K'$ \\
KOI-2626 AC & 2015-06-21 & 57194.51 & 160.4 $\pm$ 0.5 & 184.39 $\pm$ 0.07 & 1.021 $\pm$ 0.025 & $K'$ \\
KOI-2626 AC & 2017-06-29 & 57933.41 & 156.3 $\pm$ 0.7 & 183.6 $\pm$ 0.3 & 0.97 $\pm$ 0.05 & $K'$ \\
KOI-2626 AC & 2019-06-12 & 58646.61 & 156.72 $\pm$ 0.21 & 184.14 $\pm$ 0.08 & 1.057 $\pm$ 0.009 & $K'$ \\
KOI-2626 BC & 2013-07-06 & 56479.55 & 99.05 $\pm$ 0.18 & 83.10 $\pm$ 0.20 & 0.564 $\pm$ 0.012 & $K'$ \\
KOI-2626 BC & 2013-07-18 & 56491.52 & 98.99 $\pm$ 0.13 & 83.22 $\pm$ 0.07 & 0.564 $\pm$ 0.013 & $K'$ \\
KOI-2626 BC & 2014-07-28 & 56866.52 & 98.4 $\pm$ 0.4 & 82.93 $\pm$ 0.16 & 0.558 $\pm$ 0.018 & $K'$ \\
KOI-2626 BC & 2014-07-29 & 56867.37 & 97.41 $\pm$ 0.27 & 83.22 $\pm$ 0.16 & 0.600 $\pm$ 0.014 & $K'$ \\
KOI-2626 BC & 2015-06-21 & 57194.51 & 97.0 $\pm$ 0.3 & 82.85 $\pm$ 0.23 & 0.547 $\pm$ 0.028 & $K'$ \\
KOI-2626 BC & 2017-06-29 & 57933.41 & 95.76 $\pm$ 0.21 & 81.0 $\pm$ 0.8 & 0.47 $\pm$ 0.08 & $K'$ \\
KOI-2626 BC & 2019-06-12 & 58646.61 & 92.21 $\pm$ 0.22 & 81.42 $\pm$ 0.15 & 0.610 $\pm$ 0.011 & $K'$ \\
\hline
KOI-2971 AB & 2015-07-25 & 57228.50 & 296.07 $\pm$ 0.21 & 273.75 $\pm$ 0.07 & 3.579 $\pm$ 0.013 & $K'$ \\
\hline
KOI-3158 AB & 2015-06-22 & 57195.52 & 1842.5 $\pm$ 0.4 & 252.833 $\pm$ 0.020 & 2.070 $\pm$ 0.027 & $K_\text{cont}$ \\
KOI-3158 AB & 2015-07-21 & 57224.45 & 1842.3 $\pm$ 0.4 & 252.823 $\pm$ 0.020 & 2.09 $\pm$ 0.06 & $K_\text{cont}$ \\
KOI-3158 AB & 2016-06-16 & 57555.64 & 1841.6 $\pm$ 0.4 & 252.833 $\pm$ 0.020 & 2.16 $\pm$ 0.03 & $K_\text{cont}$ \\
KOI-3158 AB & 2017-11-27 & 58084.18 & 1839.9 $\pm$ 0.4 & 252.912 $\pm$ 0.020 & 2.166 $\pm$ 0.019 & $K_\text{cont}$ \\
KOI-3158 AB & 2018-06-07 & 58276.47 & 1840.2 $\pm$ 0.4 & 252.913 $\pm$ 0.020 & 2.068 $\pm$ 0.020 & $K_\text{cont}$ \\
KOI-3158 AB & 2020-06-11 & 59011.63 & 1838.2 $\pm$ 0.4 & 253.022 $\pm$ 0.020 & 2.168 $\pm$ 0.024 & $K_\text{cont}$ \\
\hline
KOI-3196 AB & 2013-08-06 & 56510.43 & 126.6 $\pm$ 0.6 & 74.5 $\pm$ 0.5 & 5.10 $\pm$ 0.11 & $K'$ \\
KOI-3196 AB & 2013-08-20 & 56524.39 & 127 $\pm$ 6 & 74.2 $\pm$ 0.7 & 4.9 $\pm$ 0.1 & $K_\text{cont}$ \\
KOI-3196 AB & 2014-07-31 & 56869.30 & 134 $\pm$ 4 & 70.6 $\pm$ 2.4 & 4.97 $\pm$ 0.17 & $K'$ \\
KOI-3196 AB & 2023-06-09 & 60104.39 & 135.2 $\pm$ 1.0 & 68.6 $\pm$ 0.5 & 4.9 $\pm$ 0.1 & $K'$ \\
\hline
KOI-3444 A-BC & 2014-08-13 & 56882.29 & 1083.17 $\pm$ 0.09 & 10.240 $\pm$ 0.017 & 2.466 $\pm$ 0.010 & $K'$ \\
KOI-3444 A-BC & 2014-11-30 & 56991.22 & 1082.3 $\pm$ 0.6 & 10.212 $\pm$ 0.016 & 2.435 $\pm$ 0.020 & $K'$ \\
KOI-3444 A-BC & 2015-05-28 & 57170.58 & 1084.9 $\pm$ 0.9 & 10.24 $\pm$ 0.07 & 2.502 $\pm$ 0.020 & $K'$ \\
KOI-3444 A-BC & 2015-07-26 & 57229.50 & 1085.10 $\pm$ 0.20 & 10.256 $\pm$ 0.011 & 2.476 $\pm$ 0.019 & $K'$ \\
KOI-3444 A-BC & 2016-06-16 & 57555.61 & 1087.70 $\pm$ 0.18 & 10.208 $\pm$ 0.008 & 2.418 $\pm$ 0.009 & $K'$ \\
KOI-3444 BC & 2020-08-29 & 59090.43 & 53.66 $\pm$ 0.18 & 186.45 $\pm$ 0.28 & 0.232 $\pm$ 0.006 & $K'$ \\
\hline
KOI-3497 AB & 2013-08-06 & 56510.51 & 843.4 $\pm$ 0.5 & 176.127 $\pm$ 0.024 & 1.05 $\pm$ 0.03 & $K'$ \\
KOI-3497 AB & 2015-07-21 & 57224.43 & 840.9 $\pm$ 0.4 & 176.263 $\pm$ 0.029 & 1.041 $\pm$ 0.016 & $K'$ \\
KOI-3497 AB & 2017-06-29 & 57933.41 & 837.5 $\pm$ 0.9 & 176.46 $\pm$ 0.06 & 1.118 $\pm$ 0.023 & $K'$ \\
KOI-3497 AB & 2019-06-12 & 58646.41 & 835.6 $\pm$ 0.5 & 176.499 $\pm$ 0.024 & 1.13 $\pm$ 0.01 & $K'$ \\
KOI-3497 AB & 2021-07-19 & 59414.43 & 831.6 $\pm$ 0.5 & 176.64 $\pm$ 0.04 & 1.054 $\pm$ 0.015 & $K'$ \\
KOI-3497 AC & 2013-08-06 & 56510.51 & 767.12 $\pm$ 0.20 & 173.680 $\pm$ 0.025 & 1.679 $\pm$ 0.020 & $K'$ \\
KOI-3497 AC & 2015-07-21 & 57224.43 & 770.6 $\pm$ 0.3 & 173.695 $\pm$ 0.029 & 1.643 $\pm$ 0.003 & $K'$ \\
KOI-3497 AC & 2017-06-29 & 57933.41 & 773.3 $\pm$ 0.4 & 173.76 $\pm$ 0.04 & 1.682 $\pm$ 0.014 & $K'$ \\
KOI-3497 AC & 2019-06-12 & 58646.41 & 777.5 $\pm$ 0.4 & 173.805 $\pm$ 0.027 & 1.742 $\pm$ 0.022 & $K'$ \\
KOI-3497 AC & 2021-07-19 & 59414.43 & 780.9 $\pm$ 0.4 & 173.88 $\pm$ 0.03 & 1.687 $\pm$ 0.009 & $K'$ \\
KOI-3497 BC & 2013-08-06 & 56510.51 & 83.7 $\pm$ 0.8 & 19.17 $\pm$ 0.21 & 0.624 $\pm$ 0.013 & $K'$ \\
KOI-3497 BC & 2015-07-21 & 57224.43 & 79.0 $\pm$ 0.3 & 22.17 $\pm$ 0.12 & 0.602 $\pm$ 0.015 & $K'$ \\
KOI-3497 BC & 2017-06-29 & 57933.41 & 74.6 $\pm$ 0.8 & 25.7 $\pm$ 0.5 & 0.564 $\pm$ 0.023 & $K'$ \\
KOI-3497 BC & 2019-06-12 & 58646.41 & 69.4 $\pm$ 0.4 & 28.3 $\pm$ 0.4 & 0.613 $\pm$ 0.021 & $K'$ \\
KOI-3497 BC & 2021-07-19 & 59414.43 & 63.9 $\pm$ 0.4 & 32.74 $\pm$ 0.27 & 0.633 $\pm$ 0.014 & $K'$ \\
\hline
KOI-4329 AB & 2013-08-21 & 56525.37 & 1845.4 $\pm$ 0.5 & 118.230 $\pm$ 0.018 & 2.738 $\pm$ 0.029 & $K_\text{cont}$ \\
KOI-4329 AB & 2019-06-13 & 58647.56 & 1843.79 $\pm$ 0.19 & 118.279 $\pm$ 0.005 & 2.945 $\pm$ 0.018 & $K'$ \\
KOI-4329 AB & 2021-07-19 & 59414.55 & 1843.6 $\pm$ 0.7 & 118.293 $\pm$ 0.022 & 2.918 $\pm$ 0.020 & $K'$ \\
KOI-4329 AB & 2023-06-09 & 60104.46 & 1844.7 $\pm$ 0.1 & 118.322 $\pm$ 0.003 & 2.828 $\pm$ 0.004 & $K'$ \\
\hline
KOI-4528 AB & 2020-08-29 & 59018.59 & 68.1 $\pm$ 1.5 & 264.9 $\pm$ 1.0 & 0.34 $\pm$ 0.04 & $K'$ \\
KOI-4528 AC & 2020-08-29 & 59018.59 & 182.4 $\pm$ 1.6 & 46.1 $\pm$ 0.4 & 1.207 $\pm$ 0.019 & $K'$ \\
\hline
KOI-4661 AB & 2014-08-18 & 56887.32 & 3851.7 $\pm$ 0.6 & 198.014 $\pm$ 0.004 & 1.240 $\pm$ 0.022 & $K'$ \\
KOI-4661 AB & 2021-06-29 & 59394.42 & 3848.6 $\pm$ 0.5 & 197.976 $\pm$ 0.004 & 1.443 $\pm$ 0.018 & $K'$ \\
KOI-4661 AC & 2014-08-18 & 56887.32 & 153.6 $\pm$ 0.6 & 86.34 $\pm$ 0.17 & 2.238 $\pm$ 0.003 & $K'$ \\
KOI-4661 AC & 2021-06-29 & 59394.42 & 156.4 $\pm$ 0.4 & 85.48 $\pm$ 0.15 & 2.257 $\pm$ 0.020 & $K'$ \\
KOI-4661 BC & 2014-08-18 & 56887.32 & 3911.09 $\pm$ 0.06 & 20.1060 $\pm$ 0.0020 & 0.999 $\pm$ 0.025 & $K'$ \\
KOI-4661 BC & 2021-06-29 & 59394.42 & 3911.1 $\pm$ 0.6 & 20.093 $\pm$ 0.005 & 0.814 $\pm$ 0.022 & $K'$ \\
\hline
KOI-5581 AB & 2022-07-05 & 59765.48 & 167 $\pm$ 7 & 127.6 $\pm$ 1.4 & 3.75 $\pm$ 0.07 & $K'$ \\
\hline
KOI-5930 AB & 2019-07-16 & 58680.37 & 1411.35 $\pm$ 0.24 & 148.859 $\pm$ 0.010 & 1.436 $\pm$ 0.023 & $K'$ \\
KOI-5930 AC & 2019-07-16 & 58680.37 & 74 $\pm$ 4 & 87 $\pm$ 4 & 2.69 $\pm$ 0.27 & $K'$ \\
KOI-5930 BC & 2019-07-16 & 58680.37 & 1379 $\pm$ 5 & 331.58 $\pm$ 0.19 & 1.25 $\pm$ 0.27 & $K'$ \\
\hline
KOI-7842 AB & 2020-08-29 & 59090.39 & 78 $\pm$ 13 & 149.7 $\pm$ 1.1 & 2.12 $\pm$ 0.22 & $K'$ \\
\end{longtable}
\clearpage

\appendix

\section{Orbit plots}
\label{sec:appendix}

\noindent\begin{minipage}{\linewidth}
    \centering
    \captionsetup{type=figure}
    \vspace{0.75cm}
    \includegraphics[ width=1\linewidth]{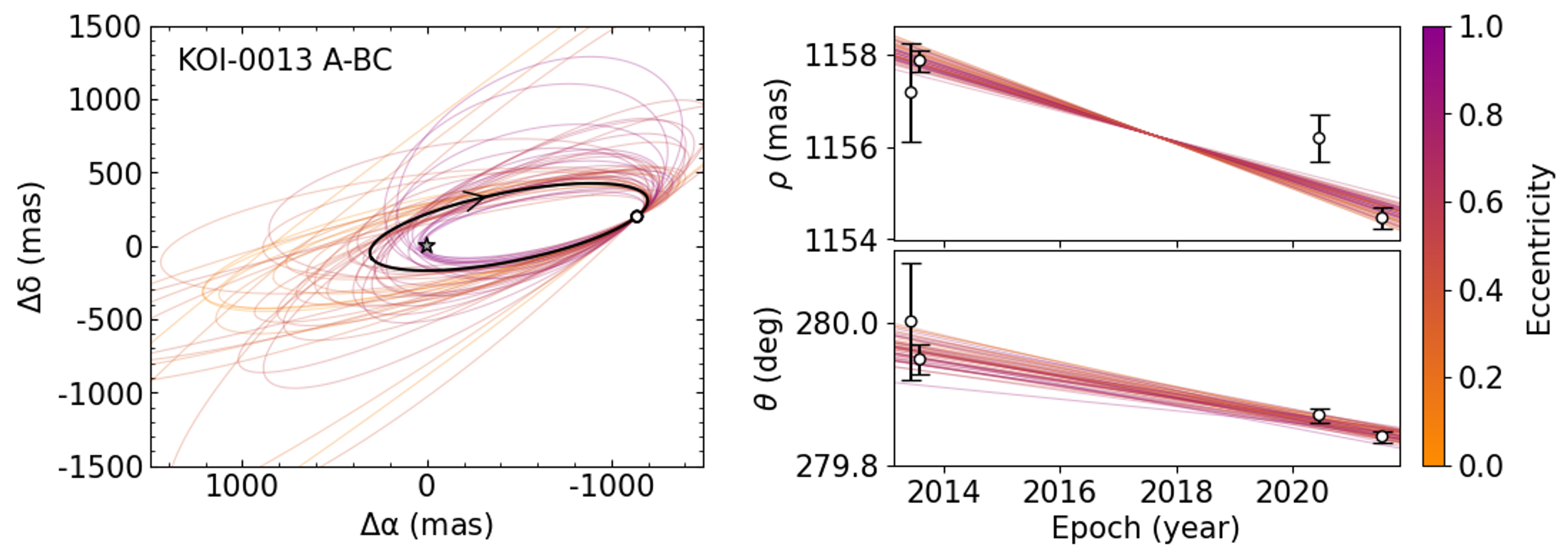}
    \captionof{figure}{\textit{Left:} 50 random orbits from the posterior sample for the \textsc{orvara} fit for the visual components of KOI-0013. The colour of the orbit indicates the eccentricity and the positions of the unresolved binary companion KOI-0013 BC, relative to the primary KOI-0013 A (shown with a black star) are marked with white circles. \textit{Right:} The measured astrometry for the position angle and relative separation over time, overlaid by 50 possible orbital solutions.}
    \label{fig:orvara_13}
\end{minipage}

\noindent\begin{minipage}{\linewidth}
    \centering
    \captionsetup{type=figure}
    \vspace{0.75cm}
    \includegraphics[ width=1\linewidth]{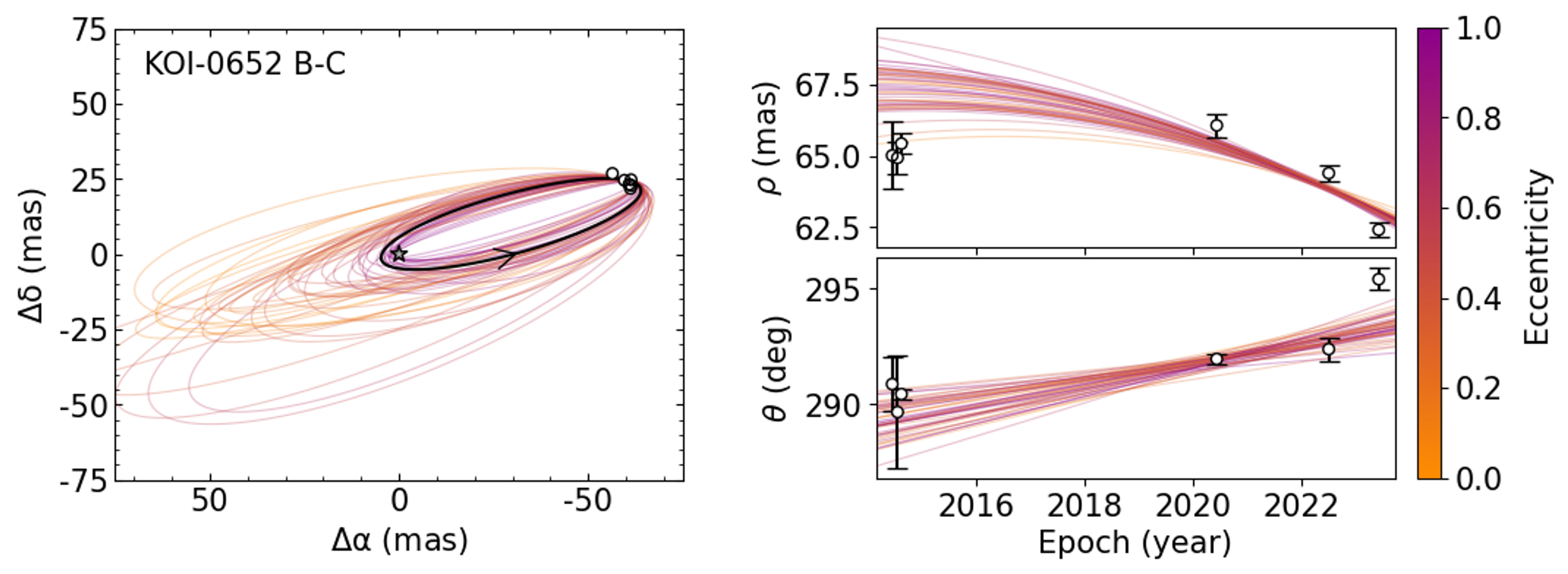}
    \captionof{figure}{\textit{Left:} 50 random orbits from the posterior sample for the \textsc{orvara} fit for the inner binary of KOI-0652. The colour of the orbit indicates the eccentricity and the positions of the companion KOI-0652 C, relative to KOI-0652 B (shown with a black star) are marked with white circles.\textit{Right:} The measured astrometry for the position angle and relative separation over time, overlaid by 50 possible orbital solutions.}
    \label{fig:orvara_0652}
\end{minipage}

\clearpage

\noindent\begin{minipage}{\linewidth}
    \centering
    \captionsetup{type=figure}
    \vspace{0.75cm}
    \includegraphics[width=1\linewidth]{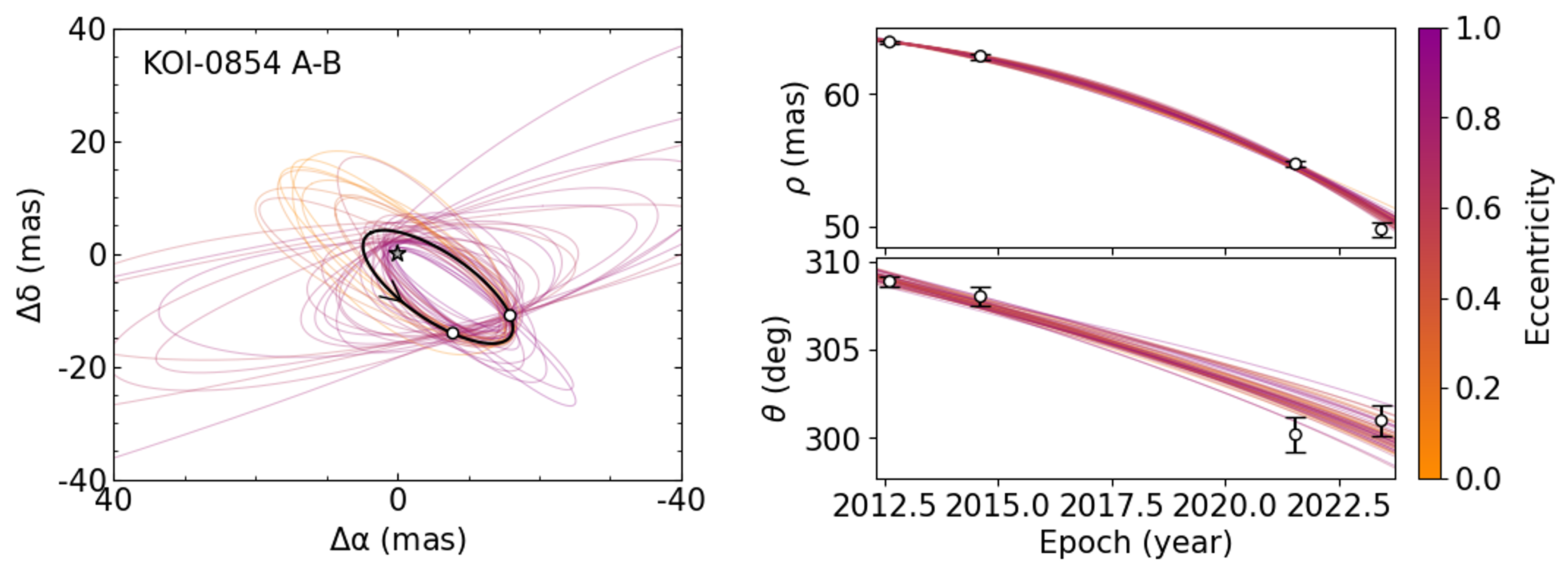}
    \captionof{figure}{\textit{Left:} 50 random orbits from the posterior sample for the \textsc{orvara} fit for the inner binary of KOI-0854. The colour of the orbit indicates the eccentricity and the positions of the companion KOI-0854 B, relative to KOI-0854 A (shown with a black star) are marked with white circles. \textit{Right:} The measured astrometry for the position angle and relative separation over time, overlaid by 50 possible orbital solutions.}
    \label{fig:orvara_0854}
\end{minipage}

\noindent\begin{minipage}{\linewidth}
    \centering
    \captionsetup{type=figure}
    \vspace{0.75cm}
    \includegraphics[width=1\linewidth]{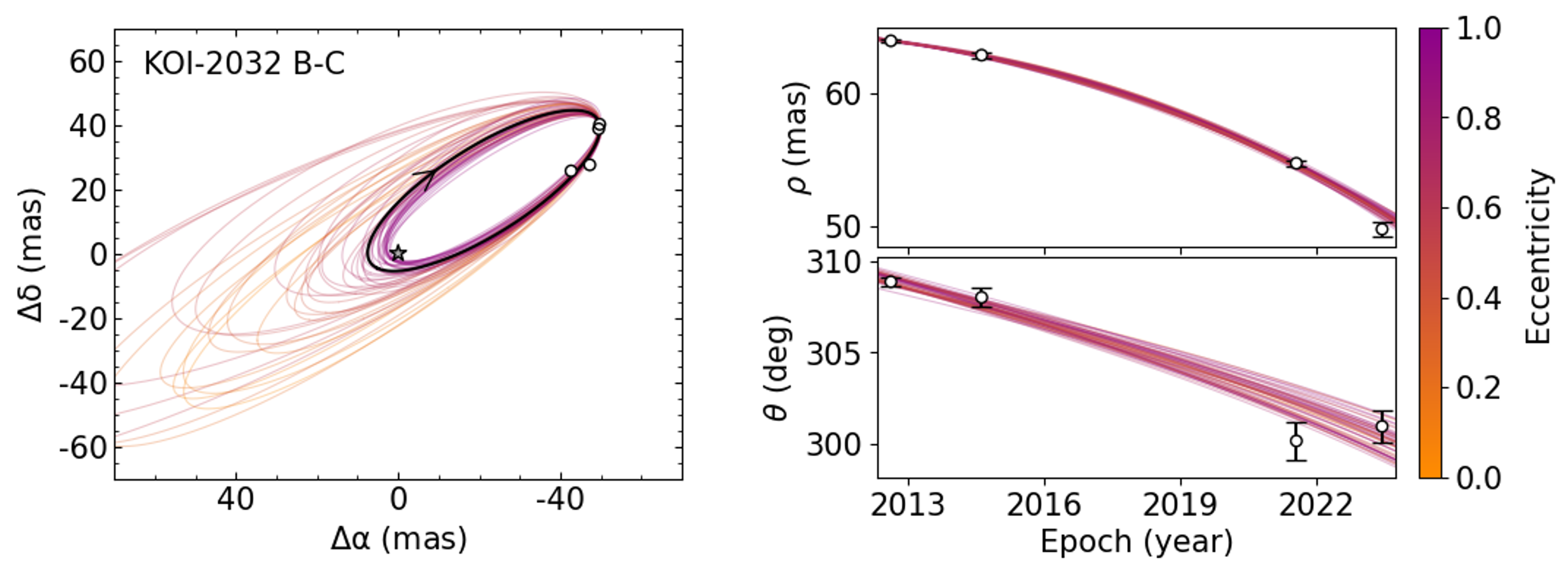}
    \captionof{figure}{\textit{Left:} 50 random orbits from the posterior sample for the \textsc{orvara} fit for the inner binary of KOI-2032. The colour of the orbit indicates the eccentricity and the positions of the companion KOI-2032 C, relative to KOI-2032 B (shown with a black star) are marked with white circles. \textit{Right:} The measured astrometry for the position angle and relative separation over time, overlaid by 50 possible orbital solutions.}
    \label{fig:orvara_2032}
\end{minipage}

\clearpage

\noindent\begin{minipage}{\linewidth}
    \centering
    \captionsetup{type=figure}
    \vspace{0.75cm}
    \includegraphics[width=1\linewidth]{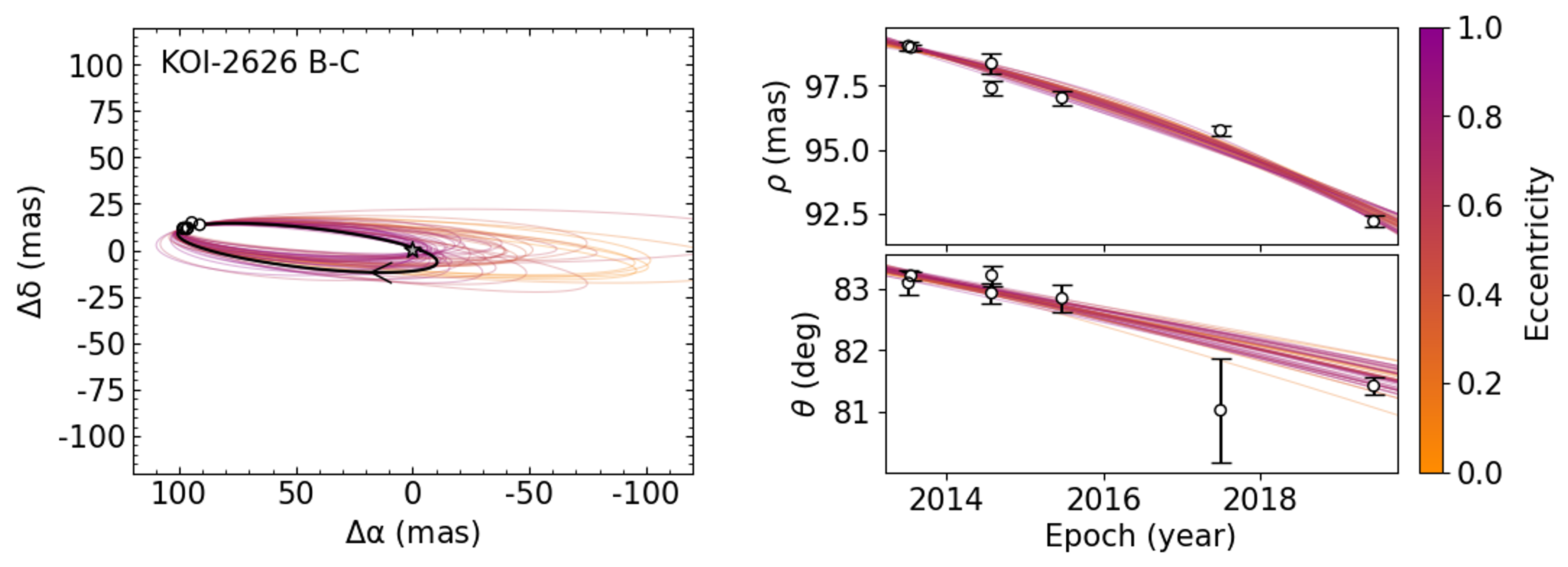}
    \captionof{figure}{\textit{Left:} 50 random orbits from the posterior sample for the \textsc{orvara} fit for the inner binary of KOI-2626. The colour of the orbit indicates the eccentricity and the positions of the companion KOI-2626 C, relative to KOI-2626 B (shown with a black star) are marked with white circles.\textit{Right:} The measured astrometry for the position angle and relative separation over time, overlaid by 50 possible orbital solutions.}
    \label{fig:orvara_2626}
\end{minipage}

\noindent\begin{minipage}{\linewidth}
    \centering
    \captionsetup{type=figure}
    \vspace{0.75cm}
    \includegraphics[width=1\linewidth]{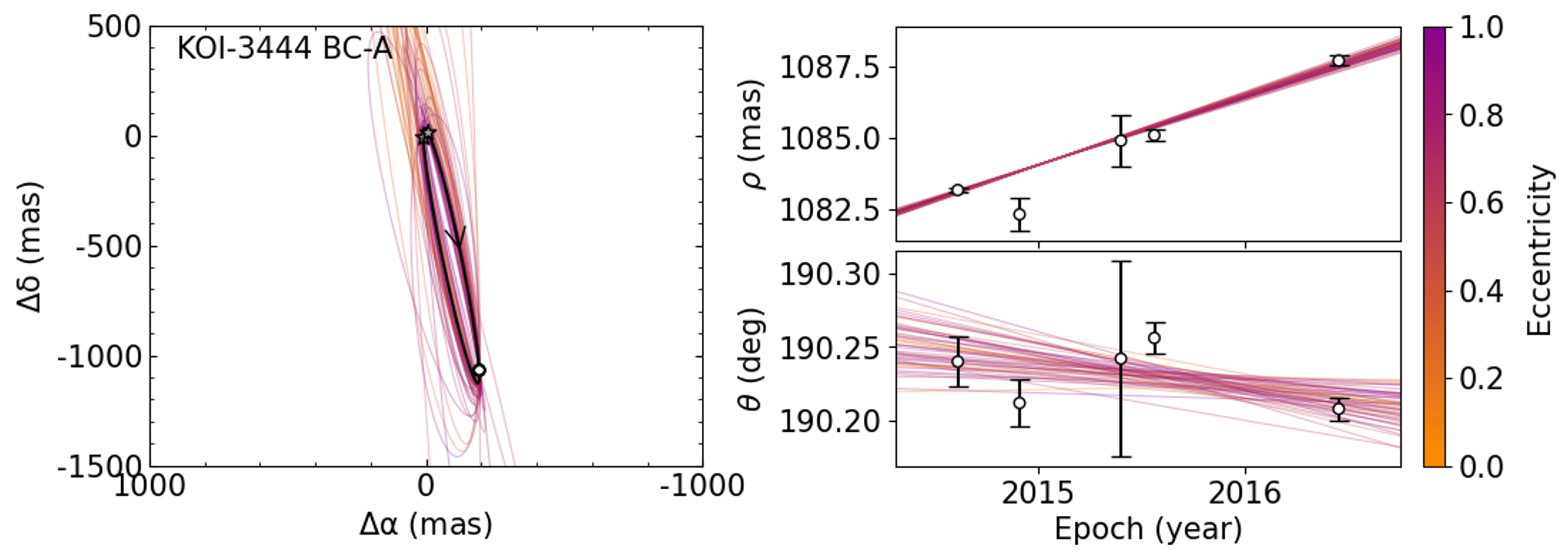}
    \captionof{figure}{\textit{Left:} 50 random orbits from the posterior sample for the \textsc{orvara} fit for the outer companion of KOI-3444. The colour of the orbit indicates the eccentricity and the positions of the primary KOI-3444 A, relative to the barycenter of the inner binary KOI-3444 BC (shown with black stars; separation of the binary stars relative to the companion is not to scale) are marked with white circles. \textit{Right:} The measured astrometry for the position angle and relative separation over time, overlaid by 50 possible orbital solutions.}
    \label{fig:orvara_3444}
\end{minipage}

\clearpage

\noindent\begin{minipage}{\linewidth}
    \centering
    \captionsetup{type=figure}
    \vspace{0.75cm}
    \includegraphics[width=1\linewidth]{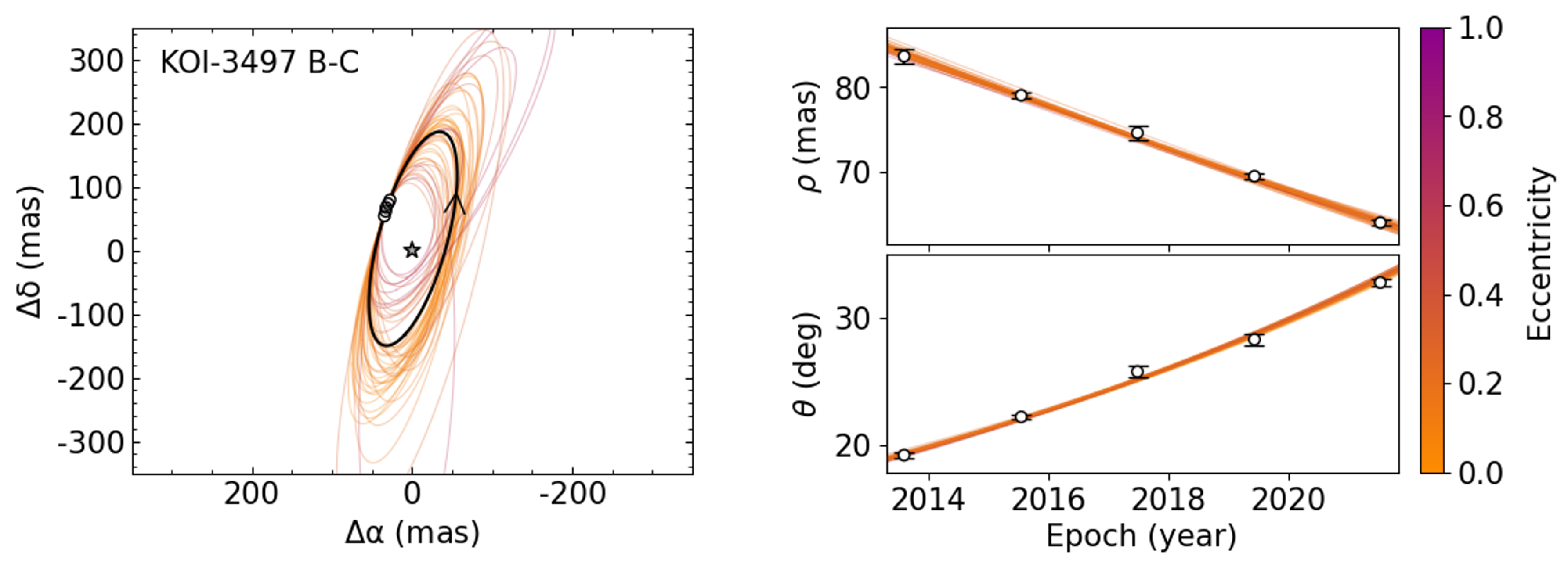}
    \captionof{figure}{\textit{Left:} 50 random orbits from the posterior sample for the \textsc{orvara} fit for the inner binary of KOI-3497. The colour of the orbit indicates the eccentricity and the positions of the companion KOI-3497 C, relative to KOI-3497 B (shown with a black star) are marked with white circles. \textit{Right:} The measured astrometry for the position angle and relative separation over time, overlaid by 50 possible orbital solutions.}
    \label{fig:orvara_3497}
\end{minipage}

\clearpage

\noindent\begin{minipage}{\linewidth}
    \centering
    \captionsetup{type=figure}
    \vspace{0.75cm}
    \includegraphics[width=1\linewidth]{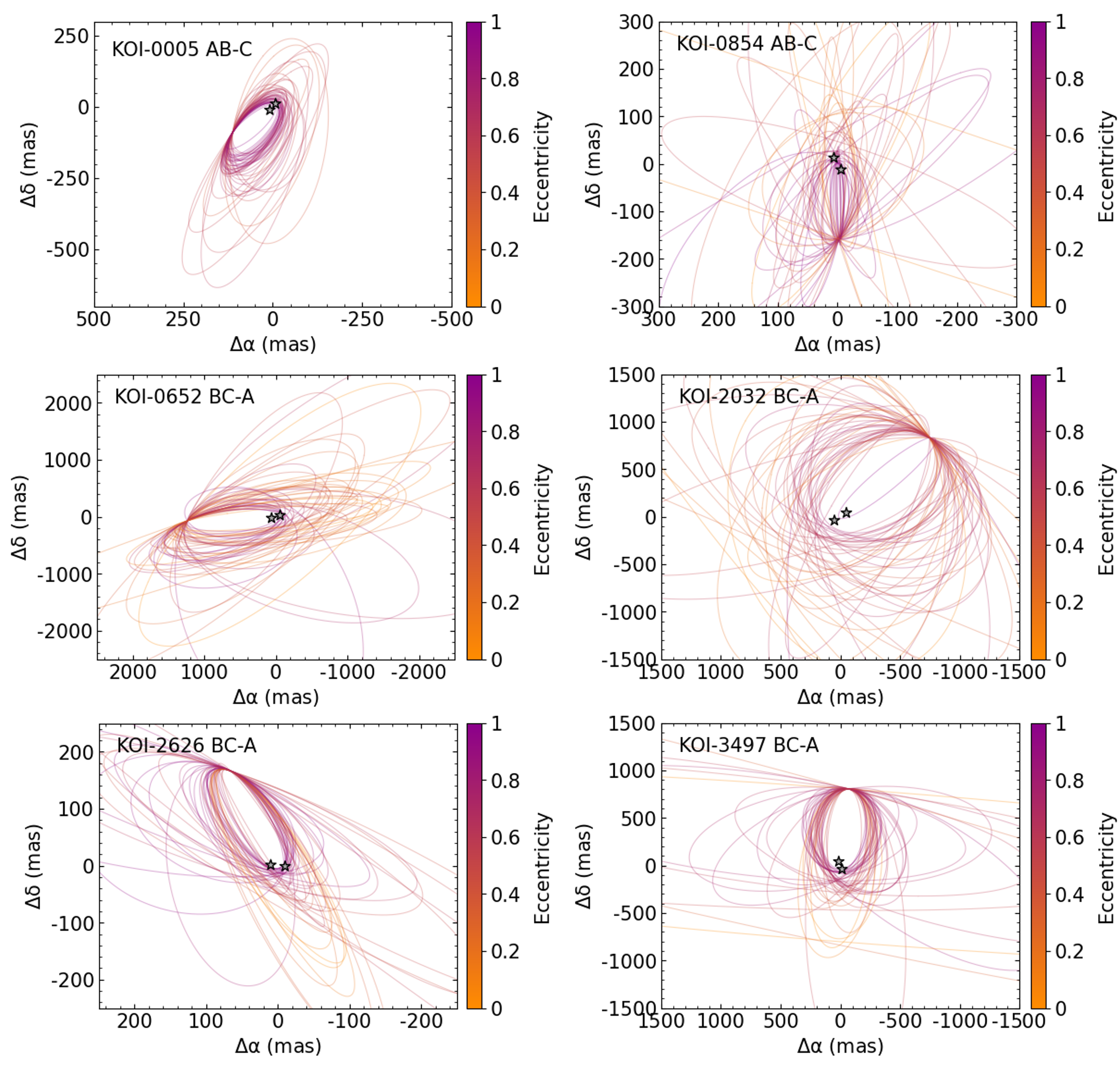}
    \captionof{figure}{Complete set of orbital fits for the outer companions in the 6 visual triples. For each system, 50 orbits from the posterior sample for the \textsc{lofti} fit for the outer companion are shown relative to the inner binary (black stars; separation of the binary stars relative to the companion is not to scale). The colour of the orbit indicates the eccentricity.}
    \label{fig:lofti_all_plots}
\end{minipage}

\clearpage

\section{Corner plots}

\noindent\begin{minipage}{\linewidth}
    \centering
    \captionsetup{type=figure}
    \vspace{0.75cm}
    \includegraphics[width=1\linewidth]{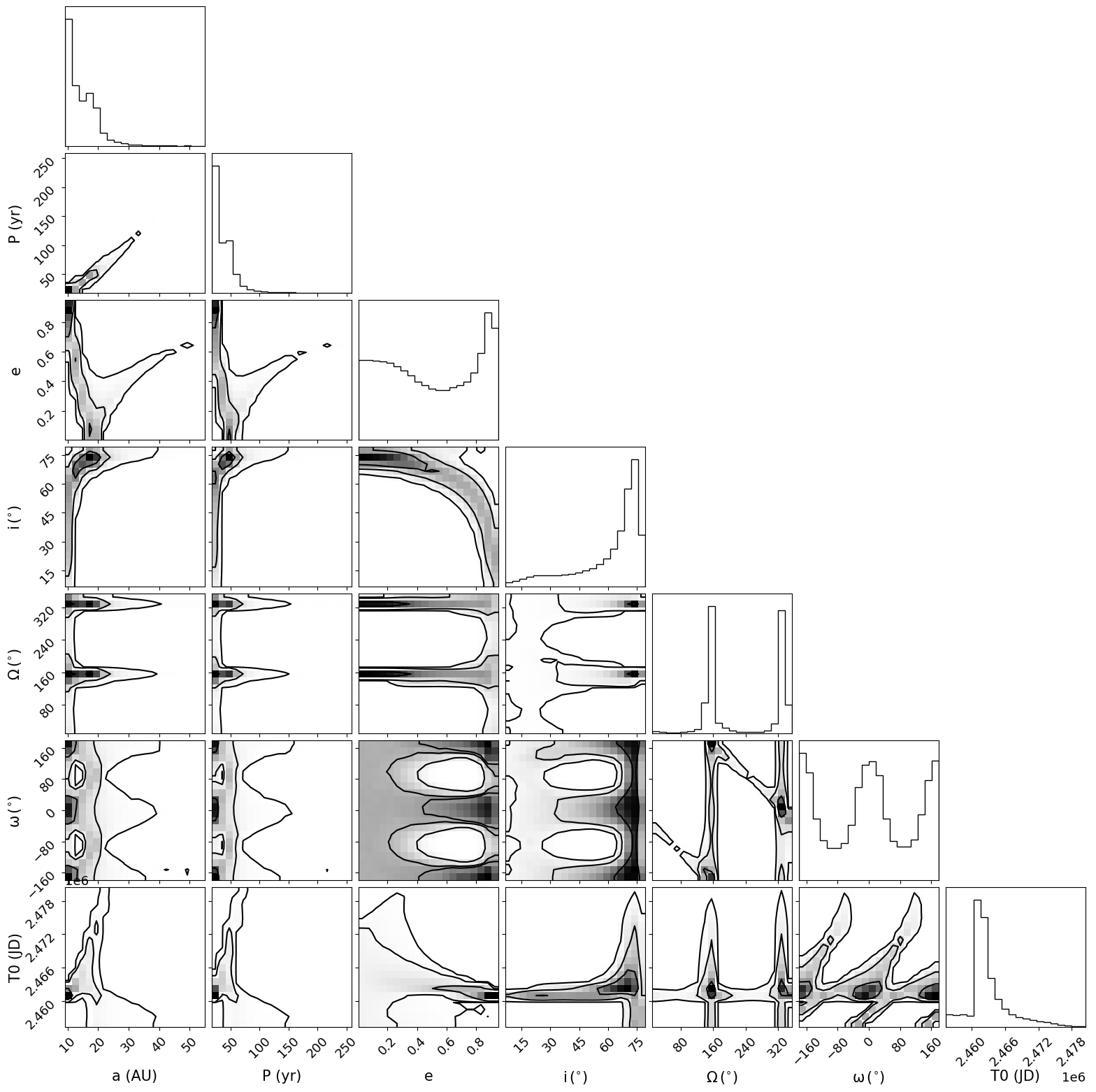}
    \captionof{figure}{Posteriors from our \textsc{orvara} orbital fit for the inner binary of KOI-0005. Details about each parameter, including credible intervals and the best-fit values of these parameters, are listed in Table \ref{tab:orbits}.}
    \label{fig:corner_0005}
\end{minipage}

\clearpage

\noindent\begin{minipage}{\linewidth}
    \centering
    \captionsetup{type=figure}
    \includegraphics[width=1\linewidth]{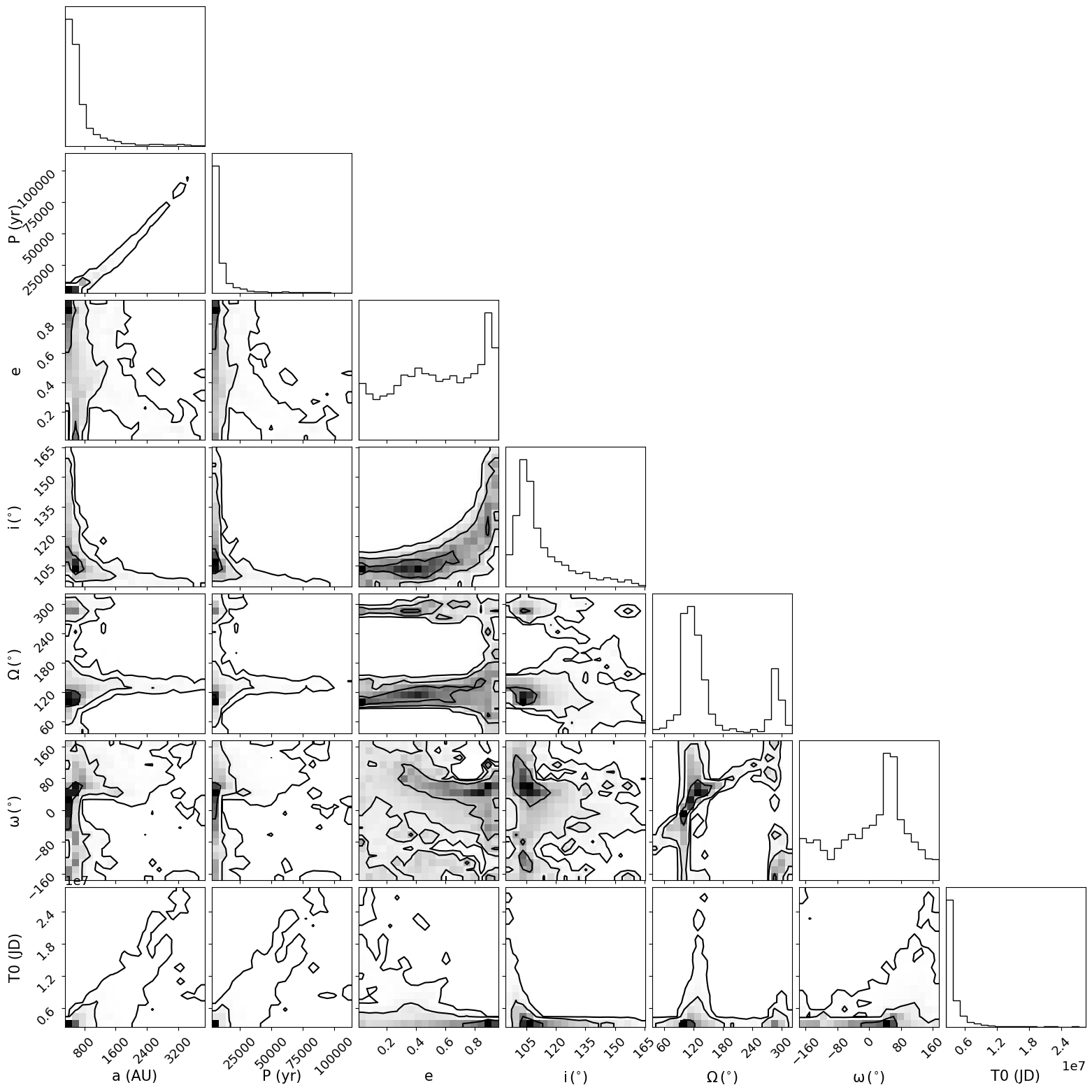}
    \captionof{figure}{Posteriors from our \textsc{orvara} orbital fit for the visual components of KOI-0013. Details about each parameter, including credible intervals and the best-fit values of these parameters, are listed in Table \ref{tab:orbits}.}
    \label{fig:corner_00013}
\end{minipage}

\clearpage

\noindent\begin{minipage}{\linewidth}
    \centering
    \captionsetup{type=figure}
    \includegraphics[width=1\linewidth]{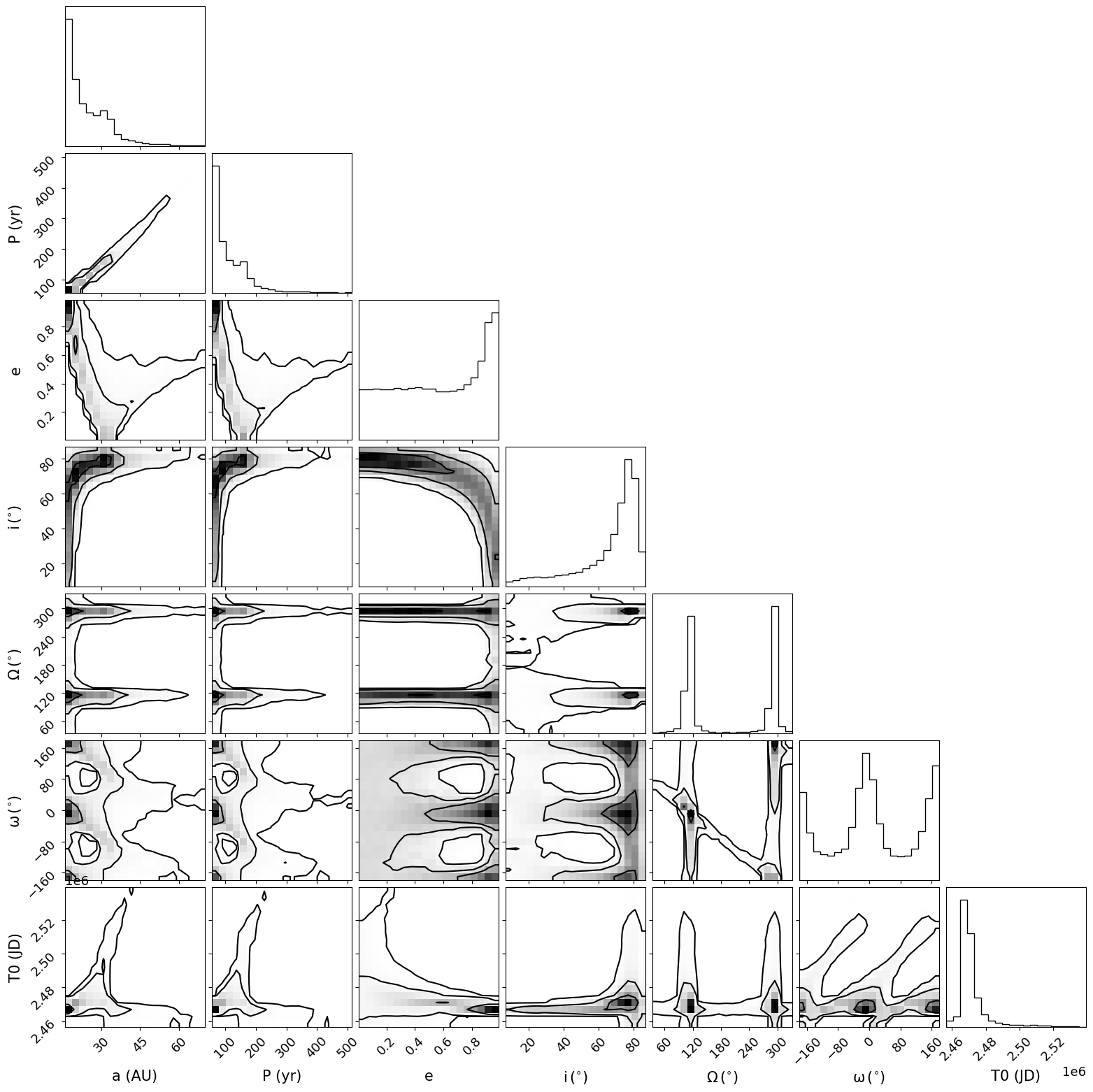}
    \captionof{figure}{Posteriors from our \textsc{orvara} orbital fit for the inner binary of KOI-0652. Details about each parameter, including credible intervals and the best-fit values of these parameters, are listed in Table \ref{tab:orbits}.}
    \label{fig:corner_0652}
\end{minipage}

\clearpage

\noindent\begin{minipage}{\linewidth}
    \centering
    \captionsetup{type=figure}
    \includegraphics[width=1\linewidth]{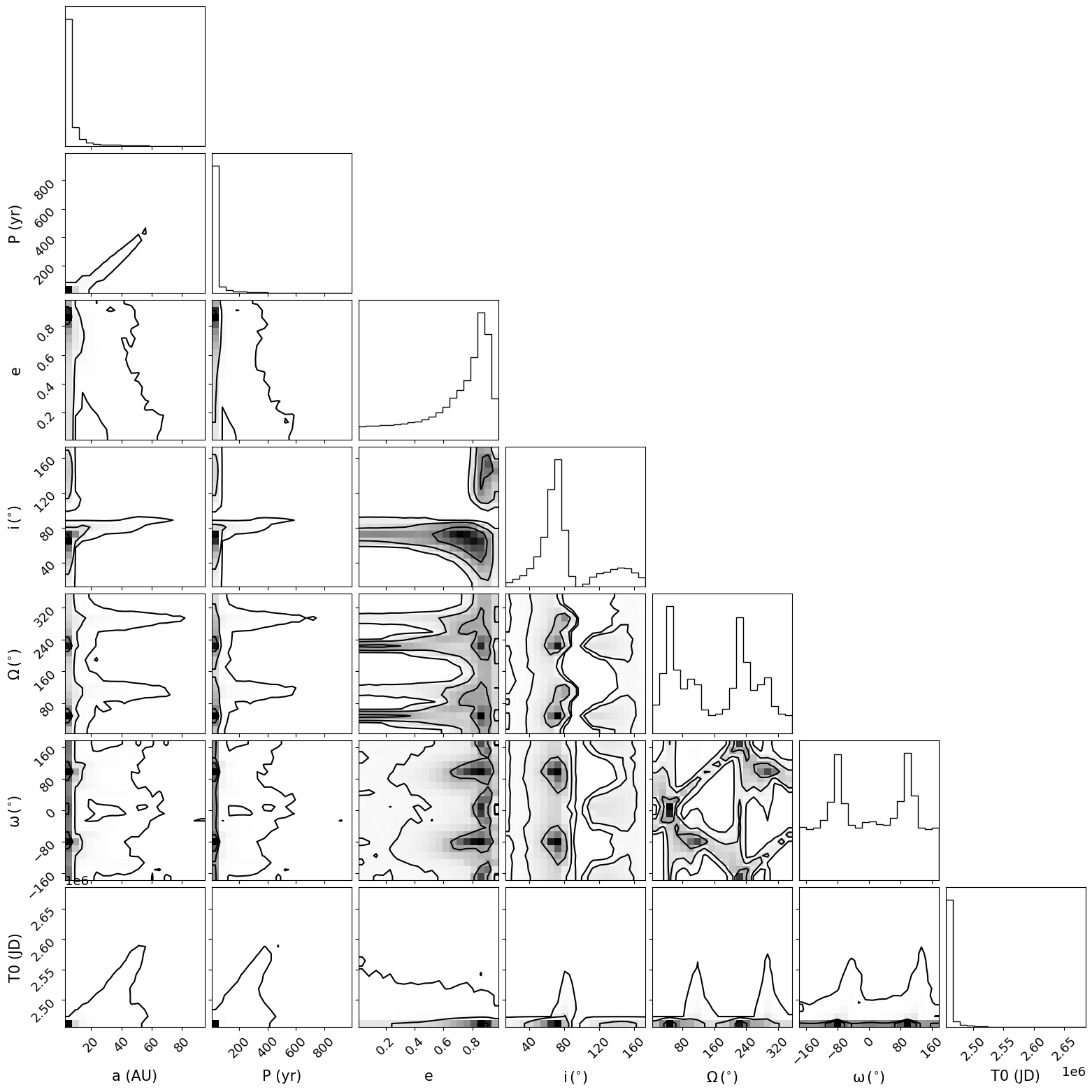}
    \captionof{figure}{Posteriors from our \textsc{orvara} orbital fit for the inner binary of KOI-0854. Details about each parameter, including credible intervals and the best-fit values of these parameters, are listed in Table \ref{tab:orbits}.}
    \label{fig:corner_0854}
\end{minipage}

\clearpage

\noindent\begin{minipage}{\linewidth}
    \centering
    \captionsetup{type=figure}
    \includegraphics[width=1\linewidth]{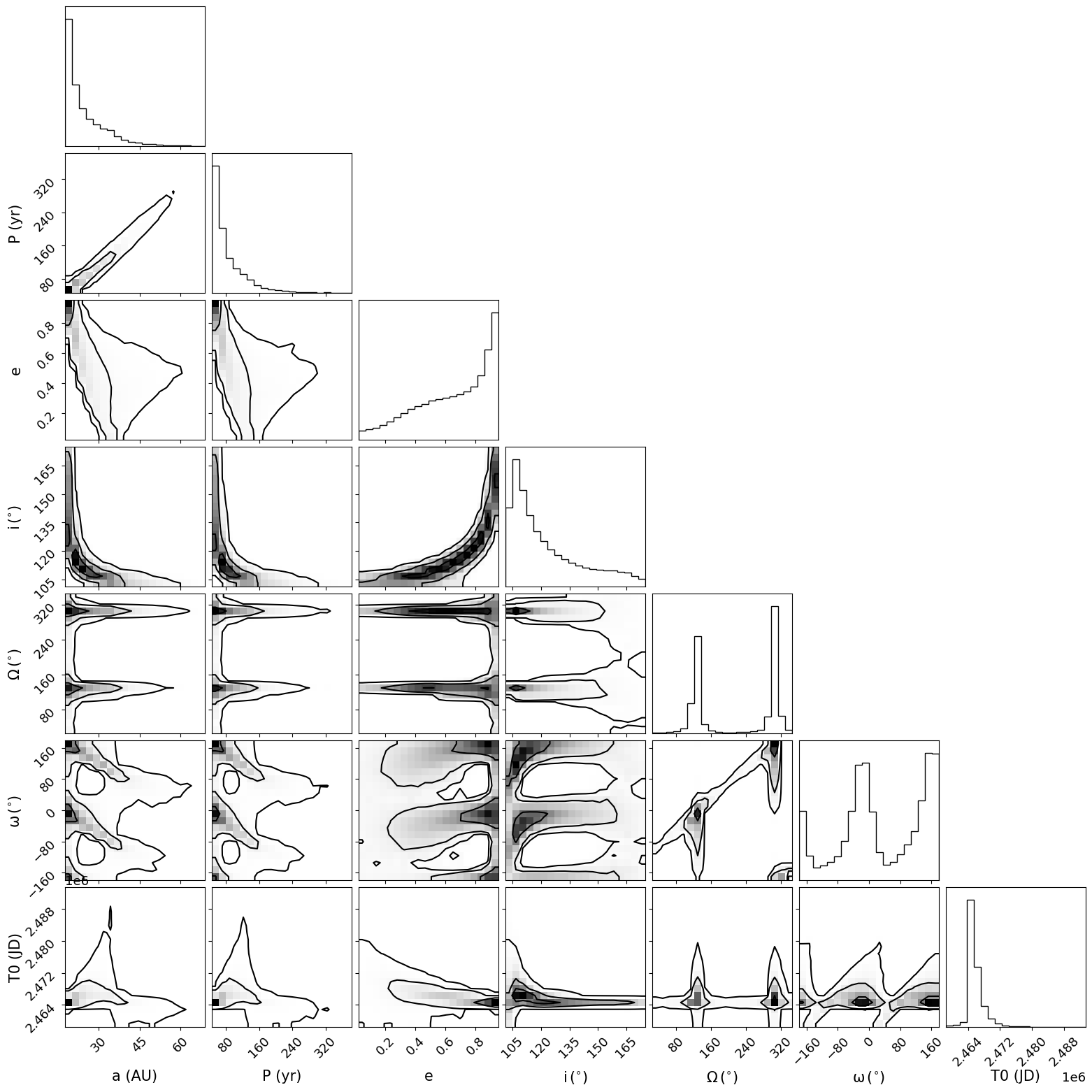}
    \captionof{figure}{Posteriors from our \textsc{orvara} orbital fit for the inner binary of KOI-2032. Details about each parameter, including credible intervals and the best-fit values of these parameters, are listed in Table \ref{tab:orbits}.}
    \label{fig:corner_2032}
\end{minipage}

\clearpage

\noindent\begin{minipage}{\linewidth}
    \centering
    \captionsetup{type=figure}
    \includegraphics[width=1\linewidth]{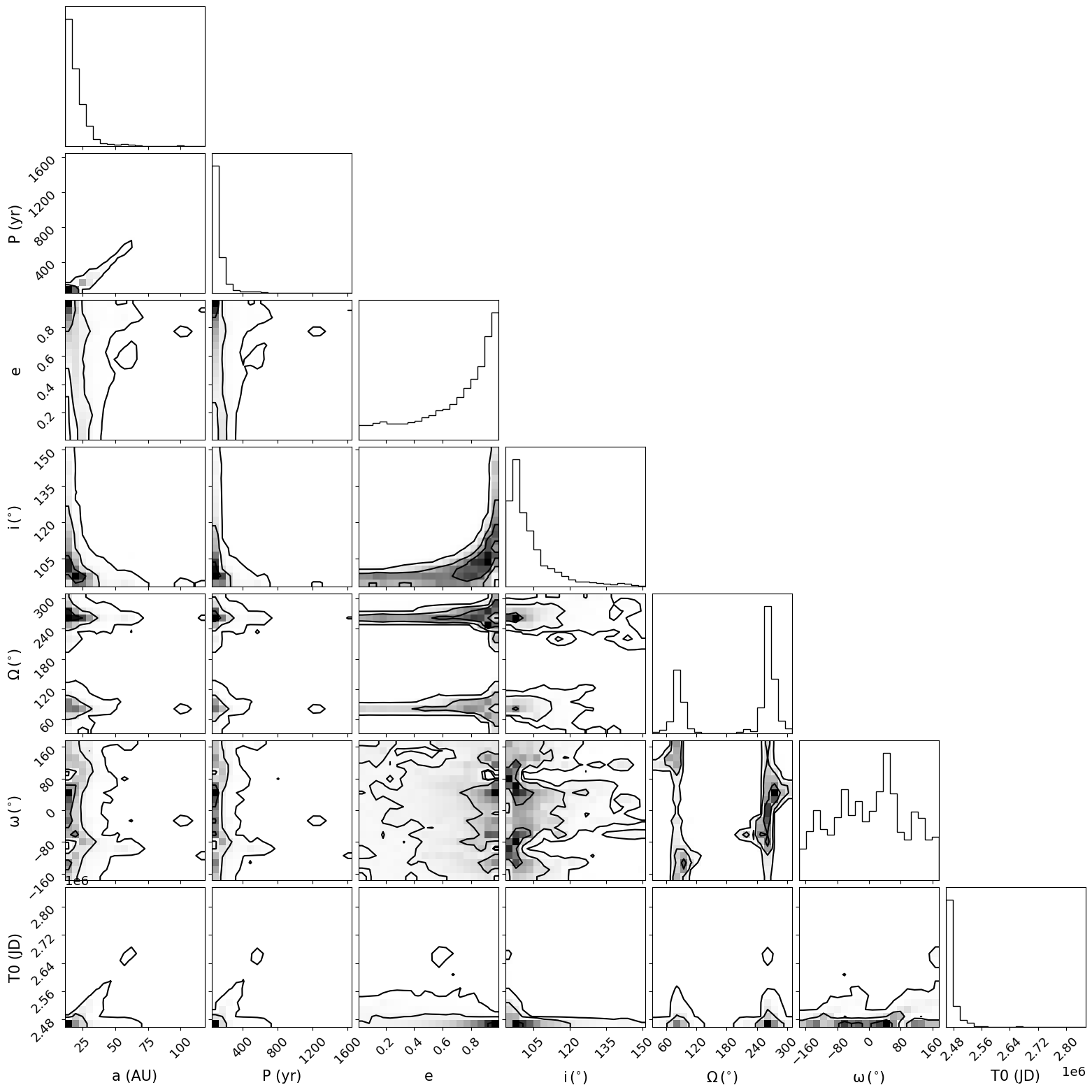}
    \captionof{figure}{Posteriors from our \textsc{orvara} orbital fit for the inner binary of KOI-2626. Details about each parameter, including credible intervals and the best-fit values of these parameters, are listed in Table \ref{tab:orbits}.}
    \label{fig:corner_2626}
\end{minipage}

\clearpage

\noindent\begin{minipage}{\linewidth}
    \centering
    \captionsetup{type=figure}
    \includegraphics[width=1\linewidth]{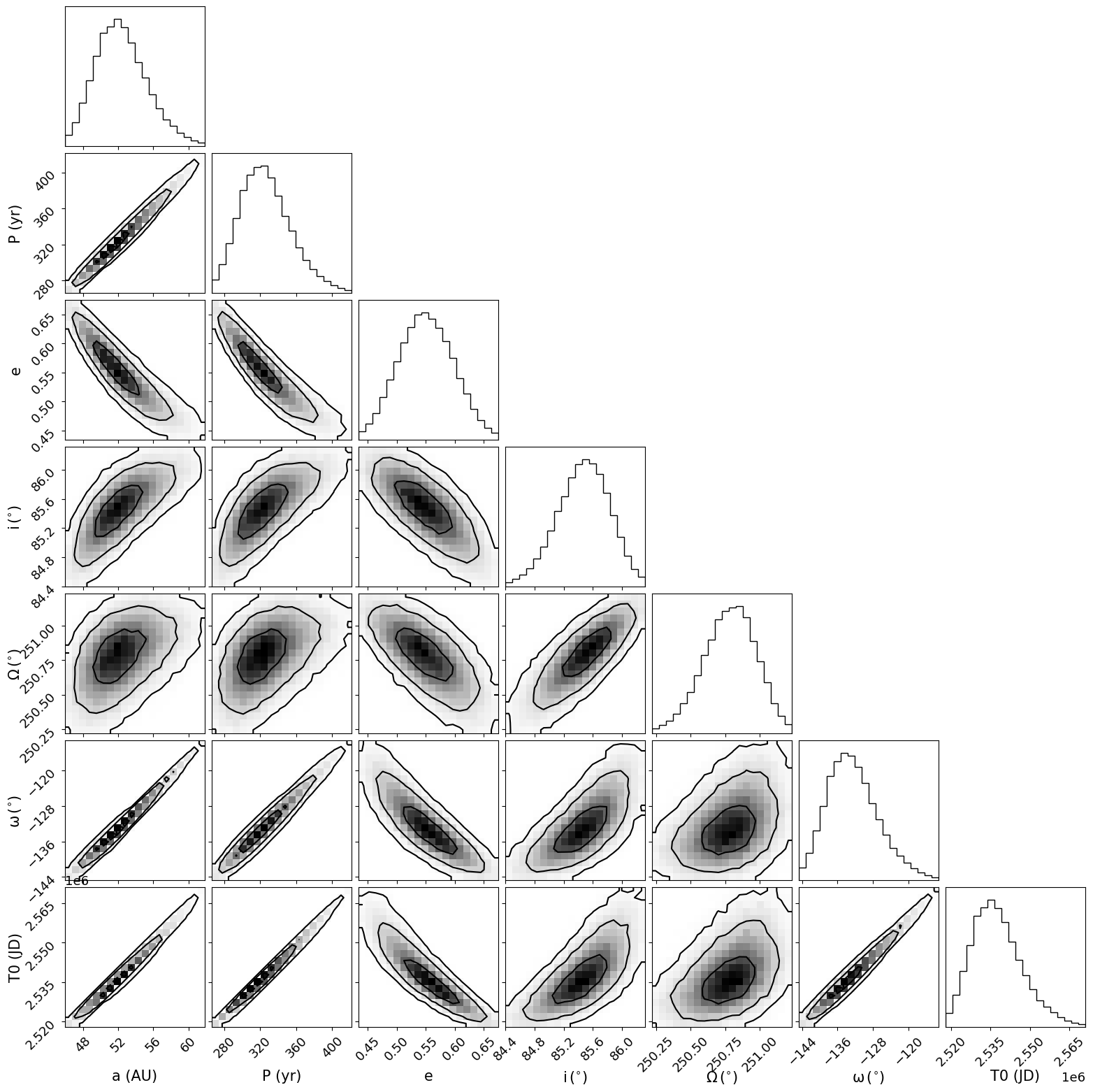}
    \captionof{figure}{Posteriors from our \textsc{orvara} orbital fit for the visual components of KOI-3158. Details about each parameter, including credible intervals and the best-fit values of these parameters, are listed in Table \ref{tab:orbits}.}
    \label{fig:corner_3158}
\end{minipage}

\clearpage

\noindent\begin{minipage}{\linewidth}
    \centering
    \captionsetup{type=figure}
    \includegraphics[width=1\linewidth]{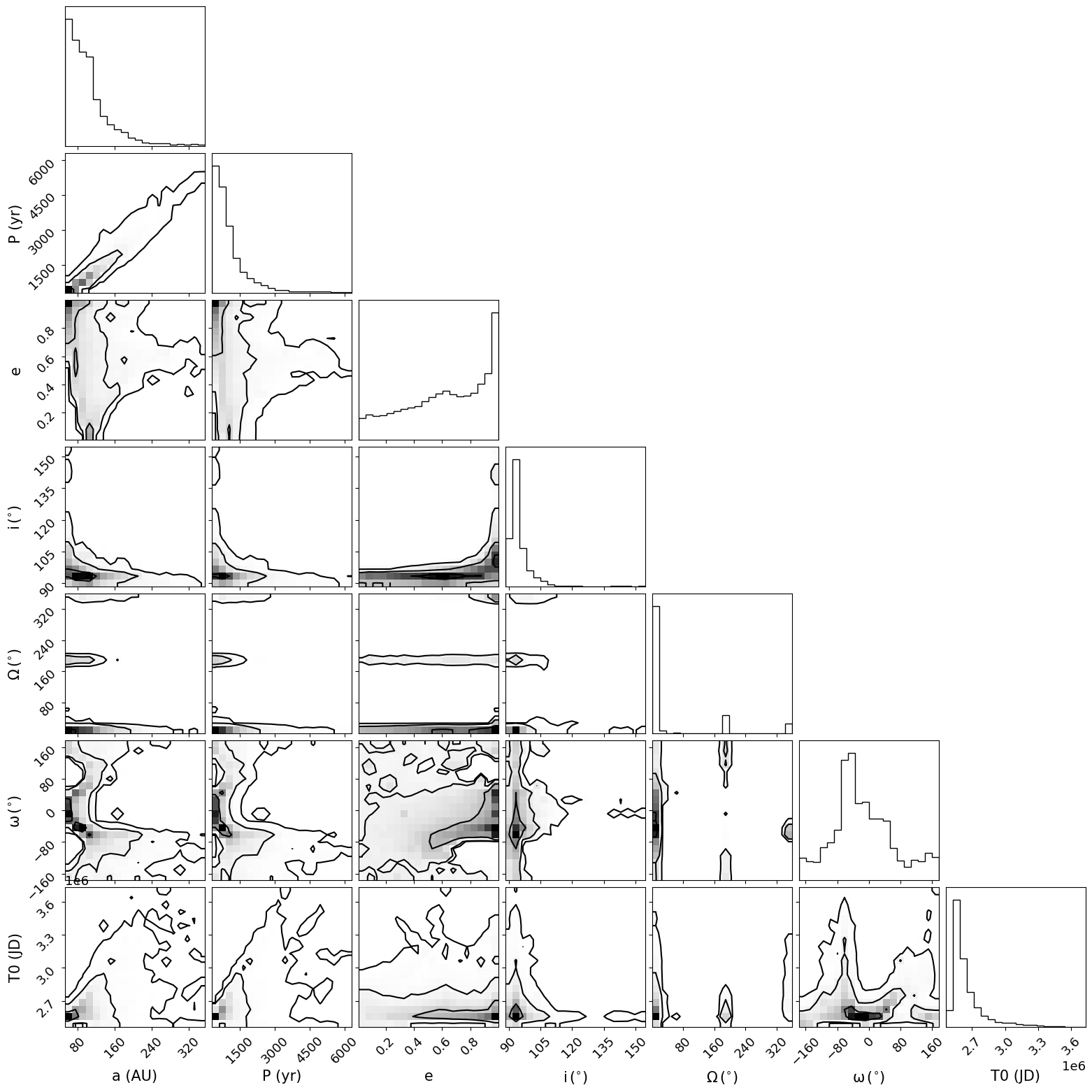}
    \captionof{figure}{Posteriors from our \textsc{orvara} orbital fit for the outer companion relative to the inner binary of KOI-3444. Details about each parameter, including credible intervals and the best-fit values of these parameters, are listed in Table \ref{tab:orbits}.}
    \label{fig:corner_3444}
\end{minipage}

\clearpage

\noindent\begin{minipage}{\linewidth}
    \centering
    \captionsetup{type=figure}
    \includegraphics[width=1\linewidth]{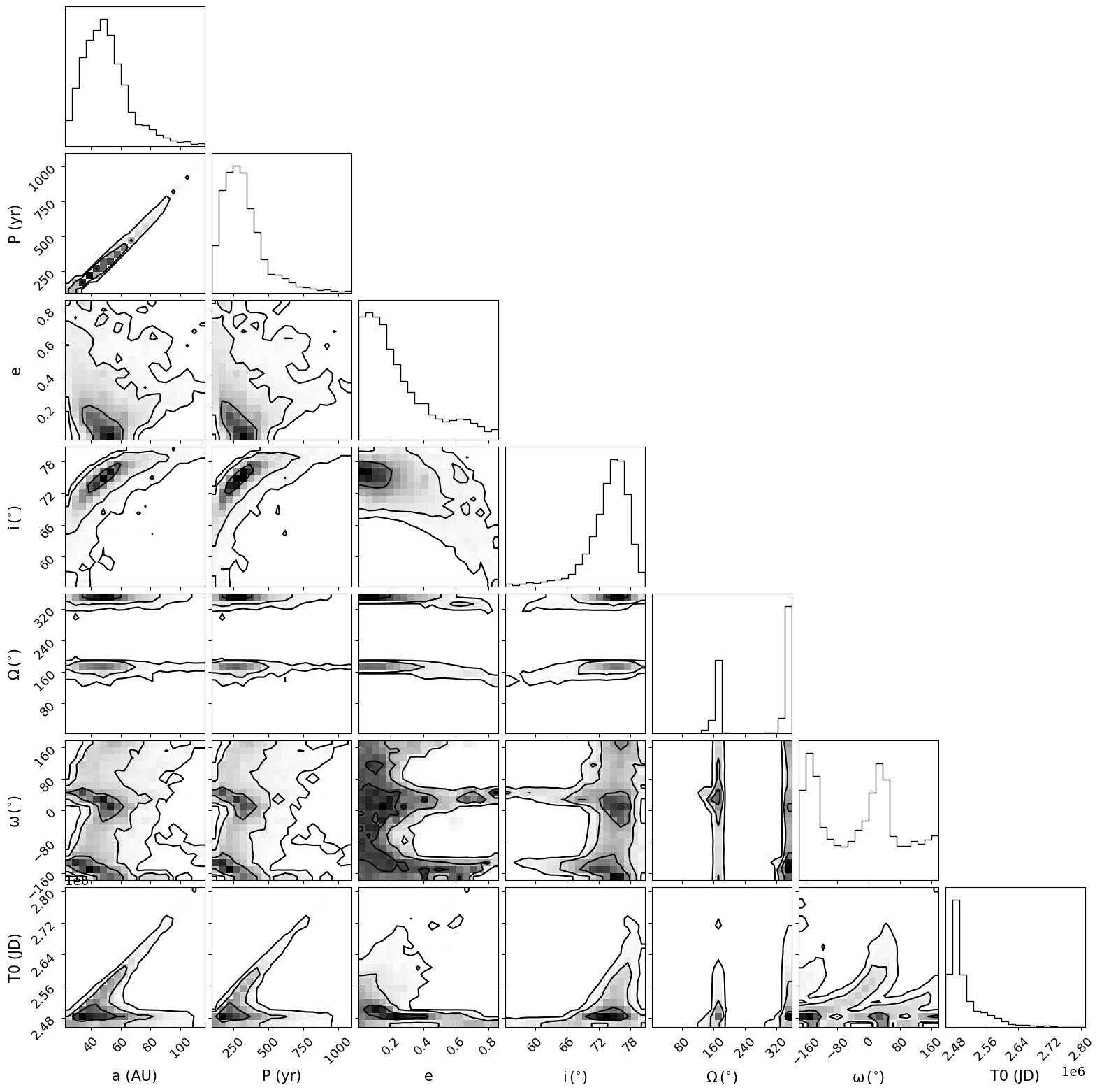}
    \captionof{figure}{Posteriors from our \textsc{orvara} orbital fit for the inner binary of KOI-3497. Details about each parameter, including credible intervals and the best-fit values of these parameters, are listed in Table \ref{tab:orbits}.}
    \label{fig:corner_3497}
\end{minipage}

\bsp	
\label{lastpage}
\end{document}